\documentclass[a4paper,12pt,numbered,print,custombib,twoside]{Classes/PhDThesisPSnPDF}

\input{Preamble/preamble}
\title{Some theoretical and experimental aspects of axion physics}

\author{Albert Renau Cerrillo}

\dept{Departament d'Estructura i Constituents de la Mat\`eria
      \\ \& \\
      Institut de Ci\`encies del Cosmos\\
      }

\university{Universitat de Barcelona\\
      \vspace{2em}
      Advisors:\\
      Dom\`enec Espriu Climent\\
      Federico Mescia}

\crest{\includegraphics[width=0.9\textwidth]{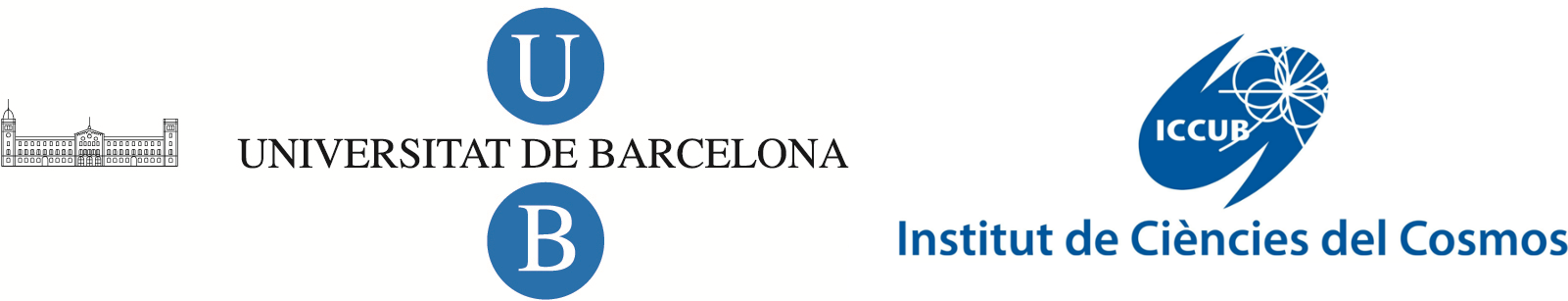}}

\degreetitle{Philosophi\ae~Doctor (PhD) in Physics\\
(Programa de Doctorat en F\'isica)}

\degreedate{November 2015}

\begin{document}

\frontmatter

\begin{titlepage}
  \maketitle
\end{titlepage}


\begin{abstract}

In this thesis we study axions, a byproduct of the Peccei-Quinn solution to the strong CP problem, which are also a viable candidate for the
dark matter content of the Universe.\\

In the first part of the thesis, we revisit the Dine-Fischler-Srednicki-Zhitnisky axion model in light of the recent Higgs LHC results and electroweak precision data. 
This model is an extension of the two-Higgs-doublet model incorporating a PQ symmetry which leads to a physically acceptable axion.
For generic values of the couplings, the model reproduces the minimal Standard Model, with a massless axion and all the other degrees of freedom at a very high scale.
However, in some scenarios, the extra Higgses could be relatively light. 
We use the oblique corrections, in particular $\Delta\rho$, to constrain the mass spectrum in this case.
Finally, we also work out the non-linear parametrization of the DFSZ model in the generic case where all scalars except the lightest Higgs and the axion have 
masses at or beyond the TeV scale.\\

In the second part, we study the relevance of a cold axion background (CAB) as a responsible for the dark matter in the Universe. 
We examine indirect consequences of its presence through its effects on photon and cosmic ray propagation. 

First, we study the axion-photon system under the joint influence of two backgrounds: an external magnetic field and a CAB. 
Their effect consists in producing a three-way mixing of the axion with the two polarizations of the photon. 
We determine the proper frequencies and eigenvectors as well as the corresponding photon ellipticity and induced rotation of the polarization plane 
that depend both on the magnetic field and the local density of axions. 
We also comment on the possibility that some of the predicted effects could be measured in optical table-top experiments.

Then, we consider the case in which no magnetic field is present. Here, circularly polarized photons are energy eigenstates, with a modified dispersion relation.
This enables the emission of a photon by a charged particle, such as a cosmic ray, which is forbidden in regular QED due to energy-momentum conservation. We study
the energy loss of a cosmic ray due to this process and compute the energy flux of photons emitted in this way, which depends on the cosmic ray spectrum.

\end{abstract}

\begin{dedication} 

Als meus pares i germana.\\

\vspace*{2em}

I a la memòria del meu avi. Tant de bo fos aquí per veure-ho.

\end{dedication}


\begin{acknowledgements}      

Voldria donar les gràcies als meus directors de tesi, Domènec Espriu i Federico Mescia, per donar-me l'oportunitat de fer recerca en aquest camp i guiar-me durant
el procés.
També vull agrair al Javier Redondo la seva ajuda i les converses mantingudes, llàstima que no hagin estat més llargues.
Gràcies a tots els professors que he tingut durant la carrera i el màster, en especial al Pere Serra i al Josep Taron, 
per transmetre'm, tots ells, la seva manera d'entendre la Física. 
I també al Ramon Sala, el meu professor de l'institut, per encendre la flama. \\

\noindent Gràcies a tothom amb qui he compartit aquest camí. Heu fet que sigui inoblidable. Als meus companys de despatx Xumeu i Vicente, per fer que les tardes hagin estat
una mica més agradables i per prestar-me els seus inestimables talents pictòrics, ja sigui en diagrames de Feynman o de gravitació.
A les nenes, Eli, Marina i Carla, per recordar-me que no tot és treballar.
A l'Adriana, per intentar, amb molta insistència i poc encert, fer-me veure el costat bell de les coses.
Als \emph{sexy lunchers}, per ser allà cada migdia i per les enriquidores converses sobre els temes importants de la vida.
I gràcies als meus companys de carrera, Blai, Blanca, Blai, Axel i Cris, per tots els moments viscuts durant aquest trajecte.\\

\noindent Gràcies a tots els meus companys de futbol sala, per aportar equilibri a la meva vida. \emph{Mens sana in corpore sano}, que diuen. 
Gràcies també als membres de la banda, la camerata i l'orquestra, per aguantar amb admirable estoicisme el so del meu clarinet baix.  \\

\noindent Gràcies als meus pares, Antonio i Rosario, pel seu suport silenciós i per contribuir a formar la persona que sóc avui en dia.
I a ma germana, Marina, amb qui comparteixo aquella \emph{intuición científica que se hereda de la madre}. \\

\noindent I, finalment, gràcies a qui ha estat al meu costat durant tot aquest temps, tant quan ha estat fàcil com quan ha estat difícil.
\emph{Who would've thought that dreams come true?}

\end{acknowledgements}

\tableofcontents


\mainmatter

\graphicspath{{1-introduction/figures/}{figures/}}

\chapter{Introduction}\label{chap:intro}

The Standard Model of particle physics is a quantum field theory of subatomic particles and their interactions. 
It describes the strong and weak nuclear forces as well as electromagnetism through the gauge principle: a continuous global symmetry group is made local (``gauged'') 
by introducing additional fields, which serve as mediators for the interactions among particles. The particular symmetry group of the Standard Model
is $SU(3)\times SU(2)\times U(1)$.\\

In its current form, the Standard Model contains 17 fundamental particles and their corresponding antiparticles.
Some particles are different from their antiparticles, while the rest are their own antiparticle.
Twelve are fermions with spin $1/2$. Half of them carry colour charge (so they feel the strong nuclear force) and are called quarks. The up, charm and top quarks
have electrical charge $+2/3$ while the down, strange and bottom quarks have $-1/3$ charge. The other half of the fermions do not feel the strong interaction and are
known as leptons. Three leptons have charge $-1$. They are kown as the electron, the muon and the tauon. The remaining leptons are known as neutrinos and, having no
electrical charge, only feel the weak nuclear force.

Next are four gauge bosons, which are carriers of forces and have spin one. The gluon, responsible for the strong nuclear force, is massless and binds quarks
together to form hadrons. Photons are the gauge bosons of electromagnetism and are also massless. A fundamental
difference between photons and gluons is that gluons carry colour, while photons are electrically neutral. This is so because the strong interaction is based
on the symmetry group $SU(3)$, which is non-commutative, while $U(1)$, the symmetry group of electromagnetism, is commutative. The weak nuclear force is mediated by
a pair of massive bosons: $W$ and $Z$, with electrical charges $1$ and $0$, respectively. The weak nuclear force is best known for its role in radioactive
nuclear decay, which only needs the $W$ boson. The neutral Z boson plays its part, for instance, in the interaction between two neutrinos.

The last particle is known as the Higgs boson. It has no spin and is crucial in understanding why some of the previous particles have mass at all. Indeed, gauge symmetry
requires massless gauge bosons, as a mass term breaks the symmetry. Morevover, the $SU(2)$ part of the Standard Model symmetry group is also broken by fermion masses.
However, if the symmetry is spontaneously broken by the vacuum expectation value of a scalar field,
a mass term is generated without explicitly breaking any symmetries. This is known as the Brout-Englert-Higgs mechanism~\cite{higgs,be}.\\

Despite the enormous theoretical and experimental success of the Standard Model, it still leaves some questions unexplained. These are a few examples:
\begin{itemize}
 \item{It does not incorporate gravity.}
 \item{Neutrinos are described as having no mass. However, experiments point out that there is a difference in their masses. Therefore, they can not all be massless.}
 \item{Matter is much more abundant than antimatter, yet the Standard Model is unable to explain this asymmetry.}
 \item{The strong CP problem: why does QCD not break the CP symmetry?}
 \item{The standard cosmological model requires the existence of dark matter and dark energy. As will be discussed, the Standard Model of particle physics provides
 no good candidate to explain these phenomena.}
\end{itemize}
This thesis is concerned with the last two points, as we entertain the notion that axions, a possible solution to the strong CP problem, are also responsible for the 
dark matter content of the Universe. Any discussion of dark matter involves some cosmological considerations. Next, we review the essential facts.\\

The cosmological principle states that our Universe is homogeneous and isotropic. This restricts its metric to the Friedmann-Robertson-Walker form~\cite{weinberg}:
\be\label{frw}
ds^2=dt^2-a(t)^2\left[\frac{dr^2}{1-kr^2}+r^2\left(d\theta^2+\sin^2\theta d\phi^2\right)\right].
\ee
This metric is written in comoving coordinates. The proper distance between two points held at constant $(r,\theta,\phi)$ is proportional to the scale factor $a(t)$.
The parameter $k=+1,0,-1$ distinguishes between different spatial geometries: 
\begin{itemize}
 \item {A universe with $k=+1$ is spatially closed. It has positive curvature, like a sphere.}
 \item {A universe with $k=0$ is spatially flat. It has no curvature, like a plane.}
 \item {A universe with $k=-1$ is spatially open. It has negative curvature, like a hyperboloid.}
\end{itemize}

By examining the Doppler shift of light from extragalactic objects, it can be deduced that they are moving away from us, a fact known as Hubble's law~\cite{hubble}. This
is interpreted as a global expansion of the Universe: an increase of $a(t)$ over time.
The rate of expansion of the Universe is known as the Hubble parameter
\be\label{hubble}
H\equiv\frac{\dot a}a.
\ee
By using Einstein's equations of General Relativity on the metric described in Eq.~\eqref{frw} we obtain Friedmann's equations, which describe the evolution 
of the scale factor
\begin{align}
\left(\frac{\dot a}{a}\right)^2&=\frac{8\pi G}{3}\sum_i\rho_i-\frac{k}{a^2},\\
\frac{\ddot a}{a}&=-\frac{4\pi G}{3}\sum_i(\rho_i+3p_i),\label{feq2}
\end{align}
where the sum runs over the different contributions to the energy density. Each of these contributions has a particular relation between its density and its pressure,
described by its equation of state
\be
p=w\rho.
\ee
The equation of state is characterized by the dimensionless number $w$. 
Non-relativistic matter has $w=0$. Its energy density depends on the scale parameter as $\rho\propto a^{-3}$.
For ultrarelativistic matter (radiation or very fast massive particles) we get instead $w=1/3$ and its energy density goes as $\rho\propto a^{-4}$. This different behaviour is
easily understood. For non-relativistic particles, their total number is constant as volume changes, so they are simply diluted over a larger volume $V=a^3$. For radiation,
we have the additional effect of red-shift: its wavelength expands with the scale factor, so its momentum decreases by an additional power of $a$. 

By examining Eq.~\eqref{feq2} we can see that expansion can be accelerated ($\ddot a>0$) for sufficiently negative pressures. Any component with $w<-1/3$ contributes to
accelerating the expansion of the Universe. These components are collectively known as dark energy. A cosmological constant, a concept first introduced 
by Einstein to ensure a static universe, is a form of dark energy with $w=-1$, 
for which the energy density remains constant as $a(t)$ changes. The case $w<-1$ is known as phantom energy~\cite{phantom} and its energy density actually 
increases with the scale factor. This has dramatic effects, as expansion is eventually so strong that it starts destroying objects if phantom energy dominates. 
First, gravitationally bound systems are dissociated, but at later times even molecules and atoms are torn apart. 
Finally, the scale factor becomes infinite in a finite time, a phenomenon known as the Big Rip~\cite{bigrip}.

\section{Dark matter}

Recent measurements have revealed that ordinary matter\footnote{Since the majority of ordinary matter is in the form of protons and neutrons, the terms ordinary matter and
baryonic matter will be used indistinctively.} (the kind of matter described by the Standard Model) constitutes only 5\% of the total energy in the Universe. 
Another 27\% is in the form of dark matter, which does not significantly interact with ordinary matter, except gravitationally. 
The remaining 68\% is dark energy. Because radiation dilutes faster than the other components, it does not contribute a significant amount of energy at the present time,
although in the past it was more important.

To quantify the amount of these components, the density parameter
\be\label{densityparam}
\Omega_X=\frac{\rho_X}{\rho_c}
\ee
is defined, where $\rho_X$ is the density of a certain component $X$ (which can be ordinary matter, dark matter or dark energy) and $\rho_c$ is the critical density 
\be
\rho_c=\frac{3H^2}{8\pi G}.
\ee
The comparison between the total density and the critical density determines the spatial geometry of the Universe. 
If a universe has density larger than the critical value, it is spatially closed ($k=+1$). If the density is smaller than the critical
density, is spatially open ($k=-1$). In the limiting case, $\rho=\rho_c$, the universe is spatially flat ($k=0$).

The combination $\Omega_Xh^2$ is frequently used, where $h$ is the reduced Hubble 
parameter, defined by $H_0=100 h (\rm km/s)/\rm Mpc$.
The quantity $H_0$ is the current value of the Hubble parameter, which has been defined in Eq.~\eqref{hubble}.\\

The nature of dark matter is still unknown. It does not emit nor absorb light at any significant level, so it can not be detected directly.
Its existence is inferred by its gravitational effects: large astronomical objects behave like they contain more mass than can be calculated from its observable parts,
such as stars or interstellar gas~\cite{halo}. Although the most accepted explanation is the existence of dark matter, alternative explanations, 
such as Modified Newtonian Dynamics~\cite{mond}, have been proposed. 

\subsection{Evidences}

Numerous observations support the existence of dark matter. These are some examples.

\subsubsection{Galactic rotation curves}

The earliest evidence for the existence of dark matter came from the observation that various luminous objects move differently than they should, given 
the gravitational attraction of surrounding visible matter~\cite{freeman}. A clear example of this is found in galactic rotation curves. 
An object moving in a stable circular orbit of radius $r$ has a velocity
\be
v(r)=\sqrt{\frac{GM(r)}r},
\ee
where  $M(r)$ is the mass enclosed inside the orbit. Stars in a spiral galaxy follow nearly circular orbits. Thus, the velocity of stars that lie in the external 
region of a galaxy should fall off as $v(r)\propto1/\sqrt r$, because $M(r)$ should be approximately constant.
However, this is not the case for most galaxies: the velocity profile $v(r)$ becomes approximately flat for large $r$.

A solution to this discrepancy is the existence of a ``dark halo'' of matter that can not be seen, with density $\rho(r)\propto1/r^2$, so that $M(r)\propto r$ and $v(r)$ 
is constant\footnote{Of course, at some point $\rho(r)$ has to fall off faster than $1/r^2$, so the total mass of the galaxy is finite.}.\\

\subsubsection{Gravitational lensing}

According to General Relativity, light is affected by gravity, despite having no mass~\cite{einstein}. 
When light passes near a massive object its path is bent, and the apparent position of the source that produced it is distorted. This can lead to the phenomenon known
as gravitational lensing~\cite{lens}.
If a very massive object (the lens) is located between an observer and a luminous object (the source),
light rays coming from the source will be bent by the lens and will reach an observer as if coming from a different position.
The angular separation between the different images is~\cite{lens, einstein}
\be
\theta=2\sqrt{4GM\frac{d_{LS}}{d_Ld_S}},
\ee
where $M$ is the mass of the object acting as the lens, $d_{LS}$ is the distance from the lens to the source and
$d_L$, $d_S$ are the distances from the observer to the lens and the source, respectively.

Years ago, two identical astrophysical objects where observed, only 5.6 arc seconds apart. It was then concluded that they were in fact the same object, 
seen twice because of lensing. 
The mass of the lensing object can be calculated using this expression and then it can be compared to the mass inferred from its luminosity, which is found to be smaller.
Again, a solution is to assume that there is some dark matter in the volume of the object acting as the lens, which provides the missing mass.\\

A particularly compelling case is found in the Bullet Cluster (1E 0657-56). It consists of two clusters of galaxies that have undergone a collision. Observation at different
wavelengths reveals that stars behaved differently from interstellar gas during the collision. Interstellar gas, observed in X-rays, interacts electromagnetically and is 
slowed significantly more than stars, observed in visible light, which only interact gravitationally. Although most of the luminous mass of the Bullet Cluster is in the
form of hot interstellar gas, most of the mass contributing to gravitational lensing is found to be separated from the gas, following the stars
instead. This fact not only gives evidence of the existence of dark matter, but also reveals its collisionless behaviour. Therefore, dark matter self-interactions must be
very weak~\cite{bullet}.

\subsubsection{Cosmic Microwave Background}

The Cosmic Microwave Background (CMB) is radiation coming from the time of recombination\footnote{Recombination refers to the event when 
electrons and protons combine to form (neutral) hydrogen, allowing photons to travel freely.} that fills our Universe. 
It follows a thermal, black-body spectrum  with an average temperature  $\langle T\rangle=2.7$~K. 
However, the temperature of different regions of the sky shows small variations of one part in $10^5$.
These anisotropies are studied in terms of the temperature fluctuation
\be
\Theta(\hat n)=\frac{T(\hat n)-\langle T\rangle}{\langle T\rangle},
\ee
where $\hat n=(\theta,\phi)$ is a direction in the sky. 
As they are defined on a two-dimensional spherical surface, temperature fluctuations are best described in terms of spherical harmonics
\be
\Theta(\hat n)=\sum_{l=0}^{\infty}\sum_{m=-l}^{l}a_{lm}Y_{lm}(\hat n)
\ee
From the coefficients in this expansion we can define the power spectrum
\be
C_l=\frac{1}{2l+1}\sum_{m=-l}^{l}|a_{lm}|^2.
\ee
A plot the $C_l$ coefficients as a function of $l$ shows peaks and valleys. A lot of information is codified in their position and height. 
In particular, the amount of baryonic matter (b) and the total amount of matter (m) can be extracted
\be
\Omega_{\rm b}h^2=0.0226,\qquad \Omega_{\rm m}h^2=0.133.
\ee
Since these numbers do not match, baryonic matter must be only a fraction of the total amount of matter. There is mass in the form of non-baryonic dark matter (dm), 
with density parameter
\be\label{dmdensity}
\Omega_{\rm dm}h^2=0.112.
\ee

\subsection{Dark matter candidates}

Structure formation (the formation of galaxies and clusters of galaxies) depends on whether dark matter is hot or cold~\cite{lss},
as structures smaller than the free-streaming length of the particles involved can not form gravitationally.

Hot dark matter refers to ultrarelativistic particles, whose momentum is much larger than their mass. It leads to top-down structure formation: large 
structures such as superclusters are created first and later are fragmented into smaller components, like galaxies.

On the other hand, cold dark matter is comprised of slow-moving particles and leads to bottom-up structure formation. 
Smaller objects are formed first and then group together to form larger structures. This is the paradigm that shows a better agreement with observations.\\

A successful dark matter candidate should be electrically neutral (or it would interact with radiation and not be dark at all) and its self-interactions should be small, 
since dark matter is essentially collisionless. It also should be stable or have a very long lifetime.

Among particles in the Standard Model, neutrinos are the only possibility. However, their density parameter can be estimated to be~\cite{dodelson}
\be
\Omega_\nu h^2=\frac{\sum_\nu m_\nu}{93\text{ eV}},
\ee
where $\sum_\nu m_\nu$ is the sum of the masses of the three mass eigenstates. Tritium $\beta$-decay experiments~\cite{weinheimer} place the bound
\be
m_\nu<2.05\text{ eV},
\ee
which applies to all three eigenvalues. Thus, their density parameter has an upper bound
\be
\Omega_\nu h^2<\frac{3\cdot2.05\text{ eV}}{93\text{ eV}}=0.066,
\ee
which is insufficient to satisfy Eq.~\eqref{dmdensity}.
Moreover, neutrinos are relativistic and constitute hot dark matter. The contribution of hot dark matter can not be too large to comply with observations. These
cosmological considerations give even more stringent bounds~\cite{thomas}.\\

Having no viable candidate within the Standard Model, dark matter must come from Beyond the Standard Model physics. There are numerous possibilities, 
so we will comment only on a few.

\subsubsection{Sterile neutrinos}

Although in the Standard Model they are described as massless, we know that neutrinos have a tiny mass, since they oscillate~\cite{conchamaltoni}. 
A common way to give them mass involves expanding the Standard Model to include right-handed neutrinos. Because neutrinos only interact through the weak force, and it 
affects only the left-handed component of particles, right-handed neutrinos are said to be sterile.
They are massive particles and can provide a form of warm dark matter (an intermediate case, between hot and dark matter).

\subsubsection{Weakly Interacting Massive Particles}

Weakly Interacting Massive Particles (WIMPs) are particles in the $100$ GeV or higher mass range and are among the most popular dark matter candidates. 

Their present density, $\Omega_\chi$, is due to the leftover density from the time when they fell out of thermal equilibrium with the hot plasma~\cite{kolb}
\be
\Omega_\chi h^2\simeq\frac{0.1\text{ pb}}{\langle\sigma_Av\rangle},
\ee
where $\sigma_A$ is their self-annihilation cross-section. This density turns out to be the correct one (cf. Eq.~\eqref{dmdensity}) when $\sigma_A$ is similar 
to that of particles interacting via the weak nuclear force (hence the name WIMP). This fact is known as the ``WIMP miracle''.\\

Supersymmetry is an example of a Beyond the Standard Model theory in which a WIMP can be found~\cite{susydm}. It states that for each particle in the Standard Model
there is an accompanying particle, with spin differing in $1/2$, known as its superpartner. If supersymmetry is exact, superpartners have exactly the same mass as the 
corresponding SM particle. If it is broken, however, they can have much larger masses.

A supersymmetric theory can have a $\mathbb{Z}_2$ symmetry called R-parity. Standard Model particles have R-parity of +1, while their superpartners have
R-parity -1. This implies that a process with an even (odd) number of supersymmetric particles in the initial state must have an even (odd) number of such particles
in the final state. As a consequence, the Lightest Supersymmetric Particle (LSP) has no lighter particles to decay into, so it must be stable.

In the Minimal Supersymmetric Standard Model, the superpartners of the neutral gauge bosons and the Higgs mix to form neutral mass eigenstates, called neutralinos.
In a substantial region of parameter space, the lightest of the neutralinos turns out to be the LSP. Being neutral, it is a good dark matter candidate.\\

Another example of WIMP comes from theories with Universal Extra Dimensions~\cite{ued}, which have additional spatial dimensions beyond the usual three. 
These extra dimensions are compactified and as a result there appear resonances of the Standard Model 
fields, separated in mass by a scale related to the inverse compactification radius. These are known as Kaluza-Klein (KK) towers of states.

Momentum conservation in the extra dimensions leads to conservation of KK number. In the same vein as with R-parity, this KK-parity implies the stability of the
Lightest Kaluza-Klein Particle (LKP). If the LKP is neutral (for example, it belongs in the KK tower of the photon or the neutrino), it can be a good dark matter candidate.

\subsubsection{Axions and other light particles}

A new light neutral scalar or pseudoscalar boson is also a possibility. It can arise when a $U(1)$ symmetry is spontaneously broken, which produces a massless
particle with no spin, known as a Goldstone boson~\cite{goldstone}. If the symmetry is only approximate, the Goldstone boson acquires a small mass (and is known as
pseudo-Goldstone boson).

The most popular of these light dark matter candidates is the axion, which appears as a byproduct of the solution to the Strong CP Problem proposed 
by R. Peccei and H. Quinn in 1977~\cite{PQ1}. It is a very light pseudoscalar particle, the pseudo-Goldstone boson of a spontaneously broken chiral $U(1)$ symmetry.
Despite having a very small mass, axions produced non-thermally constitute cold dark matter, as they have very small momentum, a possibility known as 
the cold axion background.

The rest of this Chapter is devoted to axions. In Sec.~\ref{sec:cpproblem} we describe in detail the strong CP problem and see how axions provide a solution to it.
Next, in Sec.~\ref{sec:axions} we examine the properties of the axion particle and explore different specific models in which it appears. We also describe how a cold axion
background can be the dark matter in the Universe and explain current constraints on axion parameters. Finally, we review some experiments that look for axions.

\section{The strong CP problem and a solution}\label{sec:cpproblem}

Axions appear in one of the solutions to the strong CP problem: the fact that, apparently, quantum chromodynamics (QCD) does not break the CP symmetry.
This section will be concerned with the description of this problem and its possible solutions.

\subsection{The chiral anomaly}

The QCD Lagrangian density for N quark flavours is~\cite{peskin}
\be\label{qcdlag}
\mathcal{L}_{\rm QCD}=-\frac14G^a_{\mu\nu}G^{a\mu\nu}+\sum_{j=1}^N\bar q^j(\gamma^\mu D_\mu-m_j)q^j,
\ee
where $q_j$ are the quark fields and $G^a_{\mu\nu}$ is the gluon field-strength tensor.
In the limit of vanishing quark masses, $m_j\rightarrow0$, this Lagrangian has a global $U(N)_V\times U(N)_A$ symmetry. Since $m_u,m_d\ll\Lambda_{\rm QCD}$,
sending the masses of the up and down quarks to zero should be a good limit, so we could expect $U(2)_V\times U(2)_A$ to be a good approximate symmetry.

Experimentally, one sees that the vector subgroup $U(2)_V=SU(2)_I\times U(1)_B$ (isospin times baryon number) is indeed a good approximate 
symmetry because of the appearance of multiplets in the hadron spectrum: pions have approximately the same mass, and the same happens with nucleons.
The axial symmetry, however, is spontaneously broken by the non-vanishing value of the QCD quark condensate, $\langle\bar qq\rangle\neq0$.
According to the Goldstone theorem, for any spontaneously broken symmetry in a relativistic theory 
there will be as many massless particles (Goldstone bosons) as broken symmetry generators.
However, the quark mass terms explicitly break axial symmetry, and the pseudo-Goldstone bosons acquire a mass. Nevertheless, this mass is small 
compared to the rest of the masses in the spectrum.\\ 

$U(2)_A=SU(2)_A\times U(1)_A$ has 4 (broken) generators\footnote{$U(N)_A$ is not a subgroup of $U(N)_V\times U(N)_A$ (although $U(N)_V$ is). 
However, we can separate the generators into those of $U(N)_V$ and those of $U(N)_A$. It is in this sense that we say  $U(N)_A=SU(N)_A\times U(1)_A$ },
so we should expect an equal number of pseudo-Goldstone bosons. Although the three pions ($\pi^0$ and $\pi^\pm$) are light, 
the next particle in the hadron spectrum, the $\eta$ meson, is much heavier. 

This situation persists if we include the strange quark with the light quarks ($m_u,m_d,m_s\approx0$). 
Now the symmetry group is $U(3)_V\times U(3)_A$ (it is a worse symmetry than $U(2)_V\times U(2)_A$, because $m_s$ generates larger breakings than $m_u,m_d$). 
The group is spontaneously broken to $U(3)_V$, so there are 9 broken generators. We have 8 ``light'' mesons: $\pi^0$, $\pi^\pm$, $K^\pm$, $K^0$, $\bar K^0$ and $\eta$. 
The next particle, the $\eta'$, is again much heavier. We are still missing a pseudo-Goldstone boson.\\ 

The resolution of this apparent problem consists in realising that $U(1)_A$ is not an actual symmetry of the theory. 
Although in the massless quark limit the QCD Lagrangian~\eqref{qcdlag} is invariant under a $U(1)_A$ transformation
\be\label{chirot}
q_j\to e^{i\alpha\gamma_5/2}q_j
\ee
the divergence of the associated current, $J_5^\mu$ gets quantum corrections from the triangle diagram and turns out to be non-vanishing~\cite{adler}
\be\label{anomaly}
\partial_\mu J_5^\mu=\frac{g_s^2N}{32\pi^2}G_a^{\mu\nu}\tilde G_{a\mu\nu},
\ee
where the dual field-strength tensor is
$\displaystyle\tilde G_{a\mu\nu}=\frac12\epsilon_{\mu\nu\alpha\beta}G_a^{\alpha\beta}$. The antisymmetric Levi-Civita symbol is $\epsilon_{\mu\nu\alpha\beta}$ and,
in four-dimensional Minkowski space, $\epsilon_{0123}=-\epsilon^{0123}=+1$.

Although the r.h.s. of Eq.~\eqref{anomaly} can be written as total divergence, in QCD there are vacuum field configurations, called instantons,
which make its integral non-zero and thus the corresponding charge is not conserved. 
Therefore, the actual approximate symmetry is just $SU(2)_A$ (or $SU(3)_A$) which leads to three (or eight) pseudo-Goldstone bosons
when it is spontaneously broken: the three pions (or the eight light mesons mentioned earlier).\\

\subsection{Strong CP problem}

The existence of the vacuum field configurations mentioned in the previous section complicates the structure of the QCD vacuum~\cite{thetavacuum}. 
For our purposes it is sufficient to know that it effectively adds an additional term to the Lagrangian of QCD, called the $\theta$-term:
\be
\mathcal{L}_\theta=\theta\frac{g_s^2}{32\pi^2}G_a^{\mu\nu}\tilde G_{a\mu\nu}.
\ee
This term respects the symmetries of QCD, but violates time reversal (T) and parity (P), while conserving charge conjugation (C). 
Therefore, it violates CP and induces a neutron electric dipole moment (EDM) of order~\cite{nedm}
\be
d_n\simeq\frac{e\theta m_q}{m_N^2}\simeq10^{-16}\theta \,e\,\text{cm}.
\ee
It is proportional to both $\theta$ and the quark mass, since CP violation disappears if either vanishes (more on this at the end of this section). An EDM has 
dimensions of length (or $\text{mass}^{-1}$), which are provided by the nucleon mass squared in the denominator.

The experimental bound on the neutron EDM~\cite{baker} implies a bound on the $\theta$ parameter:
\be\label{nedm}
|d_n|<2.9\cdot10^{-26} e\,\text{cm}\quad\longrightarrow\quad\theta<10^{-9}.
\ee
Here lies the strong CP problem: why is this parameter, in principle arbitrary, so small?

Apart from the QCD term, the $\theta$ parameter gets an additional contribution coming from the quark masses. Under a chiral 
rotation~\eqref{chirot} of angle $\alpha/2$, the action for just one quark species changes as
\be\label{chitrans}
\int d^4x\left[-m\bar qq-\theta\frac{g_s^2}{32\pi^2}G\tilde G\right]\longrightarrow
\int d^4x\left[-m\bar qe^{i\alpha\gamma_5}q-(\theta-\alpha)\frac{g_s^2}{32\pi^2}G\tilde G\right],
\ee
where we have adopted the shorthand $G\tilde G\equiv G_a^{\mu\nu}\tilde G_{a\mu\nu}$. A contribution to the $\theta$-term proportional to $\alpha$ has appeared, 
due to the anomaly~\eqref{anomaly}. Now consider all quark flavours. Their mass term reads
\be
\mathcal{L}_{\text{mass}}=-\bar q_{iR}M_{ij}q_{jL}+\text{h.c.},
\ee
where the $M_{ij}$ are, in general, complex. A physical basis for the quarks is one in which the mass matrix is diagonal and real, which can be achieved in two steps.
First, we perform a chiral $SU(N)$ rotation to bring the matrix to a diagonal form. This step gives rise to the CKM matrix of the weak interactions but, since there is no anomaly
associated to $SU(N)$, it does not affect the $\theta$-term\footnote{The CKM matrix is complex and its phases can not be removed by a chiral $U(1)$ transformation.
It provides CP violation through the Jarlskog invariant~\cite{jarl}.}. 

Once the mass matrix is diagonal (but still complex in general), its coefficients can be made real with a $U(1)_A$ transformation~\eqref{chirot} for each quark field, with angle 
$\alpha_j=-\beta_j$, where $\beta_j$ is the phase of the corresponding coefficient. These transformations are anomalous, as seen in Eq.~\eqref{chitrans} and they give a
contribution to the $\theta$-term equal to $-\sum_j\alpha_j=\sum_j\beta_j=\arg\det M$.

Therefore, the final coefficient of the $G\tilde G$ term is
\be
\bar\theta=\theta+\arg\det M.
\ee
The strong CP problem is a fine-tuning problem: why do these two apparently unrelated contributions combine to give such a small number?

It should be noted that if at least one quark is massless the problem disappears. Indeed, if there is no mass term, we can perform a chiral rotation on the massless quark
to compensate all the other contributions. However, the ratio $m_u/m_d\sim0.5$, coming from Lattice QCD~\cite{pdg}, seems well established. Therefore, no quark is massless.

\subsection{A solution to the problem}

A natural solution to the strong CP problem is to introduce a global chiral $U(1)$ symmetry~\cite{PQ1,PQ2,PQ3}, known as $U(1)_{\text {PQ}}$, for R. Peccei and H. Quinn.
This symmetry is spontaneously broken when one or more scalar fields develop vacuum expectation values (vevs). 
The associated Goldstone boson is called the axion and, because $U(1)_{\text {PQ}}$ is anomalous, couples to two gluons. 

The minimal ingredients needed are a scalar field ($\phi$) that couples to a quark field ($q$), which can be one of the SM quarks or a different one. 
This coupling can generically be written as 
\be\label{yukpq}
-y\bar q_L\phi q_R+\text{h.c.}.
\ee
In order to be anomalous, $U(1)_{\text{PQ}}$ has to be chiral. For instance, we can choose that only the right component of the quark field transforms. In this case, the
symmetry is
\be
q_R\to e^{-i\alpha}q_R,\quad \phi\to e^{i\alpha}\phi,
\ee
which leaves Eq.~\eqref{yukpq} invariant. Through the minimisation of some potential, $V(\phi)$, the scalar field acquires a vev $\langle\phi\rangle=f_a/\sqrt2$.

At low energies, the terms of the Lagrangian relevant for our discussion are
\be\label{Ltheta+a}
\mathcal L_{\bar\theta+a}=\bar\theta\frac{g_s^2}{32\pi^2}G^{\mu\nu}\tilde G_{\mu\nu}+\frac12\partial_\mu a\partial^\mu a+
\xi\frac{a}{f_a}\frac{g_s^2}{32\pi^2}G^{\mu\nu}\tilde G_{\mu\nu},
\ee
where $\xi$ depends on the specific axion model. Since $a/f_a$ is the phase of the scalar field, $U(1)_{\text {PQ}}$ acts on it as $a\to a+f_a\alpha$. 
The last term of Eq.~\eqref{Ltheta+a} codifies the anomaly of the current associated with $U(1)_{\text {PQ}}$
\be
\partial_\mu J^\mu_{\text {PQ}}=\xi\frac{g_s^2}{32\pi^2}G^{\mu\nu}\tilde G_{\mu\nu}.
\ee
The parameter $f_a$ is called axion decay constant and is analogous to the pion decay constant, $f_\pi$, and is defined 
by $\langle0|J^\mu_{\rm PQ}|a(p)\rangle=ip^\mu f_a$.

The coefficient of the $G\tilde G$ term is now
\be\label{lta}
\left(\bar\theta+\xi\frac{a}{f_a}\right)\frac{g_s^2}{32\pi^2}G\tilde G.
\ee
At low energies, below the QCD transition, this term represents an effective potential for the axion field, with minimum at $\langle a\rangle=-\bar\theta f_a/\xi$. 
Equation~\eqref{lta} written in terms of the physical field $a_\text{phys}=a-\langle a\rangle$ is
\be\label{la2g}
\xi\frac{a_\text{phys}}{f_a}\frac{g_s^2}{32\pi^2}G\tilde G
\ee
and no longer contains $\bar\theta$. Thus, the Peccei-Quinn (PQ) solution to the strong CP problem effectively replaces $\theta$ by the dynamical axion 
field and the Lagrangian no longer has a CP-violating $\theta$-term.

\section{Axions}\label{sec:axions}

The Peccei-Quinn solution to the strong CP problem implies the existence of a pseudoscalar particle, the axion. 
It mixes with the $\pi^0$ and $\eta$ mesons, since it has the same quantum numbers as them, and gets a mass
\be\label{axionmass1}
m_a=\frac{z^{1/2}}{1+z}\frac{f_\pi}{f_a}m_\pi,\quad z=\frac{m_u}{m_d}.
\ee
Using the values $m_\pi=135\text{ MeV}$, $f_\pi=92\text{ MeV}$ and $z=0.56$, we see that the axion mass is 
\be\label{axionmass2}
m_a=6\left(\frac{10^{9}\text{ GeV}}{f_a}\right)\text{ meV},
\ee
inversely proportional to its decay constant.

Regardless of the model, axions always couple to two photons with a term analogous to the $G\tilde G$ coupling of Eq.~\eqref{la2g}
\be\label{prima}
\mathcal L_{a\gamma\gamma}=g_{a\gamma\gamma}\frac\alpha{2\pi}\frac a{f_a}F^{\mu\nu}\tilde F_{\mu\nu}=\frac g4aF^{\mu\nu}\tilde F_{\mu\nu},
\ee
where the electromagnetic field-strength tensor $F_{\mu\nu}$ appears instead of $G^a_{\mu\nu}$. The coupling constant $g_{a\gamma\gamma}$ depends on the axion model, but
is always of $\mathcal{O}(1)$.\\

In this section we will review the different ways to implement the $U(1)_{\text {PQ}}$ symmetry, which lead to different axion models. We will also discuss the 
current experiments searching for axions and see how axions can account for the dark matter component of the Universe.

\subsection{Models}\label{models}

The PQ solution necessitates physics Beyond the Standard Model, as it can not be implemented in the Minimal Standard Model. To see this, consider the Yukawa terms for one
quark generation in the SM
\be\label{yuksm}
\mathcal{L}_{\text{Y}}=-\bar q_LY_d\Phi d_R-\bar q_LY_u\tilde\Phi u_R+\text{h.c.},\quad \tilde\Phi=i\tau_2\Phi^*,
\ee
where $\Phi$ is the Higgs doublet and $\tau_2$ denotes the second Pauli matrix. A chiral transformation would act on these fields as
\be
\Phi\to e^{iX\alpha}\Phi,\quad u_R\to e^{iX_u\alpha}u_R,\quad d_R\to e^{iX_d\alpha}d_R,
\ee
leaving the left-handed fields unchanged.
The different PQ charges $X$, $X_u$ and $X_d$ are chosen so that Eq.~\eqref{yuksm} remains invariant under the transformation. Because $d_R$ couples to $\Phi$ but $u_R$
couples to its conjugate, the charge assignments have to be
\be
X_u=X,\quad X_d=-X.
\ee
Such a transformation does not modify the $\theta$-term, as the contributions from the up and down quarks cancel:
\be\label{cancel}
\bar\theta\frac{g_s^2}{32\pi^2}G\tilde G\to\left[\bar\theta-(X_u+X_d)\alpha\right]\frac{g_s^2}{32\pi^2}G\tilde G=
\left[\bar\theta-(X-X)\alpha\right]\frac{g_s^2}{32\pi^2}G\tilde G.
\ee
In other words, one Higgs doublet contains the physical Higgs particle, plus the three Goldstone bosons that give mass to the gauge bosons. There are not enough degrees
of freedom to accommodate the axion.

If the PQ transformation involves the Standard Model quarks, we need at least two Higgs doublets which transform independently. 
Otherwise, it must involve other new quark fields.

\subsubsection{PQWW}

The original model is known as the Peccei-Quinn-Weinberg-Wilczek (PQWW) axion model~\cite{PQ1,PQ2,PQ3}.
The symmetry is realised with two Higgs doublets and the usual quarks:
\be\label{yuk2hdm}
\mathcal{L}_{\text{Y}}=-\bar q_LY_d\Phi_2 d_R-\bar q_LY_u\tilde\Phi_1 u_R+\text{h.c.},
\ee
where the Higgs fields $\Phi_i$ develop vacuum expectation values $v_i$. The chiral PQ transformation takes the form
\be
\Phi_i\to e^{iX_i\alpha}\Phi_i,\quad u_R\to e^{iX_u\alpha}u_R,\quad d_R\to e^{iX_d\alpha}d_R,
\ee
where the charges have to satisfy $X_u=X_1$ and $X_d=-X_2$. Since in this case the up and down quarks couple to different doublets which transform independently,
the cancellation seen in Eq.~\eqref{cancel} does not take place.

Two Higgs doublets contain eight degrees of freedom, three of which are eaten by the gauge bosons. 
The axion is one of the remaining five (more specifically, it is the pseudoscalar state among the additional Higgs). 
Because the axion is entirely contained in the Higgs fields, its decay constant, $f_a$, in the PQWW model is proportional to the weak scale: 
\be
f_a\propto\sqrt{v_1^2+v_2^2}\equiv v=246\text{ GeV}.
\ee

Axion interactions are proportional to $1/f_a$ and, therefore, a lighter axion interacts more weakly than a heavier one. 
The value $f_a\approx v$ turns out to be too small, experimentally. For instance, the following branching ratio can be estimated
\be
\text{Br}(K^+\to\pi^++a)\sim10^{-4}\left(\frac{v}{f_a}\right)^2
\ee
but is bounded to be smaller than about $10^{-8}$. Therefore, the PQWW axion is excluded~\cite{asano}. Moreover, astrophysical constraints place an even stronger
bound on $f_a$ (see Sec.~\ref{sec:astrocosmo}).

Other models, with $f_a\gg v$, called ``invisible'' axion models, are still viable and will be reviewed next.

\subsubsection{KSVZ}

One type of invisible axion model is the Kim-Shifman-Vainshtein-Zakharov (KVSZ) model~\cite{models3,models4}. In this model a new complex scalar singlet ($S$) 
and a new heavy quark ($Q$) are introduced. They couple with a term
\be
-y\bar Q_L S Q_R+\text{h.c.}
\ee
and they are the only fields that carry a PQ charge. 
The PQ symmetry is spontaneously broken when $S$ acquires a vev due to a ``Mexican-hat'' type potential.

Since the Standard Model fields are PQ-blind, there is no coupling of the axion to quarks or leptons at tree level.

\subsubsection{DFSZ}

The other popular invisible axion model is the Dine-Fischler-Srednicki-Zhitnisky (DFSZ) model~\cite{dfs,models2}. This model augments the original PQWW model (a two
Higgs-doublet model, or 2HDM) with a complex scalar singlet. The new singlet, the two Higgs doublets and the quarks all transform under the PQ symmetry, which is 
spontaneously broken by the vevs of the scalars. Since the axion is shared by the Higgs and the singlet, it has interactions with quarks and leptons at tree level. 

This model will be studied in Ch.~\ref{ch-2hdm} extensively.

\subsection{Axion-like particles}

Theories Beyond the Standard Model can have additional (approximate) symmetries. If they are spontaneously broken, (pseudo-)Goldstone bosons are produced. Although 
they may have nothing to do with the resolution of the strong CP problem, they can have properties similar to that of the axion and be denoted as axion-like
particles (ALPs). In particular, these new particles can couple to two photons. If the ALP is a pseudoscalar, the coupling involves $F^{\mu\nu}\tilde F_{\mu\nu}$
and has the same form as Eq.~\eqref{prima}. If it is a scalar, it couples to $F^{\mu\nu}F_{\mu\nu}$ instead, with a term
\be
\mathcal L'_{a\gamma\gamma}=g'_{a\gamma\gamma}\frac\alpha{2\pi}\frac a{f_a}F^{\mu\nu}F_{\mu\nu}.
\ee
In the case of ALPs, Eq.~\eqref{axionmass1} does not hold. The strength of their interactions is unrelated to their mass.

Examples of ALPS are familons~\cite{familon} and Majorons~\cite{majoron1, majoron2}, 
coming from the spontaneous breaking of family and lepton number symmetry, respectively.

\subsection{The cold axion background}\label{sec:cab}

Axions are introduced as part of the PQ solution to the strong CP problem. However, they have the nice property of providing a solution to another problem: the
nature of dark matter.

If axions exist, they are produced thermally in the early Universe, as are all other particles. 
However, the most interesting process for axion production in view of dark matter is non-thermal and is known as vacuum misalignment~\cite{sik1,sik2,sik3}.

After the PQ symmetry has been spontaneously broken, but at temperatures above the QCD scale, there is no potential for the axion field. Therefore, it can take any value 
(more precisely, since it is a phase, the quantity $a/f_a$ can be anywhere from $-\pi$ to $\pi$). When the temperature falls below the QCD scale, instanton effects induce
a potential on the axion, as discussed earlier. When this happens, the axion field need not be at the bottom of this instanton-induced potential. This difference is
parameterized by the misalignment angle, $\Theta_i=a_i/f_a$. Unless there is no initial misalignment, the axion field will oscillate around the minimum, according to
\be\label{axsin}
a(t)=a_0\sin(m_at).
\ee
The energy stored in this oscillations contributes to the dark matter density.

If inflation occurs at a temperature larger than the PQ scale, the spontaneous breaking of PQ symmetry will happen after it and the values of the initial misalignment 
angle may vary over regions of the Universe within the causal horizon. 
This produces topological defects in the axion field, known as strings and domain walls, which also contribute to the axion density. If inflation happens at a temperature
smaller than the PQ scale, the axion field will be homogeneous across the causal horizon and no strings nor domain walls will be present. 

Either way, the axion density parameter takes the form~\cite{axioncosmo}
\be\label{parameter}
\Omega_a h^2=\kappa_a\left(\frac{f_a}{10^{12}\text{ GeV}}\right)^{7/6}.
\ee
If the PQ breaking happens after inflation, only the realignment process is relevant and $\kappa_a\propto\Theta_i^2$. 
If it happens before, $\kappa_a$ contains contributions both from oscillations and topological defects.\\

In this thesis we will concentrate on the scenario in which inflation happens at a temperature below the PQ scale and this oscillation around the minimum 
is solely responsible for dark matter.

Because $f_a$ is forced to be large, axions are very light. They, however, constitute cold dark matter (unlike, say, neutrinos), because the oscillations carry no
spatial momentum and the local velocity dispersion of axions is small~\cite{dispersion}. 
It has been suggested that axions may form a Bose-Einstein condensate~\cite{BEC,BEC2} at the galactic level, although it remains somewhat controversial.

These collective oscillations resulting from vacuum realignment will henceforth be referred as the cold axion background (CAB).

\subsection{Astrophysical and cosmological bounds}\label{sec:astrocosmo}

Apart from the early constraints that ruled out the PQWW axion, more stringent constraints can be obtained from astrophysical considerations. 
Axions can be produced in stars and be subsequently radiated away. 
The energy loss due to this process is inversely proportional to $f_a^2$.
Axions have to interact weakly enough so as not to affect the observed stellar evolution. This implies the following lower bound on the axion decay constant~\cite{stars}
\be
f_a>10^7\text{ GeV}, 
\ee
which in turn translates to an upper bound on its mass (only for proper PQ axions)
\be
m_a<0.1\text{ eV}.
\ee

In contrast, cosmology places an upper bound on the decay constant, as the production of axions by the process of vacuum misalignment increases with $f_a$. 
Although axions
need not be the main component of dark matter (or any component at all), their density can not exceed the observed dark matter density. 
Thus, we impose that Eq.~\eqref{parameter} does not yield a value higher than Eq.~\eqref{dmdensity}. For an initial misalignment $\Theta_i\sim\mathcal{O}(1)$, this means
\be
f_a<10^{11}\text{ GeV}
\ee
or, in terms of the mass
\be
m_a>10^{-5}\text{ eV}.
\ee

Combining the two set of bounds we get a window for dark matter axions, which should nevertheless not be taken too strictly as it is somewhat model-dependent
\be\label{bounds}
10^7\text{ GeV}<f_a<10^{11}\text{ GeV},\quad 10^{-5}\text{ eV}<m_a<0.1\text{ eV}.
\ee

Axions decay to two photons due to the coupling of Eq.~\eqref{prima}, with rate
\be
\Gamma_{a\to\gamma\gamma}=\frac{g^2m_a^3}{64\pi}.
\ee
This gives the axion a lifetime of
\be
\tau_a=\left(\frac{1\text{ eV}}{m_a}\right)^510^{16}\text{ yr},
\ee
which, even for the largest axion masses, is, by far, larger than the age of the Universe, so it poses no problems regarding the stability of the dark matter candidate.\\

\begin{figure}[ht]
\center
\includegraphics[width=0.9\textwidth]{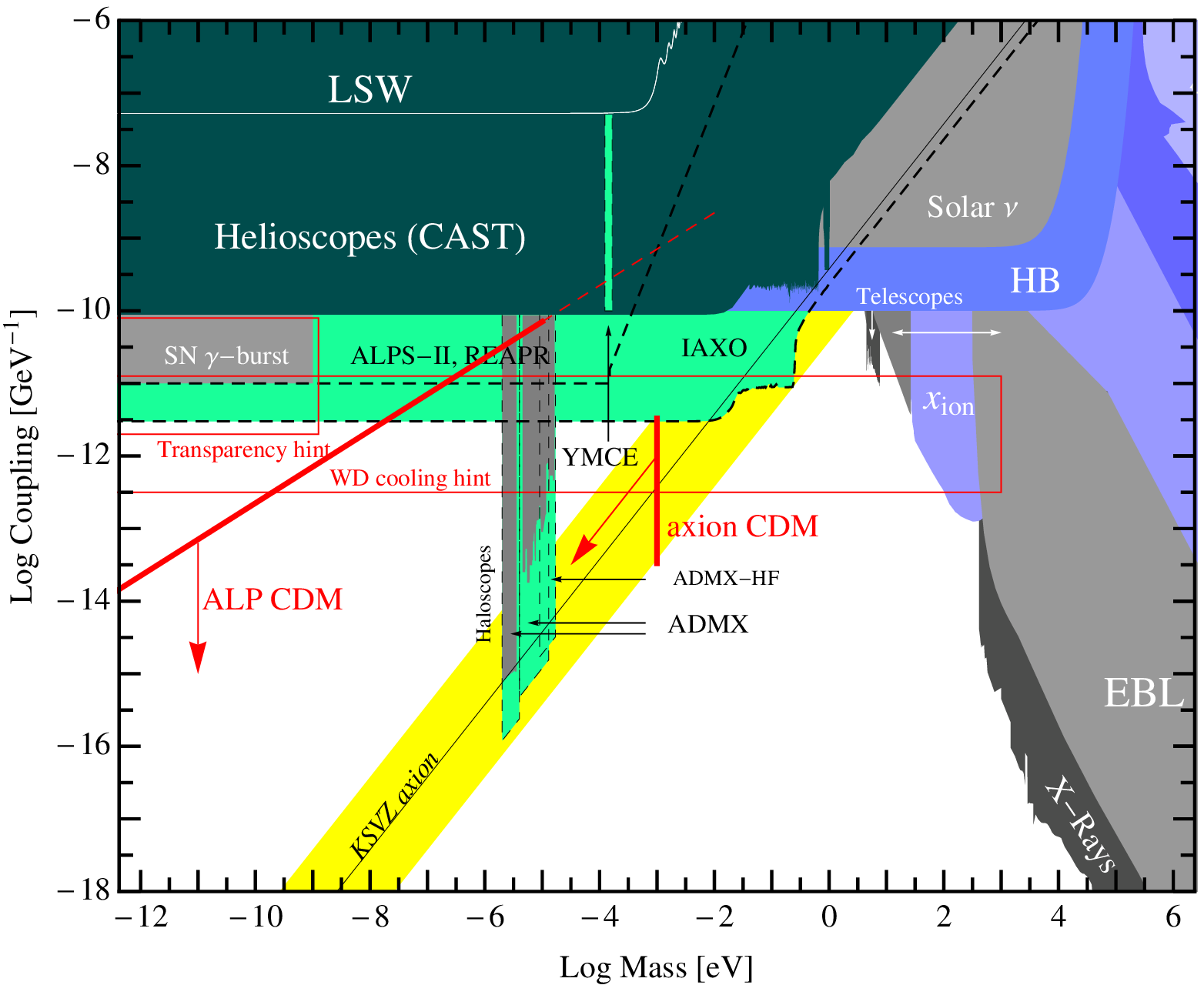}
\caption{Axion/ALP coupling to two photons vs. its mass (taken from Ref.~\cite{ringwald}).
The coupling on the vertical axis is $g/4$, with $g$ defined in Eq.~\eqref{prima}. The yellow region corresponds to proper axions, for which mass and
decay constant are related by Eq.~\eqref{axionmass1}. It is a band rather than a line because the two-photon coupling relation to $f_a$ is model-dependent.
Dark green regions are excluded by experimental searches (see Sec.~\ref{sec:exp}). Constraints from astronomical observations are in gray, while blue regions are
excluded from astrophysical and cosmological arguments (see Sec.~\ref{sec:astrocosmo}). Sensitivity of planned experiments is shown in light green. The region
indicated by the red arrows corresponds to ALPs or axions accounting for all the amount of cold dark matter. ADMX and IAXO can explore the axion band, in yellow.}
\label{axionbounds}
\end{figure}

When the coupling-mass relation of Eq.~\eqref{axionmass1} is abandoned in the more general case of ALPs, bounds get more involved, as they affect $f_a$ and $m_a$ separately.
Figure~\ref{axionbounds} presents a compilation of current bounds for the axion/ALP mass and its two-photon coupling. 

\subsection{Experimental searches}\label{sec:exp}

Their coupling with two photons provides the best chance to detect axions directly in an experiment. Equation~\eqref{prima} can be rewritten as
\be\label{edotb}
\mathcal{L}_{a\gamma\gamma}=-g\vec E\cdot\vec Ba.
\ee
A photon with polarisation parallel to an external magnetic field can be converted into an axion and vice versa, 
while photons with polarisation perpendicular to the external magnetic field are unaffected. This is the basis of most axion detection experiments.\\

Interestingly, a common feature of most axion experiments is the fact that their signal is enhanced when a certain combination of parameters is tuned to the axion mass.
Thus, their strategy is to perform a sweep, scanning the possible values of $m_a$.

Direct axion searches involve axions coming from the Sun, axions present in the galactic halo or axions produced in a laboratory.

\subsubsection{Solar axions}

Solar axion experiments attempt to directly detect axions that are produced at the core of the Sun via their coupling to two photons. 
Once produced, such axions would freely escape the Sun and reach the Earth.
These ``axion helioscopes'' try to turn solar axions into photons by means of a magnetic field. An X-ray detector then collects the reconverted photons.
The probability of conversion is~\cite{solaraxions}
\be\label{conversion}
P_{a\rightarrow\gamma}=2\left(\frac{gB}{2}\right)^2\frac{1-\cos(qL)}{(qL)^2}L^2,
\ee
where $L$ is the length of the region containing the magnetic field, $B$ is the field's strength and $q$ is the momentum transfer between the axion and the photon.
In vacuum, this momentum transfer is 
\be 
q=\frac{m_a^2}{2E_\gamma}.
\ee
This probability of conversion is maximised for small values of $qL$. By filling the magnetic field region with a gas, 
which has the effect of giving a mass to the photon ($m_\gamma$), the momentum transfer becomes
\be
q=\frac{m_a^2-m_\gamma^2}{2E_\gamma}.
\ee
By tuning the pressure of the gas, the effective photon mass can be modified. This results in an increase of the probability for a small window, when $m_\gamma$ and $m_a$
are close.

The CERN Axion Solar Telescope (CAST~\cite{cast}) and the International Axion Observatory (IAXO~\cite {iaxo}) are axion helioscopes.

\subsubsection{Cosmic axions}

Axion haloscopes are designed to detect the presence of dark matter axions in the galactic halo by using a resonant cavity.
Said cavity is filled with a magnetic field, which can convert dark matter axions into photons~\cite{axionsexp}. 
By adjusting the position of two rods inside the cavity, its resonant frequency can be modified. When it matches the axion mass, conversion is enhanced, 
which results in a signal that can be collected by a microwave receiver.

This is what the Axion Dark Matter Experiment (ADMX~\cite{admx}) is trying to do. This experiment tries to discover not only the existence of axions as particles,
but their role as dark matter as well.\\

Recently a new experiment to search for dark matter axions has been proposed: the Cosmic Axion Spin Precession Experiment (CASPEr~\cite{casper}). It relies on the
detection of an oscillating electric dipole moment induced by the axion background.

In Sec.~\ref{sec:cpproblem} we saw that the $\theta$-term of QCD induces an electric dipole moment for a nucleon, given by Eq.~\eqref{nedm}. We subsequently argued that the 
PQ resolution of the strong CP problem replaces the static $\theta$ parameter by the dynamical quantity $a/f_a$. Finally, in Sec.~\ref{sec:cab} we learned that a cold axion
background given by Eq.~\eqref{axsin} can be responsible for dark matter. Putting it all together results on an axion-induced electric dipole moment that oscillates in
time~\cite{graham}
\be\label{dn}
d_n\sim10^{-16}\frac{a_0\sin(m_at)}{f_a}e\,\text{cm}.
\ee
This is exploited by placing a magnetized sample in an external magnetic field, with its magnetization parallel to this $\vec B_{\text{ext}}$ (that is,
nuclear spins are aligned with the magnetic field). Then an external electric field $\vec E\perp\vec B_{\text{ext}}$ is turned on. If a nuclear EDM is present, 
this will cause nuclear spins in the sample to tilt slightly. Since the spins are no longer aligned with $\vec B_{\text{ext}}$, they start precessing
about it, with the Larmor frequency, which is proportional to the magnetic field. This results in a transverse magnetization that can be measured with a magnetometer, 
such as a SQUID. This is where the oscillating EDM is important. If it were static, its orientation with respect to $\vec E$ would change as the spin precesses. 
Each half cycle the spin would be reversed with respect to $\vec E$ and the net effect would compensate. 
The transverse magnetization would not build up and therefore it would not be detectable. 
However, if the nuclear EDM oscillates with a frequency proportional to $m_a$, as seen in Eq.~\eqref{dn},
the situation is different, since the effect does not cancel in each cycle.
When the magnetic field is tuned so that the Larmor frequency coincides with the EDM oscillation frequency a resonant effect is achieved.

\subsubsection{Laser-induced Axions}

Another group of experiments do not rely on naturally produced axions. Instead, they try to produce them in a laboratory with a laser beam traversing a magnetic field.

Photon regeneration experiments point a laser at an absorber, which does not let any photons pass through. 
However, a magnetic field placed before the absorber can convert some of the photons into axions, which are not absorbed. 
A second magnetic field turns them back into photons that can be detected. These are also called light shining through walls (LSW) experiments.
The Optical Search for QED Vacuum Birefringence, Axions and Photon Regeneration (OSQAR~\cite{osqar}), at CERN, 
and Any Light Particle Search (ALPS~\cite{alps}), at DESY, are LSW experiments.\\

Another way to detect the presence of axions is indirectly, through its effect on the optical properties of the vacuum. 
As discussed earlier, only photons with polarisation parallel to an external magnetic field can be converted into axions. 
This can give rise to two effects: dichroism and birefringence.

We speak of dichroism when light rays having different polarisations are absorbed differently. 
If a light ray passes through a region with a magnetic field, photons with polarisation parallel to the magnetic field can be turned 
into axions (since the axion is not detected, the photon is effectively lost, i.e. absorbed by the vacuum). 
This has the effect of changing the polarisation angle of the initial ray light.

A material is birefringent when it has a different refractive index for different polarisations. 
Photons parallel to the magnetic field can be converted into axions and then re-converted into photons. Since axions have mass,
they travel slower than the speed of light. Therefore, the component of light that is polarised parallel to the magnetic field has a refractive index greater than one, 
while the refractive index for the component normal to the magnetic field remains the same. Thus, a light ray that is initially linearly polarised gets an ellipticity, 
because the parallel component is retarded.

The Polarizzazione del Vuoto con Laser experiment (PVLAS~\cite{pvlas}) employs a Fabry-P\'{e}rot cavity to search for these effects in vacuum, 
due to either nonlinear QED effects~\cite{EH1,EH2} or the presence of light neutral particles, such as axions.

\section{Overview}

Axions represent a good solution to the strong CP problem, one of the still unresolved theoretical puzzles of the Standard Model.
For a window in parameter space not currently excluded experimentally, they are a viable dark matter candidate, so the interest in investigating them is twofold.\\

Currently there are two benchmark axion models known as KSVZ and DFSZ, which have been briefly described in Section~\ref{models}. The first goal of this thesis is 
to revisit the DFSZ model in light of the recent Higgs discovery.

Many experiments dealing with axions only try to discern whether they exist or not, but do not address their relevance as dark matter. Thus, it is important to
propose new ways to explore the possibility. This our second goal.\\

In Chapter~\ref{ch-2hdm} we consider the  DFSZ axion model. In Ref.~\cite{ce} 
a two Higgs-doublet model (2HDM) was studied in the case where all scalar particles were
too heavy to be produced in accelerators. Oblique corrections were used to constrain the parameters of the 2HDM model. 

We update this analysis by including an extra scalar singlet, as demanded by the DFSZ model and also taking into account the 2012 discovery of the Higgs particle
at $126\text{ GeV}$ by the LHC.\\

Technical aspects of the discussion in Ch.~\ref{ch-2hdm} are referred to Appendix~\ref{app:2hdm}.\\

The following chapters are devoted to the study of the cold axion background (CAB), described in Section~\ref{sec:cab}, as the responsible for dark matter. We propose ways
to indirectly detect it, by exploring its effects on photons and cosmic rays.\\

In Chapter~\ref{ch-photon} we study the joint effects of a CAB and a magnetic field on photon propagation. We discuss the mixing that these two
backgrounds create and study the corresponding proper modes. Then, we consider the evolution of a photon wave that is initially linearly polarised.\\

In Chapter~\ref{ch-cosmic} we turn our attention to cosmic ray propagation. A CAB modifies the dispersion relation of photons, which enables emission of a photon by a 
charged particle, a process forbidden in regular QED due to energy-momentum conservation. We study the energy loss that a cosmic ray experiences due to this process.
We also compute the spectrum of the radiated photons, which depends on the cosmic ray momentum distribution.\\

In Appendix~\ref{app:photon}, some aspects of the discussion in Chapters~\ref{ch-photon} and \ref{ch-cosmic} are developed in more detail.\\

Finally, Chapter~\ref{ch-concl} contains our conclusions.

\graphicspath{{2-2hdm/figures/}{figures/}}

\chapter{Axion-Higgs interplay in the two Higgs-doublet model}\label{ch-2hdm}
As mentioned in Chapter~\ref{chap:intro}, there are several extensions of the Minimal Standard Model (MSM) providing a particle with the characteristics
and couplings of the axion. In our view a particularly interesting possibility is the model suggested by Dine, Fischler, Srednicki and Zhitnitsky (DFSZ) 
more than 30 years ago~\cite{dfs,models2}, which consists in a fairly simple extension of the popular two Higgs-doublet model (2HDM). As a matter of fact it could
probably be argued that a good motivation to consider the 2HDM is that it allows for the inclusion of a (nearly) invisible 
axion~\cite{Axion1,Axion2,Axion3,Axion4,2hdmNatural1,2hdmNatural2,2hdmConf}. Of course there are other reasons why the 2HDM should be considered as a possible 
extension of the MSM. Apart from purely aestethic reasons, it is easy to reconcile such models with existing phenomenology without unnatural fine tuning 
(at least at tree level). They may give rise to a rich phenomenology, including possibly (but not necessarily) flavour changing neutral currents at some level, custodial 
symmetry breaking terms or even new sources of CP violation~\cite{2HDM1,2HDM3,2HDM4,2HDM2}.

Following the discovery of a Higgs-like particle with $m_h\sim 126$ GeV there have been a number of works considering the implications of such a finding on the 2HDM,
together with the constraints arising from the lack of detection of new scalars and from electroweak precision observables~\cite{pich2,pich2,pich3,pich4,pich5}. 
Depending on the way that the two doublets couple to fermions, models are classified as type I, II or III (see e.g.~\cite{2HDM1,2HDM3,2HDM4} for details), 
with different implications on the flavour sector. Consideration of all the different types of 2HDM plus all the rich phenomenology that can be encoded 
in the Higgs potential leads to a wide variety of possibilities with different experimental implications, even after applying all the known phenomenological
low-energy constraints.

Requiring a Peccei-Quinn (PQ) symmetry leading to an axion does however severely restrict the possibilities, and this is in our view an important asset of the DFSZ model.
This turns out to be particularly the case when one includes all the recent experimental information concerning the 126 GeV scalar state and its couplings.
We will explore this model and take into account all these constraints.\\
 
This chapter is structured as follows. In Sec~\ref{sec:mas} we discuss the possible global symmetries of the DFSZ model, namely $U(1)_{\rm PQ}$ (always present) 
and $SU(2)_L\times SU(2)_R$ (the $SU(2)_R$ subgroup may or may not be present). Symmetries are best discussed by using a matrix formalism that we review and extend. 
Section~\ref{sec:mam} is devoted to the determination of the spectrum of the theory. We present up to four generic cases that range from extreme
decoupling, where the model --apart from the presence of the axion-- is indistinguishable from the MSM at low energies, to the one where there are extra light Higgses 
below the TeV scale. This last case necessarily requires some couplings in the potential to be very small, a possibility that is nevertheless natural in a technical sense
and therefore should be contemplated as a theoretical hypothesis. We also discuss the situation where custodial symmetry is exact or approximately valid in this model.  
However, even if light scalars can exist in some corners of the parameter space, the presence of a substantial gap between the Higgs at 126 GeV and the
rest of the scalar states, with masses in the multi-TeV region or even beyond, is a rather generic characteristic of DFSZ models (therefore this hierarchy could 
be claimed to be an indirect consequence of the existence of a light invisible axion). 
In Section~\ref{sec:nonlinear} we discuss the resulting non-linear effective theory emerging in this generic situation. 
Next, in Sec.~\ref{sec:higgs}, we analyze the impact of the model on the (light) 
Higgs effective couplings to gauge bosons. Finally, in Sec.~\ref{sec:deltarho}, the restrictions that the electroweak precision parameters, particularly $\Delta\rho$, 
impose on the model are discussed. These restrictions, for reasons that will become clear in the subsequent sections, are relevant only in the case where 
all or part of the additional spectrum of scalars is light.

\section{Model and symmetries}\label{sec:mas}

The DFSZ model contains two Higgs doublets and a complex scalar singlet,
\be
\phi_1=\left(
\begin{array}{c}
\alpha_+\\
\alpha_0
       \end{array}
       \right),\quad
\phi_2=\left(
\begin{array}{c}
\beta_+\\
\beta_0
       \end{array}
       \right),\quad
\phi,
\label{eq:fields}
\ee
with vacuum expectation values (vevs) 
$\langle\alpha_0\rangle=v_1$, $\langle\beta_0\rangle=v_2$, 
$\langle\phi\rangle=v_\phi$  and $\langle\alpha_+\rangle=\langle\beta_+\rangle=0$.
Moreover, we define the usual electroweak vacuum
expectation value $v=246$ GeV as
$v^2=(v_1^2+v_2^2)/2$ and $\tan\beta=v_2/v_1$. The implementation of the PQ symmetry is only possible for type II models, where the
Yukawa terms are
\be\label{yuk}
\mathcal{L}_Y=G_1\bar q_L\tilde\phi_1u_R+G_2\bar q_L\phi_2d_R+h.c.,
\ee
with $\tilde\phi_i=i\tau_2\phi^*_i$. The PQ transformation acts on the scalars as
\be
\phi_1\to e^{iX_1\theta}\phi_1,\quad\phi_2\to e^{iX_2\theta}\phi_2,\quad\phi\to e^{iX_\phi\theta}\phi
\ee
and on the fermions as
\be
q_L\to q_L,\quad l_L\to l_L,\quad u_R\to e^{iX_u\theta}u_R,\quad d_R\to e^{iX_d\theta}d_R,\quad e_R\to e^{iX_e\theta}e_R.
\ee
For the Yukawa terms to be PQ-invariant we need
\be
X_u=X_1,\quad X_d=-X_2,\quad X_e=-X_2.
\ee
Let us now turn to the potential involving the two doublets and the new complex singlet. The most
general potential compatible with PQ symmetry is 
\ba\label{potential}
V(\phi,\phi_1,\phi_2)&=&\lambda_\phi\left(\phi^*\phi-V_\phi^2\right)^2+\lambda_1\left(\phi_1^\dag\phi_1-V_1^2\right)^2+\lambda_2\left(\phi_2^\dag\phi_2-V_2^2\right)^2\cr
&&+\lambda_3\left(\phi_1^\dag\phi_1-V_1^2+\phi_2^\dag\phi_2-V_2^2\right)^2\cr
&&+\lambda_4\left[\left(\phi_1^\dag\phi_1\right)\left(\phi_2^\dag\phi_2\right)-\left(\phi_1^\dag\phi_2\right)\left(\phi_2^\dag\phi_1\right)\right]\cr
&&+\left(a\phi_1^\dag\phi_1+b\phi_2^\dag\phi_2\right)\phi^*\phi-c\left(\phi_1^\dag\phi_2\phi^2+\phi_2^\dag\phi_1\phi^{*2}\right)
\ea
The $c$ term imposes the condition $-X_1+X_2+2X_\phi=0$. If we demand that the PQ current does 
not couple to the Goldstone boson that is eaten by the $Z$, 
we also get $X_1\cos^2\beta+X_2\sin^2\beta=0$. If furthermore we choose\footnote{There is arbitrariness
in this choice. This election conforms to the conventions existing in the literature.} $X_\phi=-1/2$ the 
PQ charges of the doublets are
\be\label{Xvalues}
X_1=-\sin^2\beta,\quad X_2=\cos^2\beta.
\ee
Global symmetries are not very evident in the way fields are introduced above. 
To remedy this let us define the matrices~\cite{ce}
\begin{align}
\Phi_{12}&=(\tilde\phi_1~\phi_2)=\left(\begin{array}{cc}
\alpha_0^* & \beta_+\\
-\alpha_- & \beta_0
                                \end{array}\right),\cr
\Phi_{21}&=(\tilde\phi_2~\phi_1)=\left(\begin{array}{cc}
\beta_0^* & \alpha_+\\
-\beta_- & \alpha_0
                                \end{array}\right)=\tau_2 \Phi_{12}^* \tau_2
\end{align}
and
\be
I=\Phi_{12}^\dag\Phi_{12}=\left(\begin{array}{cc}
 \phi_1^\dag\phi_1 & \tilde\phi_1^\dag\phi_2\\
 -\phi_1^\dag\tilde\phi_2 & \phi_2^\dag\phi_2
 \end{array}
 \right),\quad
J=\Phi_{12}^\dag\Phi_{21}=\phi_2^\dag\phi_1\mathbf{I},
\label{eq:IandJ}
\ee
with $\mathbf{I}$ being the identity matrix.
Defining also the constant matrix
$W=(V_1^2+V_2^2)\mathbf{I}/2+(V_1^2-V_2^2)\tau_3/2$, we can write the potential 
as
\begin{align}
V(\phi,I,J)&=\lambda_\phi\left(\phi^*\phi-V_\phi^2\right)^2+\frac{\lambda_1}4\left\{{\rm Tr}\left[(I-W)(1+\tau_3)\right]\right\}^2\cr
&+\frac{\lambda_2}4\left\{{\rm Tr}\left[(I-W)(1-\tau_3)\right]\right\}^2+\lambda_3\left[{\rm Tr}(I-W)\right]^2\cr
&+\frac{\lambda_4}4 {\rm Tr}\left[I^2-(I\tau_3)^2\right]+
\frac12 {\rm Tr}\left[(a+b)I+(a-b)I\tau_3\right]\phi^*\phi\cr
&-\frac c2 {\rm Tr}(J\phi^2+J^\dag\phi^{*2}).
\label{eq:pot}
\end{align}

A $SU(2)_L\times SU(2)_R$ global transformation acts on our matrices as
\be
\Phi_{ij}\to L\Phi_{ij}R^\dag,\quad I\to RIR^\dag,\quad J\to J.
\ee
We now we are in a better position to discuss the global symmetries of the
potential. The behaviour of the different parameters under $SU(2)_R$ is shown in 
Table I. See also~\cite{wudka1,wudka2}.

\begin{table}
\begin{center}
 \begin{tabular}{|c|c|}
\hline
\textbf{Parameter} & \textbf{Custodial limit}\\
\hline
$\lambda_1,\,\lambda_2,\,\lambda_4$ & $\lambda_1=\lambda_2=\lambda$ and 
$\lambda_4=2\lambda$  \\
$\lambda_3$ & $\lambda_3$\\
$\lambda_\phi$ & $\lambda_\phi$\\
$V_1^2,\,V_2^2$  & $V_1^2=V_2^2=V^2$\\
$V_\phi$ & $V_\phi$\\
$a,\,b$ & $a=b$\\
$c$ & $c$\\
\hline
\end{tabular}
\caption{
In total, there are 11 parameters: 7 are custodially preserving and 
4 are custodially breaking. By custodially breaking we mean that the resulting potential is not invariant under $SU(2)_R$.
\label{tab:cs}}
\end{center}
\end{table}

Finally, let us establish the action of the PQ symmetry previously discussed
 in this parametrisation. Under the PQ transformation:
\be
\Phi_{12}\to\Phi_{12}e^{iX\theta},\quad\phi\to e^{iX_\phi\theta}\phi
\label{eq:phiPQ}
\ee
with
\be
X=\frac{X_2-X_1}{2}\mathbf{I}-\frac{X_2+X_1}{2}\tau_3,\quad
X_\phi=\frac{X_2-X_1}{2}
\ee
Using the values for $X_{1,2}$ given in Eq.~(\ref{Xvalues})
\be\label{pqcharge}
X=\left(\begin{array}{cc}
\sin^2\beta & 0\\
0 & \cos^2\beta
\end{array}\right),\quad X_{\phi}=-\frac12.
\ee

\section{Masses and mixings}\label{sec:mam}
We have two doublets and a singlet, so a total of $4+4+2=10$ spin-zero 
particles. Three particles are eaten by the $W^\pm$ and $Z$ 
and $7$ scalars fields are left in the spectrum: two charged Higgs, two $0^-$
states and three neutral $0^+$ states. 
Our field definitions will be worked out in full detail
in Sec.~\ref{sec:nonlinear}. Here we want only to illustrate the spectrum. 
For the  charged Higgs  mass we have~\footnote{\label{note2}
Here and in the following we introduce the short-hand notation
$s^n_{m\beta} \equiv \sin^n(m\beta)$ and
$c^n_{m\beta} \equiv \cos^n(m\beta)$.}
\be\label{chargedmass}
m_{H_\pm}^2=8\left(\lambda_4v^2+\frac{cv_\phi^2}{s_{2\beta}}\right).
\ee

The quantity $v_\phi$ is proportional to the axion decay constant. Its value
is known to be very large (at least $10^7$ GeV and probably substantially 
larger if astrophysical constraints are taken into account,
see Refs.~\cite{pdg,axiondecay2,cast,iaxo,alps,admx} for several experimental and cosmological bounds). 
It does definitely make
sense to organize the calculations as an expansion in $v/v_\phi$.

In the $0^-$ sector there are two degrees of freedom that mix with each 
other  with a mass matrix which has a vanishing eigenvalue. The eigenstate with zero mass is the axion   
and $A_0$  is the pseudoscalar Higgs, with mass
\be\label{pseudoscalarmass}
m_{ A_0}^2=8c\left(\frac{v_\phi^2}{s_{2\beta}}+v^2 s_{2\beta}\right).
\ee
Equation~(\ref{pseudoscalarmass}) implies $c\ge0$. 
For $c=0$, the mass matrix in the $0^-$ sector has a second
zero, i.e. in practice the $A_0$ field behaves as another axion.

In the $0^+$ sector, there are three neutral particles that mix with each other. 
With $h_i$  we denote the corresponding  $0^+$  mass eigenstates. The mass matrix is given in 
App.~\ref{app:B}. 
In the limit of large $v_\phi$, 
the mass matrix in the $0^+$ sector can be easily diagonalised~\cite{2hdmNatural1,2hdmNatural2} and presents one eigenvalue
 nominally of order $v^2$ and two of order $v_\phi^2$. Up to ${\cal O}(v^2/v_\phi^2)$, these masses are
\ba\label{mass1}
m_{h_1}^2&=&32v^2\left(\lambda_1c_\beta^4+\lambda_2s_\beta^4+\lambda_3\right)-16v^2\frac{\left(ac_\beta^2+bs_\beta^2-cs_{2\beta}\right)^2}{\lambda_\phi},\\
\label{mass11}
m_{h_2}^2&=&\frac{8c}{s_{2\beta}}v_\phi^2+8v^2s_{2\beta}^2(\lambda_1+\lambda_2)-4v^2\frac{\left[(a-b)s_{2\beta}+2cc_{2\beta}\right]^2}{\lambda_\phi-2c/s_{2\beta}},
\\
\label{mass111}
m_{h_3}^2&=&4\lambda_\phi v_\phi^2+4v^2\frac{\left[(a-b)s_{2\beta}+2cc_{2\beta}\right]^2}{\lambda_\phi-2c/s_{2\beta}}
+16v^2\frac{\left(ac_\beta^2+bs_\beta^2-cs_{2\beta}\right)^2}{\lambda_\phi}.
\ea
The field $h_1$ is naturally identified with the scalar boson of mass 126 GeV observed at the LHC.

It is worth it to stress that there are several situations where the above formulae are non-applicable, 
since the nominal expansion in powers of $v/v_\phi$ may fail. This may be the case where the coupling 
constants $a$, $b$, $c$ connecting
the singlet to the usual 2HDM are very small, of order say $v/v_\phi$ or $v^2/v_\phi^2$. One should also 
pay attention to the case $\lambda_\phi\to 0$ (we have termed this latter case as the `quasi-free singlet limit'). 
Leaving this last case aside, we have found that the above expressions for $m_{h_i}$ apply in the following situations:

\begin{itemize}
\item[] Case 1: The couplings $a$, $b$ and $c$ are generically of ${\cal O}(1)$,

\item[] Case 2: $a$, $b$ or $c$ are of ${\cal O}(v/v_\phi)$.

\item[] Case 3: $a$, $b$ or $c$ are of ${\cal O}(v^2/v_\phi^2)$ but $c \gg \lambda_i v^2/{v_\phi^2}$.
\end{itemize}

If $c \ll \lambda_i v^2/{v_\phi^2}$ the $0^-$ state is lighter than the lightest $0^+$ Higgs 
and this case is therefore already phenomenologically unacceptable. 
The only other case that deserves a 
separate discussion is 
\begin{itemize}
\item[] Case 4: Same as in case 3 but $c \sim \lambda_i {v^2}/{v_\phi^2}$ 
\end{itemize}
In this case, the masses in the $0^+$ sector read,  up to
${\cal O}(v^2/v^2_\phi)$, as
\be\label{mass2}
m^2_{h_1,h_2}=8v^2\left(K\mp\sqrt{K^2-L}\right)\:\text{and }\:
m^2_{h_3}=4\lambda_\phi v_\phi^2,
\ee
where 
\ba
K&=&2\left(\lambda_1c_\beta^2+\lambda_2s_\beta^2+\lambda_3\right)+\frac{c v_\phi^2}{2v^2 s_{2\beta}},\cr
L&=&4\left[\left(\lambda_1\lambda_2+\lambda_1\lambda_3+\lambda_2\lambda_3\right)s_{2\beta}^2+
\frac{c v_\phi^2}{v^2s_{2\beta}}\left(\lambda_1c_\beta^4+\lambda_2s_\beta^4+\lambda_3\right)\right].
\ea
Recall that here we assume $c$ to be of ${\cal O}(v^2/v^2_\phi)$. Note that
\be\label{sumrule}
m_{h_1}^2 +m_{h_2}^2 = 32
v^2\left(\lambda_1c_\beta^2+\lambda_2s_\beta^2+\lambda_3+\frac{c v_\phi^2}{4v^2
s_{2\beta}}\right).
\ee

In the `quasi-free singlet' limit, when $\lambda_\phi\to 0$ or more generically  $\lambda_\phi \ll a,b,c$
it is impossible to sustain the hierarchy $v\ll v_\phi$, so again this case is phenomenologically 
uninteresting (see App.~\ref{app:C} for details). 

We note that once we set $\tan\beta$ to a fixed value, the
lightest Higgs to 126 GeV and $v_\phi$ to some large value
compatible with the experimental bounds, the mass spectrum in Eqs.~\eqref{chargedmass} to \eqref{mass111}
grossly depends on the parameters $c$, $\lambda_4$ and $\lambda_\phi$, the latter
only affecting the third $0^+$ state that is anyway very heavy and definitely
out of reach of LHC experiments; therefore 
the spectrum depends on only two parameters. If case 4 is applicable, the situation is slightly
different and an additional combination of parameters dictates the mass of the second 
(lightish) $0^+$ state. This can be seen in the sum rule of Eq.~(\ref{sumrule}) after requiring that
$m_{h_1}= 126$ GeV. Actually this sum rule is also obeyed in cases 1, 2 and 3, but the
r.h.s is dominated then by the contribution from parameter $c$ alone.

\subsection{Heavy and light states}
Here we want to discuss the spectrum of the theory according to the different scenarios that we have alluded to 
in the previous discussion.  Let us remember that it is always possible to identify one of the 
Higgses as the scalar boson found at LHC, namely $h_1$.
 
\begin{itemize}
\item[] Case 1. all Higgses except $h_1$ acquire a mass of order $v_\phi$. This includes the charged
and $0^-$ scalars, too. We term this situation `extreme decoupling'. The only light states are 
$h_1$, the gauge sector and the massless axion. This is the original DFSZ scenario~\cite{dfs}.

\item[] Case 2. This situation is similar to case 1 but now the typical scale of masses 
of $h_2$, $H_{\pm}$ and $A_0$
is
$\sqrt{vv_\phi}$. This range is beyond the LHC reach but it could perhaps be
explored with an accelerator in the 100 TeV region, a possibility being
currently debated. Again the only light particles are $h_1$, the axion and the
gauge sector. This possibility is natural in a technical sense as discussed
in Refs.~\cite{2hdmNatural1,2hdmNatural2}, as an approximate extra symmetry protects the
hierarchy.

\item[] Cases 3 and 4 are phenomenologically more interesting. Here we can at last have new states at the weak scale. 
In the $0^+$ sector, $h_3$ is definitely very heavy but $m^2_{h_1}$ and $m^2_{h_2}$ are
proportional to $v^2$ once we assume that  $c\sim v^2/v_\phi^2$. Depending on
the relative size of $\lambda_i $ and $c v^2_\phi/v^2$ one  would have to use
Eq.~(\ref{mass1}) or (\ref{mass2}). Because in case 3 one assumes that  $c
v^2_\phi/v^2$ is much larger than $\lambda_i$, $
h_1$ would still be the lightest
Higgs and $m_{h_2}$ could easily be in the TeV region.  When examining case 4 it
would be convenient to use the sum rule (\ref{sumrule}).  

We note that in case 4 the hierarchy between the different couplings is quite marked: typically to be 
realised one needs $c\sim 10^{-10} \lambda_i$, where $\lambda_i$ is a generic coupling of the potential.
The smallness of this number results in the presence of light states at the weak scale. For a discussion
on the `naturalness' of this possibility, see Refs.~\cite{2hdmNatural1,2hdmNatural2}.

 \end{itemize}

\subsection{Custodially symmetric potential}
While in the usual one doublet model, if  we neglect the Yukawa couplings and set the $U(1)_Y$ interactions to zero,
custodial symmetry is automatic, the latter is somewhat unnatural in 2HDM as one can write a fairly general potential. 
These terms are generically not invariant under global $SU(2)_L\times SU(2)_R$ transformations 
and therefore in the general case after the breaking  there is no custodial symmetry to speak of. 
Let us consider now the case where a global symmetry $SU(2)_L\times 
SU(2)_R$ is nevertheless present as there are probably good reasons to consider this limit. We may refer,
somewhat improperly, to this situation as being `custodially symmetric' although, after the breaking,
custodial symmetry proper may or may not be present.
If $SU(2)_L\times SU(2)_R$ is to be a symmetry, 
the parameters of the potential have to be set according to the
custodial relations in Table~\ref{tab:cs}.
Now,  there are two possibilities  to spontaneously
break  $SU(2)_L\times SU(2)_R$ and to give mass to the gauge bosons.\\

\noindent $SU(2)_L\times SU(2)_R\to U(1)$\\
If the vevs of the two Higgs fields are different ($\tan\beta\neq 1$), custodial symmetry is 
spontaneously broken to $U(1)$. In this case, one can use the minimisation equations of 
App.~\ref{sec:min} to eliminate 
$V$, $V_\phi$ and $c$  of Eq.~(\ref{eq:pot}). 
$c$ turns out to be of order $(v/v_\phi)^2$. In this case there are two extra Goldstone 
bosons: the charged Higgs is massless
\be
m_{H_\pm}^2=0.
\ee
Furthermore, the $A_0$ is light:
\be
m_{A_0}^2=16\lambda v^2\left(1+\frac{v^2}{v_\phi^2}s^2_{2\beta}\right)
\ee
This situation is clearly phenomenologically excluded.\\

\begin{figure}[ht]
 \center
 \includegraphics[scale=0.5]{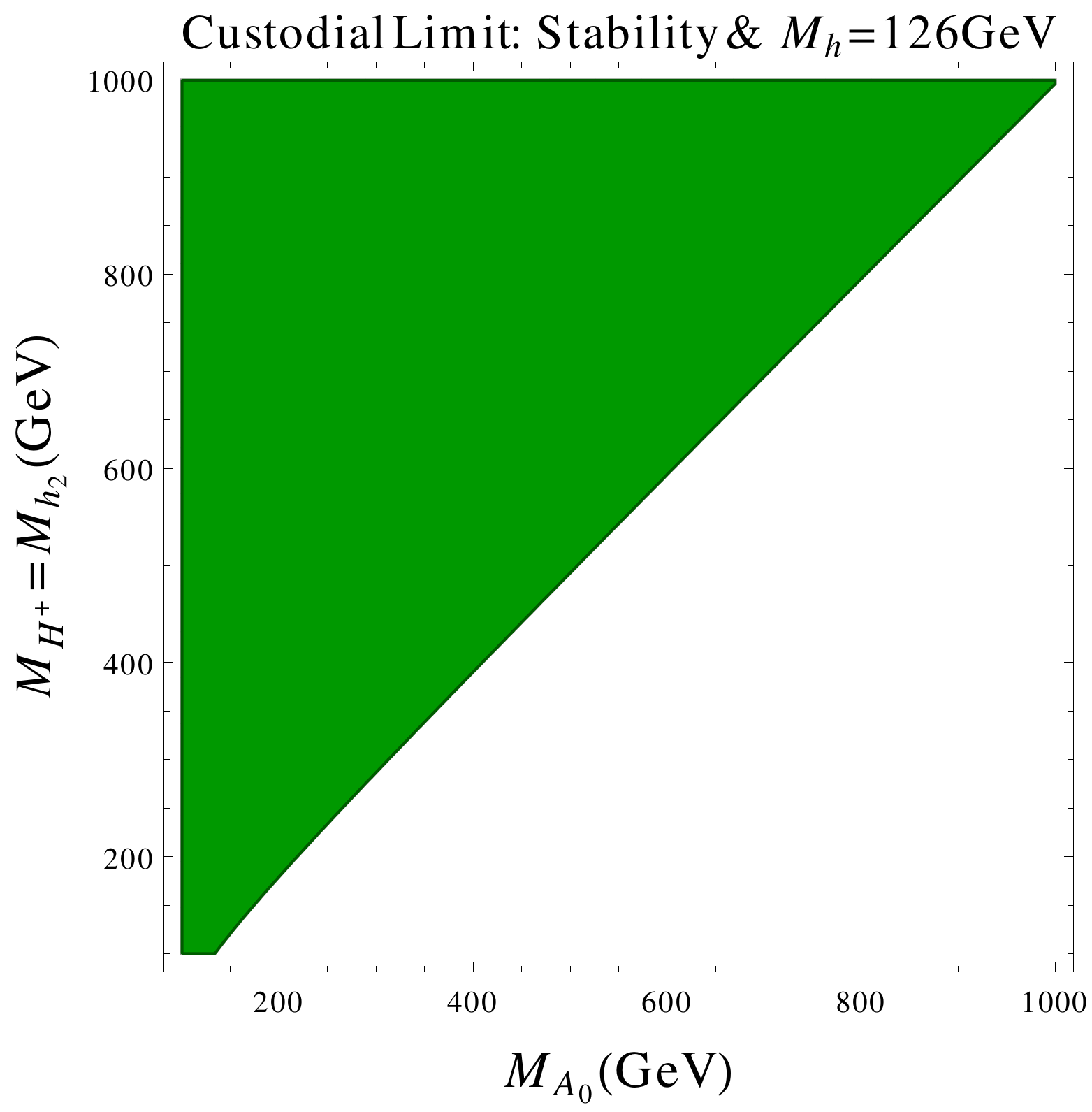}
 \caption{In green: allowed region in the custodial limit after 
 requiring vacuum stability (see e.g. App.~\ref{sec:stab} ). Each point
 in this region corresponds to a valid set of parameters in the DFSZ potential. Note that 
 $c$ is assumed to  be of order $ v^2/v_\phi^2$ and 
$c v^2_\phi/v^2$ has to be $\sim \lambda_i$ (case 4 discussed in the text).\label{fig:exclsu2}}
 \end{figure}

\noindent $SU(2)_L\times SU(2)_R\to SU(2)_V$\\
In this case, the vevs of the Higgs doublets are equal, so $\tan\beta=1$. The masses are
\be
m_{H_\pm}^2=8(2\lambda v^2+cv_\phi^2),\quad
m_{A_0}^2=8c(v^2+v_\phi^2)\quad\text{and}\quad m_{h_2}^2=m_{H_\pm}^2.
\ee
These three states are parametrically heavy, but they may be light in cases 3 and 4.\\ 
The rest of the $0^+$ mass matrix is $2\times2$ and has eigenvalues (up to second order in $v/v_\phi$)
\ba
m_{h_1}^2=16v^2\left[\lambda+2\lambda_3-\frac{(a-c)^2}{\lambda_\phi}\right],\quad
m_{h_3}^2=4\left[\lambda_\phi v_\phi^2+4v^2\frac{(a-c)^2}{\lambda_\phi}\right].
\ea
It is interesting to explore, in this relatively simple case, what sort of masses can be obtained
by varying the values of the couplings in the potential ($\lambda$, $\lambda_3$ and $c$). We
are basically interested in the possibility of obtaining a lightish spectrum (case 4 
previously discussed) and accordingly we assume that the natural scale of $c$ is $\sim v^2/v_\phi^2$.
We have to require the stability of the potential discussed in
App.~\ref{sec:stab} as well as $m_{h_1}=126$ GeV. The
allowed region is shown in Fig.~\ref{fig:exclsu2}.
Since we are in a custodially symmetric case there are no further restrictions to be obtained
from $\Delta\rho$.

\subsection{Understanding hierarchies}

In the MSM, it is well known that the Higgs mass has potentially large corrections
if the MSM is understood as an effective theory and one assumes that a larger scale
must show up in the theory at some point. This is the case, for instance, if neutrino
masses are included via the see-saw mechanism, to name just a possibility. In this case,
to keep the 126 GeV Higgs light one must do some amount of fine tuning.

In the DFSZ model such a large scale is indeed present from the outset and consequently
one has to envisage the possibility that the mass formulae previously derived may be
subject to large corrections due to the fact that $v_\phi$ leaks in the low-energy
scalar spectrum. Let us discuss the relevance of the hierarchy problem in the different
cases discussed in this section.

In case 1 all masses in the scalar sector but the physical Higgs are heavy, of order $v_\phi$,
and, due to the fact that the couplings $\lambda_i$ in the potential are generic (and
also the couplings $a,b,c$ connecting the two doublets to the singlet), the hierarchy may affect
the light Higgs quite severely and fine tuning of the $\lambda_i$ will be called for. However,
this fine tuning is not essentially different from the one commonly 
accepted to be necessary in the MSM to keep the Higgs light if a new scale is somehow present.

In cases 3 and 4 the amount of additional fine tuning needed is none or very little. In these scenarios 
(particularly in case 4) the scalar spectrum is light, in the TeV region, and the only heavy
degree of freedom is contained in the modulus of the singlet. After diagonalisation, this results in a very heavy
$0^+$ state ($h_3$), with a mass of order $v_\phi$. However, inspection of the potential reveals
that this degree of freedom couples to the physical sector with a strength $v^2/v_\phi^2$ and 
therefore may change the tree-level masses by a contribution of order $v$ ---perfectly acceptable.
In this sense the `natural' scenario proposed in Refs.~\cite{2hdmNatural1,2hdmNatural2} does not apparently lead to
a severe hierarchy problem in spite of the large scale involved.

Case 2 is particularly interesting. In this case, the intermediate masses are of order 
$\sqrt{v v_\phi}$, i.e. $\sim 100 $ TeV. There is still a very heavy mass eigenstate ($h_3$) but
again it is nearly decoupled from the lightest Higgs as in cases 3 and 4. On the contrary, the
states with masses $\sim \sqrt{v v_\phi}$ do couple to the light Higgs with strength $\sim \lambda_i$ and
thus require ---thanks to the loop suppression factor--- only a very moderate amount of fine
tuning as compared to case 1.

It is specially relevant, in the context of the hierarchy problem, to consider the previously discussed custodial case.
In the custodial limit, the $A_0$ mass is protected, as it
is proportial to the extended symmetry breaking parameter $c$. In addition,
$m_{h_2}= m_{H^\pm}$. Should one wish to keep a control on radiative corrections, doing the fine tuning
necessary to keep $h_1$ and $h_2$ light should suffice and in fact
the contamination from the heavy $h_3$ is limited, as discussed above. Of course, to satisfy present 
data we have to worry only about $h_1$.

\section{Non-linear effective Lagrangian}\label{sec:nonlinear}
We have seen in the previous section that the spectrum of scalars resulting from the potential of
the  DFSZ model is generically heavy. It is somewhat difficult to have all the scalar masses at the weak
scale, although the additional scalars can be made to have masses in the weak scale region in case
4.  The only exceptions are the three Goldstone bosons, the $h_1$ Higgs and the axion.  It is therefore somehow
natural to use a non-linear realisation to describe the low energy sector formed by gauge bosons
(and their associated Goldstone bosons), the lightest Higgs $0^+$ state  $h_1$,  and the axion.
Deriving this effective action is one of the motivations of this work.

To construct the effective action we have to identify the proper variables and in order to do so
we will follow the technique described in Ref.~\cite{ce}. In that paper the case of a generic 2HDM
where all scalar fields were very massive was considered. Now we have to modify the method to
allow for a light state (the $h_1$) and also to include the axion degree of freedom.

We decompose the matrix-valued $\Phi_{12}$ field introduced in Sec.~\ref{sec:mas} in the 
following form
\be
\Phi_{12}= {\cal U}{\cal M}_{12}.
\label{eq:um12}
\ee
${\cal U}$ is a $2\times 2$ matrix that contains the three Goldstone bosons associated to the breaking of 
$SU(2)_L$ (or, more precisely, of $SU(2)_L \times U(1)_Y\to U(1)_{\text{em}}$). We denote these
Goldstone bosons by $G^i$
\be
{\cal U}=\exp\left(i\frac{\vec G\cdot\vec\tau}{v} \right).
\ee
Note that the matrices $I$ and $J$ of Eq.~(\ref{eq:IandJ}) entering the DFSZ potential are actually independent of ${\cal U}$.
This is immediate to see in the case of $I$ while for $J$ one has to use the property $\tau_2 {\cal U}^* = {\cal U}\tau_2$ valid
for $SU(2)$ matrices. The effective potential then does depend
only on the degrees of freedom contained in ${\cal M}_{12}$,
whereas the Goldstone bosons drop from the potential, 
since, under a global  $SU(2)_L\times SU(2)_R$ rotation, $\Phi_{12}$ and ${\cal U}$  
transform as 
\be
\Phi_{12} \to L \Phi_{12} R^\dagger\quad {\cal U} \to L {\cal U} R^\dagger
\Rightarrow {\cal M}_{12} \to R {\cal M}_{12} R^\dagger.
\ee
Obviously the same applies to the locally gauged subgroup.

Let us now discuss the potential and ${\cal M}_{12}$  further. Inspection of 
the potential shows that, because of the term proportional to $c$, the phase of the 
singlet field $\phi$ does not drop automatically from the potential and thus it cannot be
immediately identified with the axion. In other words, the phase of the $\phi$
field mixes with the usual $0^-$ scalar from the 2HDM.
To deal with this let us find a suitable phase both 
in ${\cal M}_{12}$ and in $\phi$ that drops from the effective potential -- this 
will single out the massless state to be identified with the axion. 

We write ${\cal M}_{12}=M_{12}U_a$, where $U_a$ is a unitary matrix containing the axion field. 
An immediate choice is to take the generator of $U_a$ to be the identity, which
obviously can remove the phase of the singlet in the term in the 
effective potential proportional to $c$ while leaving the other terms manifestly
invariant. This does not exhaust all freedom however as we can include in the
exponent of $U_a$ a term proportional to $\tau_3$. It can be seen immediately
that this would again drop from all the terms in the effective potential,
including the one proportional to $c$ when taking into account that $\phi$
is a singlet under the action of $\tau_3$ which of course is nothing
but the hypercharge generator.
We will use the remaining freedom just discussed to properly 
normalize the axion and $A_0$ fields in the kinetic terms, to which we now turn.

The gauge invariant kinetic term is
\be\label{kinetic}
\mathcal{L}_{\text{kin}}=\frac12(\partial_\mu\phi)^*\partial^\mu\phi+\frac14\text{Tr}\left[(D_\mu\Phi_{12}^\dag)D^\mu\Phi_{12}\right],
\ee
where the covariant derivative is defined by
\be\label{covariantder}
D_\mu\Phi_{12}=\partial_\mu\Phi_{12}-i\frac g2\vec W_\mu\cdot\vec\tau\Phi_{12}+i\frac{g'}2B_\mu\Phi_{12}\tau_3.
\ee
By defining  $U_a=\exp\left(2 i a_\phi\,X /\vf\right)$ with $X$ in Eq.~\eqref{pqcharge},
all terms in the kinetic term are diagonal and exhibit the canonical normalization. 
Moreover, the field $a_\phi$ disappears from the potential.
Note that the phase redefinition implied in $U_a$ exactly coincides with 
the realisation of the PQ symmetry on $\Phi_{12}$ in Eq.~(\ref{eq:phiPQ})  as is to be expected (this
identifies uniquely the axion degree of freedom). 

Finally, the non-linear parametrisation  of $\Phi_{12}$ reads as 
\be
\Phi_{12}= {\cal U}{M}_{12} U_a,
\label{eq:phi12onL}
\ee
with 
\be
M_{12}=
\sqrt2\left(
\begin{array}{cc}
(v+H)c_\beta-\sigma^*s_\beta & \sqrt2 H_+ c_\beta\\
\sqrt2 H_- s_\beta & (v+H)s_\beta+\sigma c_\beta
\end{array}
\right),
\label{eq:m12field}
\ee
where
\be
\sigma=S+i\frac{v_\phi}\vf A_0
\ee
and 
\begin{align}
v+H&=\frac{c_\beta}{\sqrt2}\Re[\alpha_0]+\frac{s_\beta}{\sqrt2}\Re[\beta_0],\cr
S&=-\frac{s_\beta}{\sqrt2}\Re[\alpha_0]+\frac{c_\beta}{\sqrt2}\Re[\beta_0], \cr
H_\pm&=\frac{c_\beta}2\beta_\pm-\frac{s_\beta}2\alpha_\pm, 
\label{eq:HS}
\end{align}
in terms of the fields in Eq.~(\ref{eq:fields}). The singlet field $\phi$ is non-linearly parametrised as 
\be
\phi=\left(v_\phi+\rho-i\frac {v s_{2\beta}}\vf
A_0\right)\exp\left(i\frac{a_\phi}\vf\right).
\label{eq:phifield}\ee
With the parametrisations above the kinetic term is diagonal in terms of the fields of $M_{12}$ and 
$\rho$.
Moreover, the potential is independent of the axion and the Goldstone bosons. 
All the fields appearing in Eqs.~(\ref{eq:m12field}) and~(\ref{eq:phifield}) have vanishing vevs. 

Let us stress that $H$, $S$ and $\rho$ are not mass eigenstates and their relations 
with the $h_i$ mass 
eigenstates are defined through
\be
 H=\sum_{i=1}^3 R_{Hi}h_i,\quad S=\sum_{i=1}^3  R_{Si}h_i,\quad 
 \rho=\sum_{i=1}^3  R_{\rho i}h_i.
\label{eq:roth123}
\ee
The rotation matrix $R$ as well as  the corresponding mass matrix are given in App.~\ref{app:B}. 
$H$ and $S$ are the so called interaction eigenstates. 
In particular, $H$ couples to the gauge fields in the same way that the  usual MSM Higgs does. 

\subsection{Integrating out the heavy Higgs fields}
In this section we want to integrate out the heavy scalars in $\Phi_{12}$ of Eq.~(\ref{eq:phi12onL}) 
in order to build a low-energy effective theory at the TeV scale with an axion 
and a light Higgs.

As a first step, let us imagine that {{\em all} the states in $\Phi_{12}$ are heavy; 
upon their integration we will recover the Effective Chiral Lagrangian~\cite{efcl1,efcl2,efcl3,efcl4,efcl5,efcl6}
\be
\mathcal{L}= \frac{v^{2}}{4} {\rm Tr}\, D_{\mu}{\cal U}^{\dagger}D^{\mu}{\cal U} + \sum_{i=0}^{13} a_{i} \mathcal{O}_i\, ,
\ee
where the $\mathcal{O}_i$ is a set of local gauge invariant operators~\cite{ey}, and $D_\mu$ is the covariant derivative defined in
Eq.~\eqref{covariantder}. The corresponding effective couplings $a_i$ collect the 
low energy information (up to energies $E\simeq 4\pi v$) pertaining to the heavy states integrated out. 
In the unitarity gauge, the 
term $D_{\mu}{\cal U}^{\dagger}D^{\mu}{\cal U}$ would generate 
the gauge boson masses.

If a light Higgs ($h=h_1$) and an axion are present, they have to be included explicitly 
as dynamical states~\cite{heff1,heff2,heff3,heff4,heff5,heff6}, and the 
corresponding effective
Lagrangian will be (gauge terms are omitted in the present discussion)
\begin{align}\label{chiralwithhiggs}
\mathcal{L}  = {} &
\frac{v^{2}}{4} \left( 1+2 g_1\frac{h}{v}+ g_2 \frac{h^2}{v^2}+ \ldots\right) 
{\rm Tr}\, 
{\cal D}_{\mu}{\cal U}^{\dagger}{\cal D}^{\mu}{\cal U} \nn\\
&+\left(\frac{v_\phi^2}{v_\phi^2+v^2s_{2\beta}}\right)\partial_\mu a_\phi \partial^\mu a_\phi+\frac{1}{2} \partial_{\mu} h \partial^{\mu} h  - V(h)  \\
&+ \sum_{i=0}^{13} a_{i}\left(\frac{h}{v}\right)\mathcal{O}_i +\mathcal{L_{\text{ren}}},
\nn\end{align}
where~\footnote{Note that the axion kinetic term is not yet normalized in this expression. 
Extra contributions to the axion kinetic term also come from the term in the first line of Eq.~(\ref{chiralwithhiggs}). 
Only once we include these extra contributions, the axion kinetic term gets properly normalized. See also discussion below.}
\be
{\cal D}_\mu {\cal U} = D_\mu {\cal U}+ {\cal U}(\partial_\mu U_a)U^\dagger_a
\label{eq:dcalU}
\ee 
formally amounting to a redefinition of the `right' gauge field and
\be
V(h) = \frac{m^2_h}{2} h^{2} - d_{3} (\lambda v)  h^{3} -  d_{4}
\frac{\lambda}{4} h^{4},
\ee
\be
\mathcal{L_{\text{ren}}} = \frac{c_1}{v^4}\left(\partial_\mu h \partial^\mu
h\right)^2 
+ \frac{c_2}{v^2}\left(\partial_\mu h\partial^\mu h\right){\rm Tr}\,{\cal D}_{\nu}{\cal U}^{\dagger}{\cal D}^{\nu}{\cal U}
+ \frac{c_3}{v^2}\left(\partial_\mu h\partial^\nu h\right){\rm Tr}\,{\cal D}^{\mu}{\cal
U}^{\dagger}{\cal D}_{\nu}{\cal U}.
\ee
Here $h$ is the lightest $0^+$ mass eigenstate, with mass $126$ GeV  
but couplings in principle different from the ones of a MSM Higgs. 
 The terms in  $\mathcal{L_{\text{ren}}}$ are
required for renormalizability~\cite{ey,dobadollanes} 
at the one-loop level and play no role in the discussion.

The couplings $a_i$ are now functions of $h/v$, $ a_{i}({h}/{v})$, which are assumed to have 
a regular expansion and contribute
to different effective vertices. Their constant parts  $a_{i}(0)$ are related to the electroweak
precision parameters (`oblique corrections'). 

Let us see how the previous Lagrangian~(\ref{chiralwithhiggs}) can be derived. 
First, we integrate out from $\Phi_{12}= {\cal U}{M}_{12}U_a$ all heavy degrees of freedom, such as
$H^\pm$ and $A_0$, whereas we retain the $h_1$ components of $H$ and $S$, namely
 \begin{align}
 \Phi_{12}&= {\cal U}U_a {\overline M}_{12},\cr
 { \overline M}_{12} &= 
 \sqrt2\left(\begin{array}{cc}
 (v+H)c_\beta-Ss_\beta  & 0\\
 0 & (v+H)s_\beta+Sc_\beta
 \end{array}
 \right),
 \label{eq:m12bar}
\end{align}
where $H$ and $S$ stand for $R_{H1} h_1$ and $R_{S1} h_1$, respectively. 

When the derivatives of the kinetic term of Eq.~(\ref{kinetic}) act on ${\overline M}_{12}$, we get the contribution $\partial_{\mu} h \partial^{\mu} h$ in 
Eq.~(\ref{chiralwithhiggs}).
Since the unitarity matrices ${\cal U}$ and $U_a$ drop from the  potential of Eq.~(\ref{eq:pot}), only 
$V(h)$ remains.

To derive the first line of Eq.~\eqref{chiralwithhiggs}, we can use Eqs.~\eqref{eq:dcalU} and \eqref{eq:m12bar} 
to extract the following contribution from the kinetic term of Eq.~(\ref{kinetic})
\be
 {\rm Tr}\,({\cal D}_{\mu}{\cal U}  { \overline M}_{12})^{\dagger}{\cal D}^{\mu}{\cal U} { \overline M}_{12}=
 \frac{v^{2}}{4} \left( 1+2\frac{H}{v}+ \ldots\right) 
{\rm Tr}\, 
{\cal D}_{\mu}{\cal U}^{\dagger}{\cal D}^{\mu}{\cal U} + {\cal L}(a_\phi,S).
\label{mam}
\ee
Here we used that $ {\overline M}_{12}  {\overline M}_{12}^\dagger $
has a piece proportional to the identity matrix and another proportional to $\tau_3$
that cannot contribute to the 
coupling with the gauge bosons since $ {\rm Tr}  D_{\mu}{\cal U}^{\dagger} D^{\mu}{\cal U} \tau_3 $ vanishes identically.
The linear contribution  in $S$ is of this type and thus decouples from 
the gauge sector and as a result only terms linear in $H$ survive.
Using that $[U_a,{ \overline M}_{12}]= [U_a, \tau_3]=0$, the matrix $U_a$ cancels out in all traces and the 
only remains of the axion in the low energy action is the modification $D_\mu \to {\cal D}_\mu$. 
The resulting effective action is invariant under global transformations ${\cal U}\to L{\cal U}R^\dagger$ but
now $R$ is an $SU(2)$ matrix only if custodial symmetry is preserved (i.e. $\tan\beta=1$). Otherwise the
right global symmetry group is reduced to the $U(1)$ subgroup. It commutes with $U(1)_{\text{PQ}}$. 

We then reproduce Eq.~(\ref{chiralwithhiggs}) with $g_1=1$. However, this is true for the field $H$, not
$h=h_1$ and this will reflect in an effective reduction in the value of the $g_i$ when one considers the coupling
to the lightest Higgs only.

A coupling among the $S$ field, the axion and the neutral Goldstone or the neutral gauge boson survives in Eq.~(\ref{mam}). 
This will be discussed in Sec.~\ref{sec:higgs}.
As for the axion kinetic term, it is reconstructed with the proper normalization 
from the first term in Eq.~(\ref{kinetic}) together
with a contribution from the `connection' $(\partial_\mu U_a)U_a^\dagger$ in
$ {\rm Tr}\,{\cal D}_{\mu}{\cal U}{\cal D}^{\mu}{\cal U} $ (see Eq.~(\ref{eq:Lsax}) in next section).
There are terms involving two axions and the Higgs that are not very relevant phenomenologically at this point.
This completes the derivation of the $O(p^2)$ terms in
the effective Lagrangian.

To go beyond this tree level and to determine the low energy constants  $a_i(0)$ in
particular requires a one-loop integration of the heavy degrees of freedom and matching the Green's functions
of the fundamental and the effective theories.

See e.g. Refs.~\cite{efcl1,efcl2,efcl3,efcl4,efcl5,efcl6,ey} for a classification of all possible operators appearing up to ${\cal O}(p^6)$ that
are generated in this process. The information on physics beyond the MSM is encoded in the
coefficients of the effective chiral Lagrangian operators. Without including the (lightest) Higgs 
field $h$ (i.e. retaining only the constant term in the functions $a_i(h/v)$)
and ignoring the axion, there are
only two independent ${\cal O}(p^2)$ operators
\be
\label{p2}
{\cal L}^2= \frac{v^2}{4} {\rm Tr}(D_\mu {\cal U} D^\mu{\cal U}^\dagger) +
a_0(0) \frac{v^2}{4} ({\rm Tr}(\tau^3 {\cal U}^\dagger D_\mu{\cal U}))^2.
\ee
The first one is universal, its coefficient being fixed by the $W$ mass.  As
we just saw, it is flawlessly reproduced in the DFSZ model at tree level
after assuming that the additional degrees of freedom are heavy. Loop
corrections do not modify it if $v$ is the physical Fermi scale.
The other one is related to the $\rho$ parameter.
In addition there are a few ${\cal O}(p^4)$ operators with their
corresponding coefficients
\be
\label{p4}
{\cal L}^4=\frac12 a_1(0) g g^\prime {\rm Tr}({\cal U} B_{\mu \nu} {\cal U}^\dagger
W^{\mu\nu})
-\frac14a_8(0)g^2{\rm Tr}({\cal U}\tau^3{\cal U}^\dagger W_{\mu\nu}){\rm Tr}(
{\cal U}\tau^3{\cal U}^\dagger W^{\mu\nu})+ ...
\ee
In the above expression, $W_{\mu\nu}$ and $B_{\mu \nu}$ are the
field strength tensors associated to the $SU(2)_L$ and $U(1)_Y$ gauge fields, respectively.
In this paper we shall only consider the self-energy, or oblique,
corrections, which are  dominant in the 2HDM
model just as they are in the MSM.

Oblique corrections are often parametrised in terms
of the parameters $\varepsilon_1$, $\varepsilon_2$ and
$\varepsilon_3$  introduced in  Ref.~\cite{AB}.
In an effective theory such as the one described by the Lagrangian in Eqs.~(\ref{p2}) and (\ref{p4}), $\varepsilon_1$, $\varepsilon_2$
and $\varepsilon_3$ receive one-loop (universal)
contributions from the leading ${\cal O}(p^2)$ term $v^2{\rm Tr}(D_\mu
{\cal U}D^\mu {\cal U}^\dagger)$ and tree level contributions from
the $a_i(0)$. Thus
\be
\varepsilon_1= 2 a_0(0)+\ldots \qquad  \varepsilon_2= -g^2a_8(0) +\ldots
\qquad \varepsilon_3= -g^2a_1(0)+\ldots
\ee
where the dots symbolize the one-loop
${\cal O}(p^2)$ contributions. The latter
are totally independent of the specific symmetry breaking sector.
See e.g. Ref.~\cite{ce} for more details.

A systematic integration of the heavy degrees of freedom, including the
lightest Higgs as external legs, would provide the dependence of the
low-energy coefficient functions on $h/v$, i.e. the form of the
functions $a_i(h/v)$. However this is of no interest to us here.

\section{Higgs and axion effective couplings}\label{sec:higgs}
The coupling of $h_1$ to two W bosons can be worked out from the one of $H$, which is exactly as in the MSM, namely
\be
g_1^{SM} H W_\mu W^\mu =g_1^{SM} (R_{H1} h_1 + R_{H2} h_2 + R_{H3}h_3  )W_\mu W^\mu
\ee
where $R_{H1}=1- (v/v_\phi)^2 A_{13}^2/2$ and $g_1^{SM}\equiv 1$. With the expression of $A_{13}$ given in App.~\ref{app:B}, 
\be
g_1= g_1^{SM} \times\left( 1- \frac{2v^2}{v_\phi^2 
\lambda_\phi^2}\left(ac^2_\beta+bs^2_\beta c_{2\beta} -  c s_{2\beta}\right)^2\right).
\ee
It is clear that in cases 1 to 3 the corrections to the lightest Higgs couplings to the gauge bosons are 
very small, experimentally indistinguishable from the MSM case. In any case
the correction is negative and $g_1< g_1^{SM}$.

Case 4 falls in a different category. Let us remember that this case corresponds
to the situation where $c \sim \lambda_i {v^2}/{v_\phi^2}$. Then the corresponding
rotation matrix is effectively $2\times 2$, with an angle $\theta$ that is
given in App.~\ref{app:B}. Then 
\be
g_1= g_1^{SM} \cos \theta.
\ee
In the custodial limit, $\lambda_1=\lambda_2$ and $\tan\beta =1$, this angle vanishes
exactly and $g_1= g_1^{SM}$. Otherwise this angle could have any value. Note however
that when $c \gg \lambda_i {v^2}/{v_\phi^2}$, then $\theta\to 0$ and the value
$g_1\simeq g_1^{SM}$ is recovered. This is expected, as when $c$ grows case 4 moves into case 3.
Experimentally, from the LHC results we know~\cite{heff4} that 
$g_1=[0.67,1.25]$ at $95\%$ CL.

Let us now discuss the Higgs-photon-photon coupling in this type of models. First,
we consider the contribution from gauge and scalar fields in the
loop.  The diagrams
contributing to the coupling between the lightest scalar state $h_1$
and photons are exactly the same ones as in a generic 2HDM, via a loop
of gauge bosons and one of charged Higgses. In the DFSZ case the 
only change with respect to a generic 2HDM could be a modification in the $h_1WW$ (or 
Higgs-Goldstone bosons coupling) or in the $h_1 H^+ H^+$
tree-level couplings. The former has already been discussed while
the triple coupling of the lightest Higgs to two charged Higgses 
gets modified in the DFSZ model to
\begin{align}
\lambda_{h_1H_+H_-}={}& 8vR_{H1}\left[(\lambda_1+\lambda_2)s^2_{2\beta}+4\lambda_3+2\lambda_4\right]
+16vs_{2\beta} R_{S1}\left(\lambda_2 c^2_\beta-\lambda_1s^2_\beta\right)\cr
&  +8v_\phi R_{\rho1}\left(as^2_\beta+bc^2_\beta-cs_{2\beta}\right).
\end{align}
Note that the first line involves only constants that are
already present in a generic 2HDM, while the second one does involve
the couplings $a,b$ and $c$ characteristic of the DFSZ model.

The corresponding entries of the rotation matrix in the $0^+$ sector can be found 
in App.~\ref{app:B}. 
In cases 1, 2  and 3 the relevant entries are $R_{H1}\sim 1$, $R_{S1}\sim v^2/v_\phi^2$ and
$R_{\rho 1}\sim v/v_\phi$, respectively. Therefore the second term in the first line is
always negligible but the piece in the second one can give a sizeable contribution
if $c$ is of ${\cal O}(1)$ (case 1). This case could therefore be excluded or confirmed from
a precise determination of this coupling. In cases 2 and 3 this effective
coupling aligns itself with a generic 2HDM but with large (typically $\sim 100$ TeV) or moderately large 
(few TeV) charged Higgs masses.

Case 4 is slightly different again. In this case $R_{H1}=\cos\theta$ and
$R_{S1}=\sin \theta$, but $R_{\rho 1}=0$. The situation is again similar to a generic
2HDM, now with masses that can be made relatively light, but with a mixing angle
that, because of the presence of the $c$ terms, may differ slightly from the
2HDM. For a review of current experimental fits in 2HDM the interested reader can 
see Refs.~\cite{pich2,pich3,pich4,pich5}.

In this section  we will also list the tree-level couplings of the axion to the
light fields, thus completing the derivation of the effective low-energy
theory. The tree-level couplings are very few actually as the axion
does not appear in the potential, and they are necessarily 
derivative in the bosonic part. From the kinetic term we get
\be
{\cal L}(a_\phi,S)=\frac{2R_{S1}}{\vf}h_1\partial^\mu a_\phi\left(\partial_\mu G_0+m_ZZ_\mu\right) + \text{ terms with 2 axions},
\label{eq:Lsax}
\ee

From the Yukawa terms \eqref{yuk} we get
\be
{\cal L}(a_\phi,q,\bar q)=\frac{2i}{\vf}a_\phi\left(m_u s^2_\beta\bar u\gamma_5u+m_d c^2_\beta\bar d\gamma_5d\right).
\ee

The loop-induced couplings between the axion and gauge bosons 
(such as the anomaly-induced coupling $a_\phi\tilde F F$, of extreme importance for
direct axion detection~\cite{pdg,cast,iaxo,alps,admx,axiondecay2}) will not be discussed here as they are
amply reported in the literature. 

\section{Matching the DFSZ model to the 2HDM}

The most general 2HDM potential can be  
read\footnote{We have relabelled $\lambda_i\to\Lambda_i$ to avoid confusion with 
the potential of the DFSZ model.} e.g. from Refs.~\cite{2HDM2,pich1,pich2,pich3,pich4,pich5}
\begin{alignat}{5}
 V(\phi_1,\phi_2) 
&= m^2_{11}\,\phi_1^\dagger\phi_1
 + m^2_{22}\,\phi_2^\dagger\phi_2
 - \left[ m^2_{12}\,\phi_1^\dagger\phi_2 + \text{h.c.} \right] \notag \\
&+ \frac{\Lambda_1}{2} \, (\phi_1^\dagger\phi_1)^2
 + \frac{\Lambda_2}{2} \, (\phi_2^\dagger\phi_2)^2
 + \Lambda_3 \, (\phi_1^\dagger\phi_1)\,(\phi_2^\dagger\phi_2) 
 + \Lambda_4 \, |\phi_1^\dagger\,\phi_2|^2 \notag \\
&+ \left[ \frac{\Lambda_5}{2} \, (\phi_1^\dagger\phi_2)^2 
        + \Lambda_6 \, (\phi_1^\dagger\phi_1) \, (\phi_1^\dagger\phi_2)
        + \Lambda_7 \, (\phi_2^\dagger\phi_2)\,(\phi_1^\dagger\phi_2) + \text{h.c.} 
   \right] \;.
\label{eq:2hdmpotential}
\end{alignat}
This potential contains 4 complex and 6 real parameters (i.e. 14 real numbers). 
The most popular 2HDM is obtained by imposing a ${\mathbb{Z}}_2$ symmetry that is softly broken,
namely $\Lambda_6=\Lambda_7=0$  and $m_{12}\ne0$. The $\mathbb{Z}_2$ approximate 
invariance helps remove flavour changing neutral currents at tree level.
A special role is played by the term proportional to $m_{12}$. 
This term softly breaks $\mathbb{Z}_2$ but is necessary to control the decoupling limit of 
the additional scalars in a 2HDM and to eventually reproduce the MSM with a light
Higgs.
 
In the DFSZ model discussed here, $v_\phi$ is very large and
at low energies the additional singlet field $\phi$ reduces approximately
to $\phi \simeq v_\phi \exp(a_\phi/v_\phi)$. Indeed, from Eq.~(\ref{eq:phifield}) 
we see that $\phi$ has a  $A_0$ component but it can be in practice neglected
in the case of an invisible axion, since this component is proportional to $v/v_\phi$. In addition, the radial
variable $\rho$ can be safely integrated out. 

Thus, the low-energy effective theory defined by the DFSZ model
is a particular type of 2HDM model with the non-trivial
benefit of solving the strong CP problem thanks  to the appearance of an
invisible axion\footnote{Recall that mass generation due to the anomalous
coupling with gluons  has not been considered in this work.}.  
Indeed, the DFSZ model reduces at low energy to a 2HDM containing 9 parameters in practice (see below, 
note that $v_\phi$ is used as an input), instead of the 14 of the general 2HDM case. 

The constants $\Lambda_{6,7}$ are absent, as in many $\mathbb{Z}_2$-invariant 2HDMs but also $\Lambda_{5}=0$, as all these terms are   
not invariant under the Peccei-Quinn symmetry. In addition,
the $m_{12}$ term that sofly breaks $\mathbb{Z}_2$ and is necessary 
to control the decoupling to the MSM 
is dynamically generated by the spontaneous breaking of PQ symmetry. 
There is no $\mu=m_{12}$ problem here concerning the naturalness of having non-vanishing $\mu$.  

At the electroweak scale 
the DFSZ potential in Eq.~(\ref{potential}) can be matched to the 2HDM 
terms of Eq.~(\ref{eq:2hdmpotential}) by the substitutions
\begin{align}
m^2_{11}&=\left[-2\lambda_1V_1^2+2
\lambda_3(V_1^2+V_2^2)+av_\phi^2\right]/4\\
m^2_{22}&=\left[-2\lambda_1V_2^2+2
\lambda_3(V_1^2+V_2^2)+bv_\phi^2\right]/4\\
m_{12}^2&=cv_\phi^2/4\\
\Lambda_1&=(\lambda_1+\lambda_3)/8,\quad
\Lambda_2=(\lambda_2+\lambda_3)/8\\
\Lambda_3&=(2\lambda_3+\lambda_4)/16,\quad
\Lambda_4=-\lambda_4/16,\\
\Lambda_5&=0,\quad\Lambda_6=0,\quad\Lambda_7=0.
\end{align}

Combinations of parameters of the DFSZ potential can be determined from
the four masses $m_{h_1}$, $m_{h_2}$, $m_{A_0}$ and $m_{H^+}$ and the two
parameters $g_1$ (or $\theta$) and $\lambda_{h_1H_+H_-}$ that control
the Higgs-$WW$  and (indirectly) the Higgs-$\gamma\gamma$ couplings, whose expression 
in terms of the parameters of the potential have been given in Sec.~\ref{sec:higgs}. As we have seen for
generic couplings, all masses but that of the lightest Higgs decouple and the
effective couplings take their MSM values.
In the phenomenologically
more interesting cases (cases 3 and 4), two of the remaining constants ($a$ and $b$)  
drop in practice from the low-energy predictions and the effective 2HDM corresponding to the 
DFSZ model depends only on 7 parameters.  
If in addition custodial symmetry is assumed to be exact or nearly exact, 
the relevant parameters are actually completely determined by measuring three masses
and the two couplings ($m_{h_2}$ turns out to be equal to $m_{H^+}$ if custodial invariance
holds). Therefore, the LHC
has the potential to fully determine all the relevant parameters of
the DFSZ model. 
 
Eventually, the LHC and perhaps a future linear collider will hopefully be able to assess  
the parameters of the 2HDM potential and their symmetries to check the DFSZ relations. 
Of course, finding a pattern of couplings in concordance with the pattern 
predicted by the low-energy limit of the DFSZ model would not yet prove the latter to 
be the correct microscopic theory as this would require measuring the 
axion couplings, which are not present in a 2HDM. In any case,
it should be obvious that the effective theory of the DFSZ is 
significantly more restrictive than a general 2HDM.

We emphasize that the above discussion refers mostly to case 4 as discussed in 
this work and it partly applies to case 3 too. Cases 1 and 2 are in practice indistinguishable from
the MSM up to energies that are substantially larger from the ones currently 
accessible, apart from the presence of the axion itself. 
As we have seen, the DFSZ model in this case is quite predictive and it does not 
correspond to a generic 2HDM but to one where massive scalars are all decoupled 
with the exception of the 126 GeV Higgs.  

\section{Constraints from electroweak parameters}\label{sec:deltarho}
For the purposes of setting bounds on the masses of the new scalars
in the 2HDM, $\varepsilon_1=\Delta\rho$ is the most effective parameter. It can be computed by~\cite{AB}
\be
\varepsilon_1\equiv\frac{\Pi_{WW}(0)}{M_W^2}-\frac{\Pi_{ZZ}(0)}{M_Z^2},
\label{eq:deltarho}
\ee
with the gauge boson vacuum polarisation functions defined as 
\be
\Pi^{\mu\nu}_{VV}(q)=g^{\mu\nu}\Pi_{VV}(q^2)+q^\mu q^\nu~{\rm terms}\quad (V=W,Z).
\ee
We need to compute loops of the type of Fig.~\ref{vvxy}. 
\begin{figure}[ht]
\center
\includegraphics[scale=0.6]{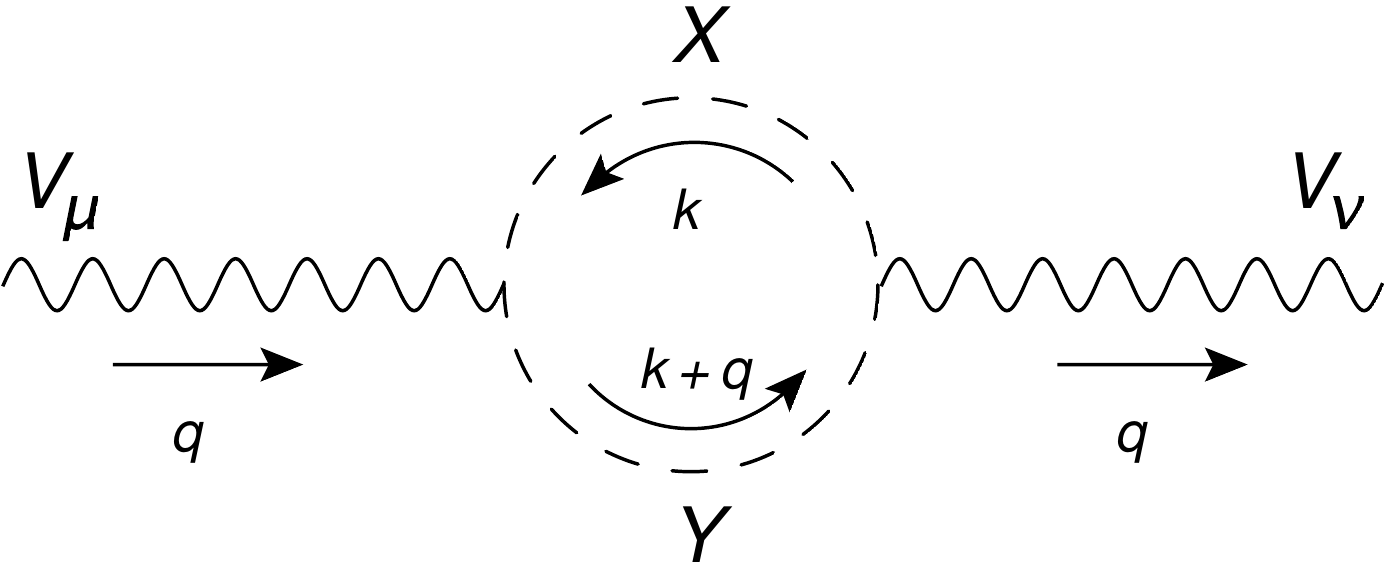}
\caption{Feynman diagram relevant for the calculation of $\Pi^{\mu\nu}_{VV}(q)$.}\label{vvxy}
\end{figure}
These diagrams produce three kinds of terms. The terms proportional to two powers of 
the external momentum, $q_\mu q_\nu$, do not enter in $\Pi_{VV}(q^2)$. The terms
proportional to just one power vanish upon integration. Only the 
terms proportional to $k_\mu k_\nu$, the momentum in the loop, survive and contribute.

Although it is an unessential approximation, to keep formulae relatively
simple we will compute $\varepsilon_1$ in the approximation $g^\prime=0$. The term
proportional to $(g^\prime)^2$ is 
actually the largest contribution in the MSM (leaving aside the 
breaking due to the Yukawa couplings) but it is only logarithmically
dependent on the masses of any putative scalar state and it can be safely omitted
for our purposes~\cite{ce}. The underlying reason is that in the 2HDM custodial
symmetry is `optional' in the scalar sector and it is natural to investigate
power-like contributions that would provide the strongest constraints. 
We obtain, in terms of the mass eigenstates and the rotation matrix of Eq.~(\ref{eq:roth123}),
\begin{align}
\varepsilon_1={}& \frac{1}{16\pi^2 v^2}\Bigg[m_{H_\pm}^2 - \frac{v^2_\phi}{\vvf} f(m_{H_\pm}^2,m_{A_0}^2) \cr
&+ \sum_{i=1}^3   R^2_{Si}\left(\frac{v^2_\phi}{\vvf} f(m_{A_0}^2,m_{h_i}^2)-f(m_{H_\pm}^2,m_{h_i}^2)  \right)
\Bigg],
\end{align}
where $f(a,b)=a b/(b-a)\log b/a$ and $f(a,a)=a$. 
Setting $v_\phi\to \infty$ and keeping the Higgs masses fixed, we formally recover the  $\Delta \rho$ 
expression in the 2HDM (see the Appendix in Ref.~\cite{ce}), namely
\be
\varepsilon_1=\frac{1}{16\pi^2 v^2}\Bigg[m_{H_\pm}^2 - f(m_{H_\pm}^2,m_{A_0}^2)+\sum_{i=1}^3   R^2_{Si}
\left(f(m_{A_0}^2,m_{h_i}^2)-f(m_{H_\pm}^2,m_{h_i}^2)  \right)
\Bigg]
\ee
Now, in the limit $v_\phi\to \infty$ and  $m_{H_\pm}\to m_{A_0}$ (cases 1, 2 or 3 previously
discussed) the  
 $\Delta \rho$ above 
will go to zero as $v/v_\phi$ at least and the experimental bound is fulfilled automatically. 

However, we are particularly interested in case 4 that allows for a light spectrum
of new scalar states. We will study this in two steps. First we assume a 
`quasi-custodial' setting whereby we assume that 
custodial symmetry is broken {\em only}
via $\lambda_{4B}=\lambda_4-2\lambda$ being non-zero. 
Imposing vacuum stability and the experimental bound of 
$(\varepsilon_1-\varepsilon_1^{SM})/\alpha=\Delta T=0.08(7)$  
from the electroweak fits in Ref.~\cite{Dtexp} 
one gets the exclusion plots shown in Fig.~\ref{fig:ecl-su2}. 
\begin{figure}[ht]
\center
\includegraphics[width=0.45\textwidth]{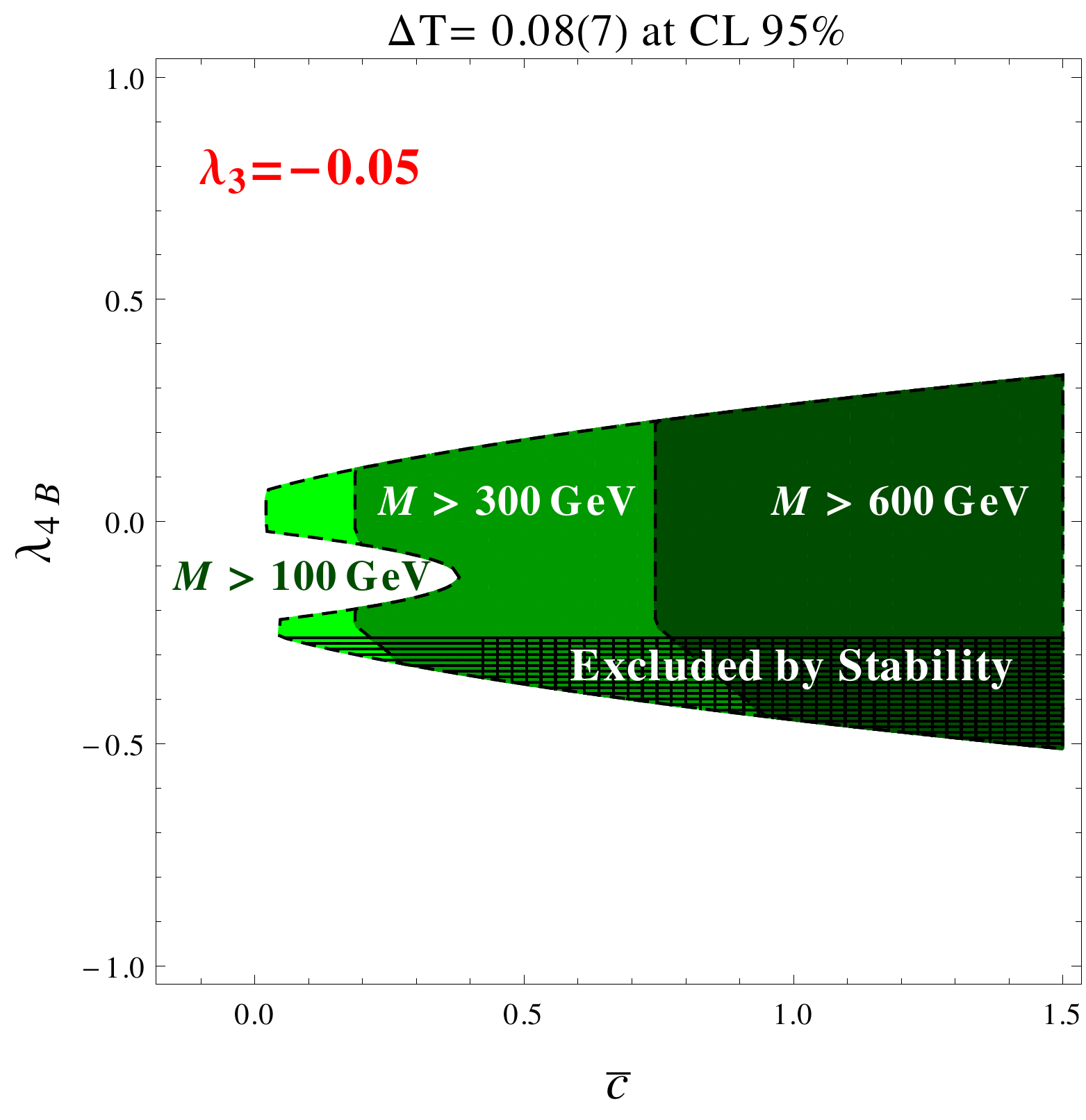}
\includegraphics[width=0.45\textwidth]{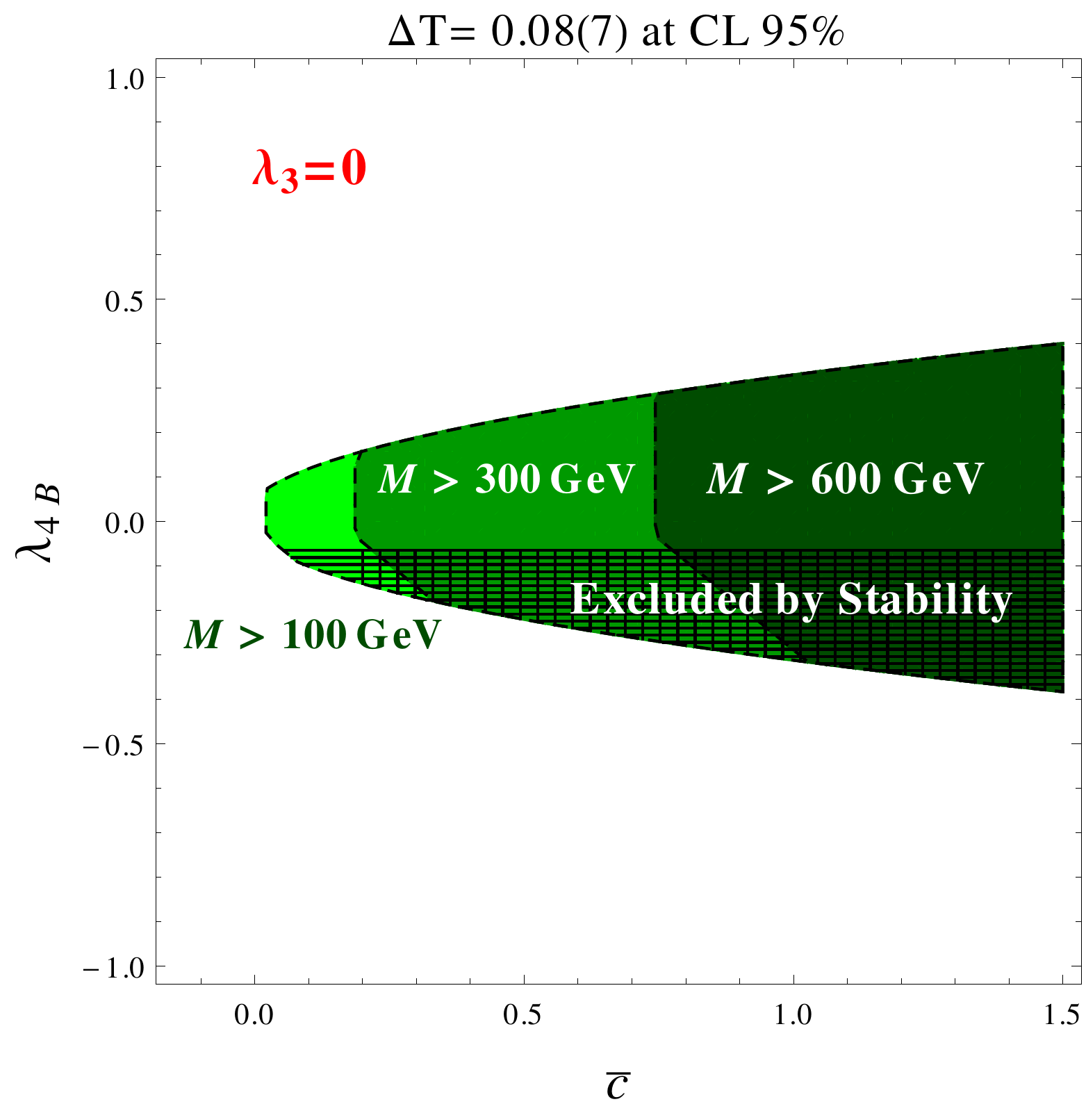}
\caption{Exclusion region for a custodial 2HDM limit 
with a $\lambda_{4B}=\lambda_4-2\lambda$ breaking term 
as a function of $\lambda_{4B}$ and $\bar c= c {v^2_\phi}/{v^2}$.
Different colour regions imply different cuts assuming that all masses 
($m_{A_0}$, $m_{H_\pm}$ and $m_{h_2}$) are greater than 100, 300 or 600 GeV (light to dark).
The potential becomes unstable for $\lambda_3 > 0.03$. }
\label{fig:ecl-su2}
\end{figure}

It is also interesting to show (in this same `quasi-custodial' limit) the range
of masses allowed by the present constraints on $\Delta T$, without any reference to
the parameters in the potential. This is shown for two reference values of $m_{A_0}$ in Fig.~\ref{eclM-su2}. Note
the severe constraints due to the requirement of vacuum stability.  
\begin{figure}[ht]
\center
\includegraphics[width=0.45\textwidth]{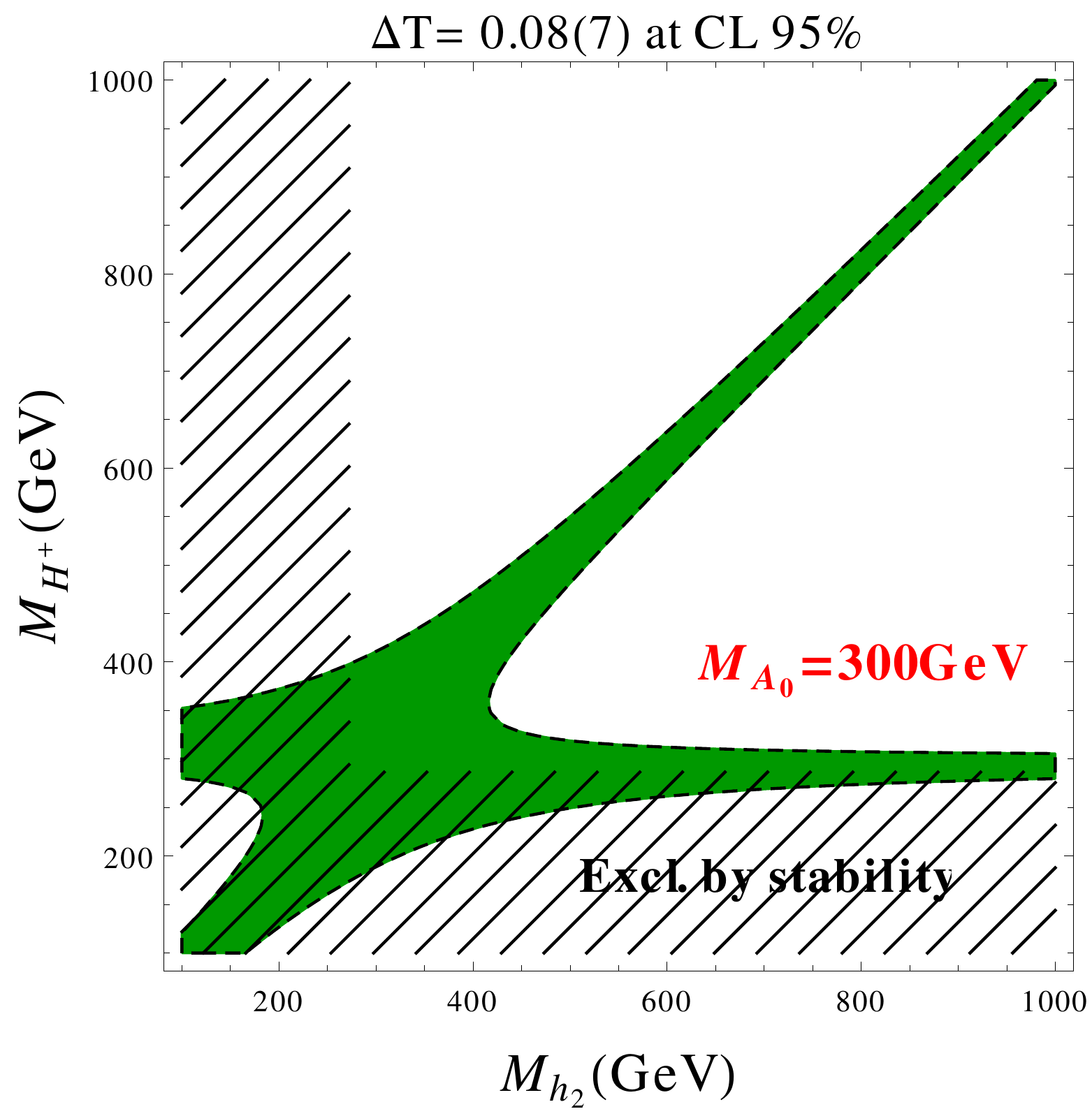}
\includegraphics[width=0.45\textwidth]{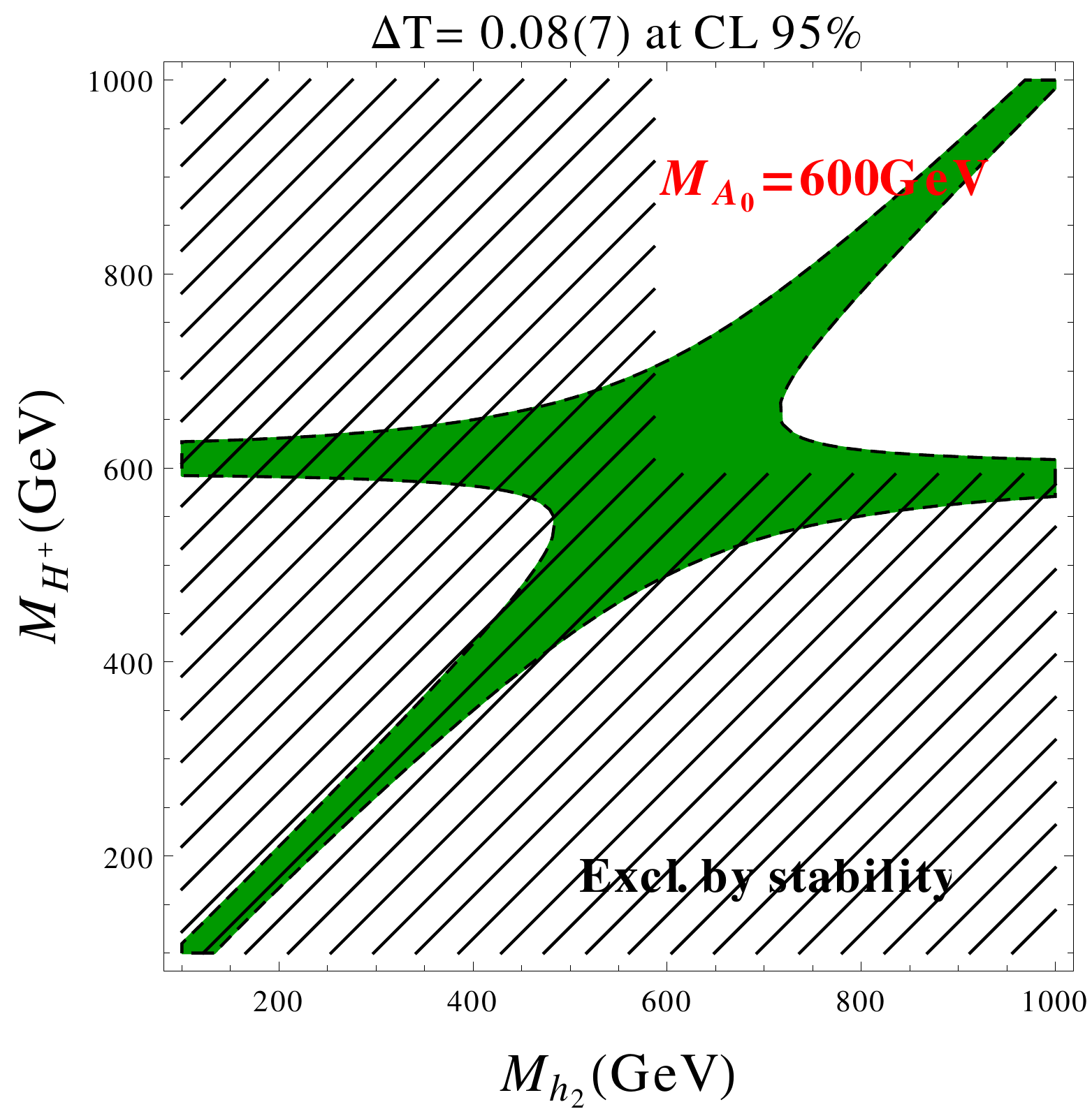}
\caption{Exclusion plot imposed by the constraint from $\Delta T$ on the second
$0^+$ state (i.e. `second Higgs') and the charged Higgs masses for two reference values of
$m_{A_0}$ in the `quasi-custodial' case. The concentration of points along approximately two axis 
is easy to understand
after inspection of the relevant formula for $\Delta T$. The regions excluded by considerations
of stability of the potential are shown.}
\label{eclM-su2}
\end{figure}

Finally let us turn to the consideration of the general case 4. We now completely give up 
custodial symmetry and hence the three masses $m_{A_0}$, $m_{H_\pm}$ and $m_{h_2}$ are unrelated, except for the eventual lack of stability of the potential. 
In this case, the rotation $R$ can be different from the identity, which was the case in the `quasi-custodial' scenario above.
In particular, $R_{S2}=\cos\theta$ from App.~\ref{app:B} and the angle 
$\theta$ is not vanishing. However, experimentally 
$\cos \theta$ is known to be
very close to one (see Sec.~\ref{sec:higgs}). If we assume that
$\cos\theta$ is exactly equal to one, we get the exclusion/acceptance regions shown
in Fig.~\ref{eclM-gen}. Finally, Fig.~\ref{eclM-theta} depicts the analogous plot for
$\cos\theta=0.95$ that is still allowed by existing constraints. We see that the allowed range
of masses is much more severely restricted in this case.

\begin{figure}[ht]
\center
\includegraphics[width=0.45\textwidth]{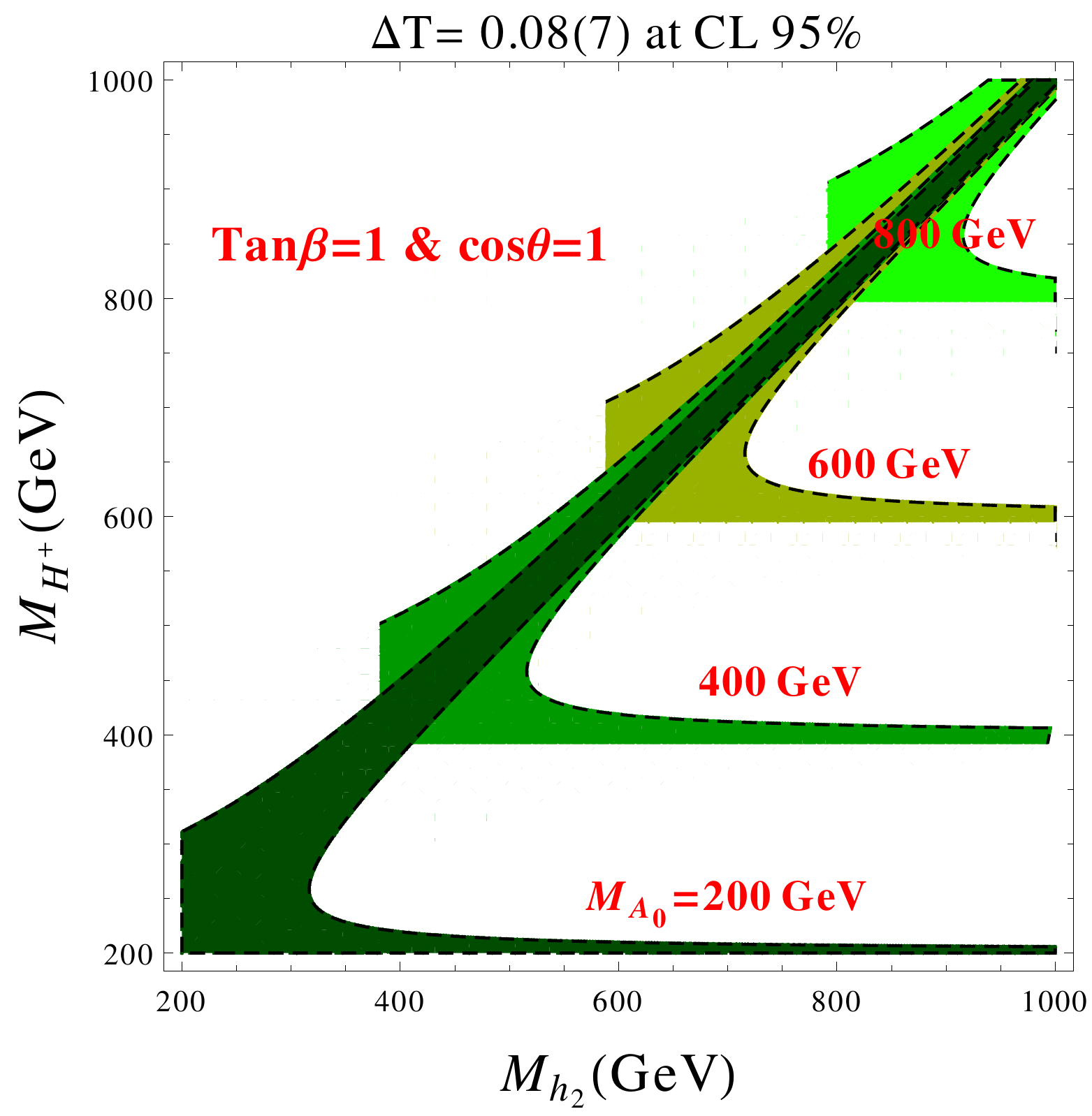}
\includegraphics[width=0.45\textwidth]{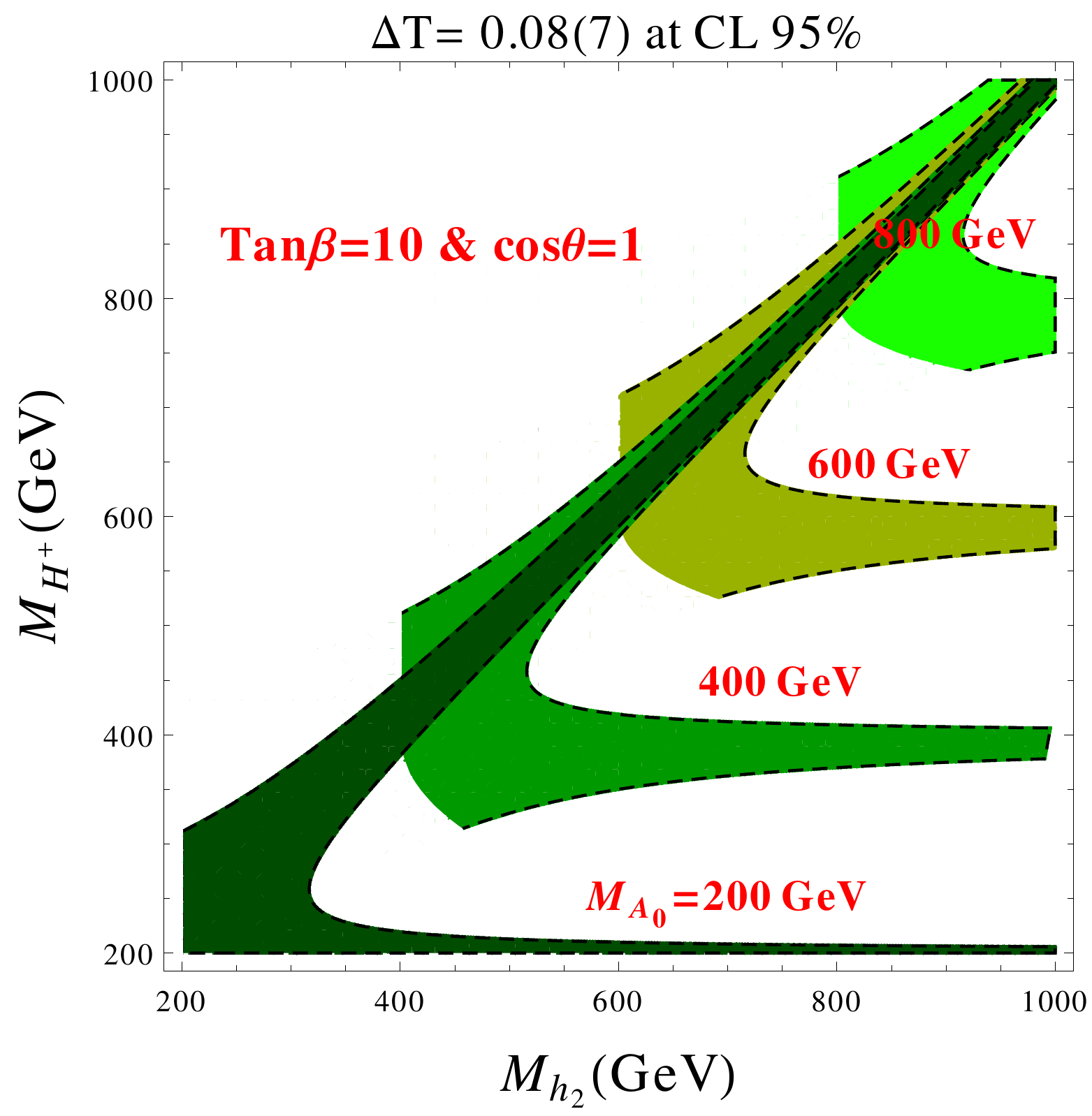}
\caption{Exclusion plot imposed by the constraint from $\Delta T$ on the second
$0^+$ state (i.e. `second Higgs') and the charged Higgs masses for several reference 
values of
$m_{A_0}$ and $\tan\beta$ in the general case. The value $\cos\theta=1$ is assumed here. The successive
horizontal bands correspond to different values of $m_{A_0}$. The stability bounds have already been implemented,  
effectively cutting off the left and lower arms of the regions otherwise acceptable, as seen in Fig.~\ref{eclM-su2}. }
\label{eclM-gen}
\end{figure}

\begin{figure}[ht]
\center
\includegraphics[width=0.45\textwidth]{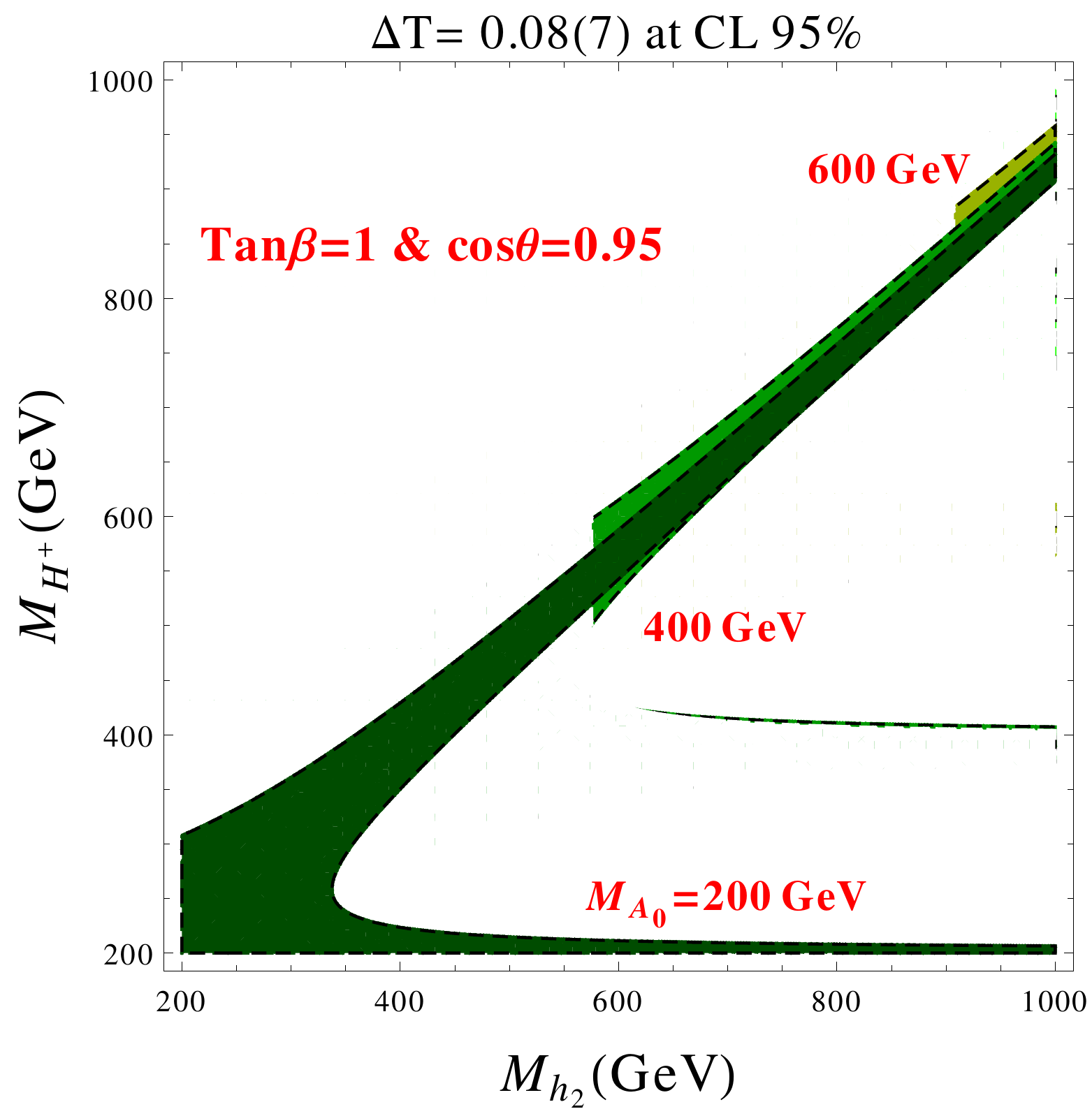}
\caption{Exclusion plot imposed by the constraint from $\Delta T$ on the second
$0^+$ state (i.e. `second Higgs') and the charged Higgs masses for several reference values of
$m_{A_0}$. Here we take $\tan\beta=1$ and allow $\cos\theta=0.95$, which is
consistent with present constraints. }
\label{eclM-theta}
\end{figure}

\section{Summary}

With the LHC experiments gathering more data, the exploration of the symmetry breaking sector of 
the Standard Model will gain renewed impetus. Likewise, it is important to search for dark matter
candidates as this is a degree of freedom certainly missing in the Minimal Standard Model. An 
invisible axion is an interesting candidate for dark matter; however trying 
to look for direct evidence of its existence at the LHC is hopeless as it is extremely weakly coupled. 
Therefore we have to resort to less direct ways to explore this sector
by formulating consistent models that include the axion and deriving 
consequences that could be experimentally tested.

In this chapter we have explored such consequences in the DFSZ model, an extension of the popular 2HDM.
A necessary characteristic of models with an invisible axion is the presence of the Peccei-Quinn
symmetry. This restricts the form of the potential. We have taken into account the recent
data on the Higgs mass and several effective couplings, and included the constraints from electroweak
precision parameters.

Four possible scenarios have been considered. In the majority of
parameter space of the DFSZ model we do not really expect to see any relevant
modifications with respect to the Minimal Standard Model predictions. The new
scalars have masses of order $v_\phi$ or $\sqrt{v v_\phi}$ in two of the cases
discussed. The latter could perhaps be reachable with a $100$ TeV circular
collider although this is not totally guaranteed. In a third case, it would be
possible to get scalars in the  multi-TeV region, making this case testable in
the future at the LHC.  Finally, we have identified a fourth situation where a
relatively light spectrum emerges. The last two cases correspond to a situation
where the coupling  between the singlet and the two doublets is of order
$v^2/v_\phi^2$; i.e. very small ($10^{-10}$ or less) and in order to get a
relatively light spectrum in addition one has to 
require some couplings to be commensurate (but not necessarily fine-tuned). 

The fact that some specific couplings are required to be very small may seem odd, but it is technically natural, as the couplings in question do break some extended
symmetry and are therefore protected. From this point of view these values are perfectly acceptable.

The results on the scalar spectrum are derived here at tree level only and are of course subject to large radiative
corrections. However one should note two ingredients that should ameliorate the hierarchy problem. The
first observation is that the mass of the $0^-$ scalar is directly proportional to $c$; it is exactly zero
if the additional symmetries discussed in \cite{2hdmNatural1,2hdmNatural2} hold. It is therefore somehow protected. On the
other hand custodial symmetry relates different masses, helping to maintain other relations. Some hierarchy
problem should still remain but of a magnitude very similar to the one already present in the Minimal Standard
Model. 

We have imposed on the model known constraints such as the fulfilment of the bounds on the $\rho$-parameter. These
bounds turn out to be automatically fulfilled in most of the parameter space and become only relevant when 
the spectrum is light (case 4). This is particularly relevant as custodial symmetry is by no means automatic
in the 2HDM. Somehow the introduction of the axion and the related Peccei-Quinn symmetry makes possible custodially
violating consequences naturally small. We have also considered the experimental bounds on the Higgs-gauge bosons and
Higgs-two photons couplings.

In conclusion, DFSZ models containing an invisible axion are natural and, in spite of the large
scale that appears in the model to make the axion nearly invisible, there is the possibility that
they lead to a spectrum that can be tested at the LHC. This spectrum is severely constrained, 
making it easier to prove or disprove such a possibility in the near future. 
On the other hand it is perhaps more likely that the new states predicted by the model
lie beyond the LHC range. In this situation the model hides itself by making 
indirect contributions to most observables quite small.

\graphicspath{{3-photon/figures/}{figures/}}

\chapter{Photon propagation in a cold axion background and a magnetic field}\label{ch-photon}

After the QCD phase transition, instanton effects induce a potential on the axion field, giving it a mass $m_a$. Astrophysical and
cosmological constraints (see Sec.~\ref{sec:astrocosmo}) force this mass
to be quite small. Yet, the axion provides cold dark matter, as it is not produced thermically.
If the axion background field is initially misaligned (not lying at the bottom of the instanton-induced potential),
at late times it oscillates coherently as
\begin{equation}\label{cos}
 a_b(t)= a_0 \sin\left( m_a t\right),
\end{equation}
where the amplitude, $a_0$, is related to the initial misalignment angle. See Sec.~\ref{sec:cab} for more details.

The oscillation of the axion field has an approximately constant (i.e. space-independent) energy
density given by
\be\label{density}
\rho=\frac12 a_0^2m_a^2,
\ee 
which contributes to the total energy of the Universe. If we assume that 
cold axions are the only contributors to the dark matter density 
apart from ordinary baryonic matter, their density must be~\cite{kolb,wmap2,wmap3}
\be
\rho\simeq 10^{-30}\text{ g\,cm}^{-3}\simeq 10^{-10}\text{ eV}^4.
\ee
Of course dark matter is not uniformly distributed, as it traces visible matter (or rather the other way round). In the  
galactic halo of dark matter (assumed to consist of axions) a typical value for the density would be~\cite{ggt}
\be\label{halodensity}
\rho_a\simeq 10^{-24}\text{ g\,cm}^{-3}\simeq 10^{-4}\text{ eV}^4
\ee
extending over a distance of 30  to 100 kpc in a galaxy such as
the Milky Way.

The mechanism of vacuum misalignment and the
subsequent redshift of momenta suggest that it is natural for axions to remain coherent (or
very approximately so) over relatively long distances, perhaps even forming a BEC as has been suggested.
Thus one should expect not only that the momentum of individual axions satisfies the condition
$k\ll m_a$ as required from cold dark matter but also all that axions oscillate in phase, rather than incoherently,
at least locally. In addition one needs that the modulus of the axion field is large
enough to account for the DM density.

Finding an axion particle with the appropriate characteristics is not enough
to demonstrate that a CAB exists. Detecting the coherence of the axion background and hence validating
the misalignment proposal is not the goal of most axion experiments.

The ADMX Phase II experiment~\cite{admx} tries to detect axions in the
Galaxy dark matter halo that, under the influence of a strong magnetic field, would convert to photons with
a frequency equal to the axion mass in a resonant cavity. This experiment is sensitive to the local axion density 
and in order to get a significant signal the axion
field has to be relatively constant at length scales comparable to the cavity size.
ADMX is therefore sensitive to the CAB.  The experiment claims sensitivity
to axions in the approximate mass range $10^{-6}$~eV to $10^{-5}$~eV and this is also the range of momenta
at which the axion background field can be significantly probed in such an experiment.

Looking for the collective effects on photon propagation resulting from the presence of a CAB is another
possible way of investigating whether a CAB is present at the scales probed by the experiment.
Of course we do not anticipate large or dramatic effects given the presumed smallness of the photon-to-axion
coupling and the low density background that a CAB would provide. However, interferometric and polarimetric
techniques are very powerful and it is interesting to explore
the order of magnitude of the different effects in this type of experiments.
Potentially, photons can also probe the CAB structure
in different ranges of momenta. In addition, precise photon measurements could in principle
check the coherence of the oscillations over a variety of distances.
Discussing in detail the effects of a CAB on photons is the purpose
of this chapter.

Several studies on the influence of axions on photon propagation at cosmological scales exist~\cite{cosmoaxions1,cosmoaxions2,cosmoaxions3,cosmoaxions4}.
The consequences are only visible for extremely low mass axions, such as the ones hypothetically
produced in string theory scenarios~\cite{axiverse}.
We do not consider very light axions here in detail as their masses do not fall into the favoured range but
exploring such small masses might be of interest too.\\
~\\
\indent The consequences
of the mere existence of axions as propagating degrees of freedom on
photon propagation have been studied for a long time and are well understood.
It is well known that photons polarised in
a direction perpendicular to a magnetic field are not affected by the existence of axions~\cite{raffelt1,raff&sto} but
photons polarised in the parallel direction mix with them. As a consequence there is a small rotation in the
polarisation plane due to photon-axion mixing as well as a change in the ellipticity~\cite{mpz}.\\
~\\
\indent Throughout this chapter we will see that the effects of the CAB on the propagation of photons are extremely small,
so it is quite pertinent to question whether these effects could be experimentally measured. The answer is surely
negative with present day experimental capabilities but some effects are not ridiculously small either to be
discarded from the outset: the effects of a coherent CAB are in some cases quite comparable
to, or even larger than, the influence of axions as mere propagating degrees of freedom, which
have been profusely studied before. They might even be comparable
to non-linear QED effects, which have also been actively sought for experimentally. Therefore we think
it is legitimate to present this study in view of the physical relevance of the presumed existence of a CAB
as a dark matter candidate.\\
~\\
\indent This chapter is structured as follows:
In Sec.~\ref{sec:eom} we review the problem and derive the equations
of motion for the axion and photon in the presence of both backgrounds, both for linear and circular polarisation
bases for the photon. We also review there the range of relevant values for the intervening parameters.
In Sec.~\ref{sec:gaps} we discuss the results for the case of no magnetic field, when there is no photon-axion
conversion but the CAB still mixes the two photon polarisations. Some
gaps in the photon momenta are present due to the time periodicity of the CAB.
We derive the precise location and width of these momentum
gaps. In Sec.~\ref{sec:magnetic} we study the consequences that the combined background
has on photon wave-numbers and polarisations. In Sec.~\ref{sec:rotation} we
explore the consequences of the change in the plane of polarisation of the photons in the presence of the CAB, making
use of the photon propagator derived in a combined CAB and constant magnetic field.

\section{Equations of motion of the axion and photon system}\label{sec:eom}
The Lagrangian density describing axions and photons consists of the usual kinetic terms plus
the interaction term of Eq.~\eqref{prima}
\begin{equation}\label{fulllagrangian}
\mathcal{L}=\frac12\partial_\mu a\partial^\mu a-\frac12m_a^2a^2-\frac14F_{\mu\nu}F^{\mu\nu}+
\frac g4aF_{\mu\nu}\tilde F^{\mu\nu},
\end{equation}
where we have rewritten the axion-photon coupling
as $g=g_{a\gamma\gamma}\frac{2\alpha}{\pi f_a}$.
We are not considering the non-linear effects due to the Euler-Heisenberg Lagrangian \cite{EH1,EH2}
that actually can provide some modifications in the polarisation plane.
Later we shall discuss their relevance.

We decompose the fields as a classical piece describing the backgrounds
(external magnetic field $\vec B$  and a CAB as given in Eq.~\eqref{cos}) plus quantum fluctuations describing
the photon and the axion particles, e.g. $a\to a_b + a$.
For the (quantum) photon field, we work in the Lorenz gauge,
$\partial_\mu A^\mu=0$, and use the remaining gauge freedom to set $A^0=0$. The resulting equations of motion are
\begin{equation}\label{EL}
\begin{array}{l}
(\partial_\mu\partial^\mu+m_a^2)a+gB^i\partial_tA_i=0,\\
\partial_\mu\partial^\mu A^i+gB^i\partial_ta+\eta\epsilon^{ijk}\partial_jA_k=0,
\end{array}
\end{equation}
where $\eta=g\partial_t a_b$. We neglect the space derivatives of $a_b$, thereby assuming homogeneity
of the axion background, at least at the scale of the photon momentum and translational invariance.
Since $\eta$ is time-dependent, we make a Fourier transform with respect to
the spatial coordinates only,
\begin{equation}
\phi(t,\vec x)=\int\frac{d^3k}{(2\pi)^3}e^{i\vec k\cdot\vec x}\hat\phi(t,\vec k),
\end{equation}
and get the equations
\begin{equation}\label{TF}
\begin{array}{l}
(\partial_t^2+\vec k^2+m_a^2)\hat a+gB^i\partial_t \hat A_i=0,\\
(\partial_t^2+\vec k^2)\hat A^i+gB^i\partial_t\hat a+i\eta\epsilon^{ijk}k_j\hat A_k=0.
\end{array}
\end{equation}
As can be seen, the presence of a magnetic field mixes the axion with the photon.
To proceed further, we write the photon field as
\begin{equation}
 \hat A_\mu(t,\vec k)=\sum_\lambda f_\lambda(t)\varepsilon_\mu(\vec k,\lambda),
\end{equation}
where $\varepsilon_\mu$ are the polarisation vectors and $f_\lambda(t)$ are the functions we will have to solve for.
If we choose a linear polarisation basis for the photon, the equations are, in matrix form,
\begin{equation}\label{linear}
\left(
\begin{array}{ccc}
 \partial_t^2+k^2+m_a^2 &   -ib\partial_t    & 0        \\
 -ib\partial_t          & \partial_t^2+k^2 & -\eta(t)k                    \\
 0      & -\eta(t)k                & \partial_t^2+k^2
\end{array}
\right)
\left(
\begin{array}{c}
 \hat a \\ i f_\parallel \\ f_\perp
\end{array}
\right)
=
\left(
\begin{array}{c}
 0 \\ 0 \\ 0
\end{array}
\right),
\end{equation}
where $k=|\vec k|$ and $b=g|\vec B^\perp|$, where $\vec B^\perp$ is the component of the magnetic field
perpendicular to the momentum (the parallel
component does not affect propagation at all if the Euler-Heisenberg piece is neglected).
The subscripts $\parallel$ and $\perp$ of the photon polarisations refer to parallel or perpendicular to this $ \vec B^\perp$.
In a circular polarisation basis, defining
\begin{equation}\label{imaginary}
 f_\pm=\frac{f_\parallel\pm if_\perp}{\sqrt2},
\end{equation}
the equations take the form
\begin{equation}\label{circulareq}
\left(
\begin{array}{ccc}
 \partial_t^2+k^2+m_a^2 &   i\frac b{\sqrt2}\partial_t    & i\frac b{\sqrt2}\partial_t        \\
 i\frac b{\sqrt2}\partial_t   & \partial_t^2+k^2+\eta(t)k & 0                    \\
 i\frac b{\sqrt2}\partial_t      & 0                & \partial_t^2+k^2-\eta(t)k
\end{array}
\right)
\left(
\begin{array}{c}
 i\hat a \\  f_+ \\ f_-
\end{array}
\right)
=
\left(
\begin{array}{c}
 0 \\ 0 \\ 0
\end{array}
\right).
\end{equation}
As we see from the previous expressions, the presence of a CAB changes in a substantial way the mixing
of photons and axions. Now all three degrees of freedom are involved.

A difference in the approach between this work and Ref.~\cite{raff&sto} is worth noting.
In going from Eq.~\eqref{EL} to Eq.~\eqref{TF} we have performed a
Fourier transform in space, but not in time, because the magnetic field is homogeneous
but $\eta(t)$ is time-dependent. Equation (4) in Ref.~\cite{raff&sto}, however,
uses a transform in time rather than in space because the CAB is not considered.

There are several ways to deal with the periodic CAB. One possibility is to try to treat it exactly. Unfortunately
this unavoidably leads to the appeareance of Mathieu functions due to the sinusoidal variation
of the background and the analysis becomes extremely involved.
On the other hand, the substantial ingredient in the problem is the existence of periodicity itself
and the fine details are not so relevant\footnote{Recall that the generic appeareance of bands in the energy levels
of a solid relies on the
periodicity of the potential and not on its precise details.}. Therefore, to
keep the discussion manageable, we approximate the sinusoidal variation of the axion background $a_b(t)$ in
Eq.~\eqref{cos} by a piecewise linear function, see Fig.~\ref{fig:tri}.
 \begin{figure}[ht]
 \center
 \includegraphics[scale=0.8]{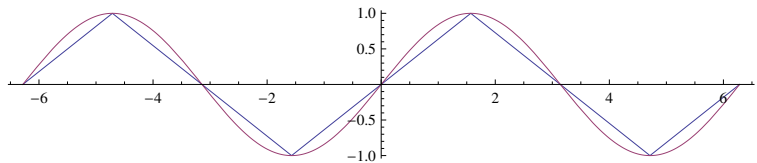}
 \caption{$a_b(m_at)/a_0$ and its approximating function.}\label{fig:tri}
 \end{figure}
Since $\eta(t)$ is proportional to the time derivative of $a_b(t)$, in this approximation it is a square-wave function,
alternating between intervals
where $\eta=\eta_0$ and $\eta=-\eta_0$ with a period $2T=2\pi/m_a$.
Here, $\eta_0=\frac2\pi ga_0m_a=g_{a\gamma\gamma}\frac{4\alpha}{\pi^2}\frac{a_0m_a}{f_a}$.

A brief numerical discussion of the parameters involved in the problem and their relative
importance is now in order.
The bound~\eqref{bounds} on $f_a$ implies one on $g$. Taking $g_{a\gamma\gamma}$ of $\mathcal{O}(1)$, the
range $f_a=10^7 - 10^{11} \text{ GeV}$  translates to $g=10^{-18} - 10^{-22}\text{ eV}^{-1}$.
Assuming a halo dark matter density of $\rho=10^{-4}\text{ eV}^4$ this means that $\eta_0=10^{-20} - 10^{-24} \text{ eV}$.
When working with natural units and magnetic fields it is useful to know that $1\text{ T}\approx195\text{ eV}^2$.
To have a reference value, a magnetic field of $10\text{ T}$ implies the
range $b=10^{-15} - 10^{-19}\text{ eV}$, for $f_a=10^7 - 10^{11} \text{ GeV}$.

Finally, let us now comment on the relevance of the contribution of the Euler-Heisenberg pieces compared
to the ones retained
in the description provided by Eq.~\eqref{fulllagrangian}. As it is known (see e.g. Refs.~\cite{raffelt1,raff&sto}), an external
magnetic field perpendicular to the photon motion contributes, via the Euler-Heisenberg terms, to the
mixing matrices,  affecting the (2,2) and (3,3) entries of Eqs.~\eqref{linear} and~\eqref{circulareq}.
They modify the $k^2$ terms with corrections of order $10^{-2}\times \alpha^2 \times (B^2/m_e^4)$, where
$m_e$ is the electron mass, leading to birefringence and therefore to ellipticity. For magnetic fields of $\sim 10$ T this
gives a contribution of order $10^{-21}$ that
may be comparable to axion-induced effects for large magnetic
fields, particularly if $f_a$ is very large, or to the effects from the CAB (which for $k\sim 1$ eV are
in the range $10^{-20} - 10^{-24}$). Since there is no new physics involved
in the contribution from the Euler-Heisenberg Lagrangian, in order to facilitate the analysis we will not consider it here.
In any case given the smallness of the Euler-Heisenberg and the axion effects, they can safely be
assumed to be additive. The relevant modifications due to
the Euler-Heisenberg term can be found in Refs.~\cite{raffelt1,raff&sto,pvlas2}.

Of course, the effects of the Euler-Heisenberg Lagrangian are absent or negligible if there is no magnetic field
or if it is relatively weak,
and we will see that for a range of parameters the effect of a CAB  might be comparable to the former.

\section{No magnetic field}\label{sec:gaps}
If there is no magnetic field ($b=0$, $\eta_0\neq0$) the axion and the photon are no longer mixed. Because $\eta(t)$ does mix the two linear polarisations,
in this case it is useful to choose the circular polarisation basis, which diagonalises the system, as can be seen from Eq.~\eqref{circulareq} by setting $b=0$. 
The equation for the two photon polarisations is
\be
\left[\partial_t^2+k^2\pm\eta(t)k\right]f_\pm(t)=0.
\ee
As mentioned,
we will approximate the sine function in $\eta(t)$ by a square wave function:
\be\label{square}
\eta(t)=\left\{\begin{array}{cc}
                 +\hh & 2nT<t<(2n+1)T \\
                 -\hh & (2n+1)T<t<2nT
               \end{array}
\right..
\ee
There is an equation for each polarisation. However, they are related.
To recover one from the other we can just make the replacement $\hh\rightarrow-\hh$. Also,
because $\eta(t)$ changes sign after a time $T$, one solution
is a time-shifted copy of the other: $f_-(t)=f_+(t+T)$.
In what follows we will work in the case $\lambda=+$. It is obvious that our conclusions also apply to the
other physical polarisation, $\lambda=-.$\\
Since $\eta(t)$ is defined piecewise, we will solve the equation in two regions:\\
-- Region 1:  $0<t<T$, $\eta(t)=\hh$
\be
\frac{d^2f_1(t)}{dt^2}+(k^2+\hh k)f_1(t)=0,
\ee
\be\label{alpha}
f_1(t)=A'e^{i\omega_+ t}+Ae^{-i\omega_+ t}~,\quad\omega_+^2= k^2+\hh k.
\ee
-- Region 2: $-T<t<0$, $\eta(t)=-\hh$
\be
\frac{d^2f_2(t)}{dt^2}+(k^2-\hh k)f_2(t)=0,
\ee
\be\label{beta}
f_2(t)=B'e^{i\omega_- t}+Be^{-i\omega_- t}~,\quad\omega_-^2= k^2-\hh k.
\ee
We impose that both functions coincide at $t=0$ and we do the same for their derivatives
\be
f_1(0)=f_2(0), \qquad f^\prime_1(0)=f^\prime_2 (0).
\ee
We now write $f(t)=e^{-i\Omega t}g(t)$ and demand that $g(t)$ have the same periodicity as $\eta(t)$
\begin{align}
g_1(t)=e^{i\Omega t}f_1(t)=A'e^{i(\Omega+\omega_+) t}+Ae^{i(\Omega-\omega_+) t},\cr
g_2(t)=e^{i\Omega t}f_2(t)=B'e^{i(\Omega+\omega_-) t}+Be^{i(\Omega-\omega_-) t},\cr
g_1(T)=g_2(-T),\quad g_1'(T)=g_2'(-T).
\end{align}
For these conditions to be fulfilled, the coefficients have to solve the linear system
\be
{\hat M}
\left(
  \begin{array}{c}
    A'  \\
    A \\
    B'  \\
    B \\
  \end{array}
\right)=
\left(
  \begin{array}{c}
    0 \\
    0 \\
    0 \\
    0 \\
  \end{array}
\right)
,\ee
with
\be
{\hat M^T}=\left(
  \begin{array}{cccc}
    1 & \omega_+ & e^{i(\Omega+\omega_+)T} & (\Omega+\omega_+)e^{i(\Omega+\omega_+)T} \\
    1 & -\omega_+ & e^{i(\Omega-\omega_+)T} & (\Omega-\omega_+)e^{i(\Omega-\omega_+)T} \\
    -1 & -\omega_- & -e^{-i(\Omega+\omega_-)T} & -(\Omega+\omega_-)e^{-i(\Omega+\omega_-)T} \\
    -1 & \omega_- & -e^{-i(\Omega-\omega_-)T} & -(\Omega-\omega_-)e^{-i(\Omega-\omega_-)T} \\
  \end{array}
\right).
\ee
~\\
\indent The problem being discussed here is formally similar to the solution of the Kronig-Penney one-dimensional
periodic potential~\cite{kp}, except the periodicity is now in time rather than in space.
In order to find a non-trivial solution one has to demand the condition of
vanishing determinant of $\hat M$, which reduces to
\be
\cos(2\Omega T)=\cos(\omega_+ T)\cos(\omega_- T)-\frac{\omega_+^2+\omega_-^2}{2\omega_+\omega_-}\sin(\omega_+ T)\sin(\omega_- T).
\label{kp}
\ee
In order to get analytical expressions we will work in
the limit of long wavelengths $k T\ll 1$, which is just the one
that is potentially problematic as discussed at the beginning of this chapter.
Expanding both sides:
\be
\Omega^2-\frac13\Omega^4T^2+...= k^2-\left(\frac13 k^4-\frac1{12}\hh^2 k^2\right)T^2+...,
\ee
which means
\be
\Omega^2\approx\left(1+\frac{\hh^2T^2}{12}\right) k^2.
\ee
If the determinant vanishes, the system to solve is
\be
\left(
  \begin{array}{ccc}
   1  & 1 & -1 \\
    0  & 1  & -\frac12(1-\frac{\omega_-}{\omega_+}) \\
     0 & 0 & 1 \\
  \end{array}
\right)\left(
         \begin{array}{c}
           A' \\
           A \\
           B' \\
         \end{array}
       \right)=\left(
                 \begin{array}{c}
                   1 \\
                   \frac12(1+\frac{\omega_-}{\omega_+}) \\
                   h(\omega_+,\omega_-,T) \\
                 \end{array}
               \right)B,
\ee
where
\be
h(\omega_+,\omega_-,T)=-\frac{\omega_+-\omega_-}{\omega_++\omega_-}\frac{e^{i\omega_+ T}-e^{-i2\Omega T}e^{i\omega_- T}}{e^{i\omega_+ T}-e^{-i2\Omega T}e^{-i\omega_- T}},
\ee
leading to
\begin{align}\label{coefficients}
\frac{A'}B={}&\left[1-\frac{\omega_+-\omega_-}{\omega_++\omega_-}\frac{e^{i\omega_+ T}e^{i2\Omega T}-e^{i\omega_- T}}{e^{i\omega_+ T}e^{i2\Omega T}-e^{-i\omega_- T}}\right.\cr
&\left. - \frac12(1+\frac{\omega_-}{\omega_+})+\frac12(1-\frac{\omega_-}{\omega_+})\frac{\omega_+-\omega_-}{\omega_++\omega_-}\frac{e^{i\omega_+ T}e^{i2\Omega T}-e^{i\omega_- T}}{e^{i\omega_+ T}e^{i2\Omega T}-e^{-i\omega_- T}}\right],\cr
\frac AB={}&\left[\frac12(1+\frac{\omega_-}{\omega_+})-\frac12(1-\frac{\omega_-}{\omega_+})\frac{\omega_+-\omega_-}{\omega_++\omega_-}\frac{e^{i\omega_+ T}e^{i2\Omega T}-e^{i\omega_- T}}{e^{i\omega_+ T}e^{i2\Omega T}-e^{-i\omega_- T}}\right],\cr
\frac{B'}B={}&\left[-\frac{\omega_+-\omega_-}{\omega_++\omega_-}\frac{e^{i\omega_+ T}e^{i2\Omega T}-e^{i\omega_- T}}{e^{i\omega_+ T}e^{i2\Omega T}-e^{-i\omega_- T}}\right].
\end{align}

In the limit $\hh\ll k$, $k T\ll1$,
\be\label{limit}
\frac{A'}B\approx-\frac{B'}B\approx\frac14\frac{\hh} k, \quad \frac{A}B\approx1-\frac{\hh}{2k}
.\ee
Finally, imposing the usual normalization,
\be
\displaystyle\int f_k(t)f_{k'}^*(t)=2\pi\delta(k-k'),
\ee
we get
\be\label{BB}
B=\left[\frac{\sqrt{ k^2+\hh k}}{2k+\hh}\left(\left|\frac AB\right|^2+
\left|\frac {A'}B\right|^2\right)+\frac{\sqrt{k^2-
\hh k}}{2k-\hh}\left(1+\left|\frac {B'}B\right|^2\right)\right]^{-1/2}\approx\left(1+\frac{\hh}{4k}\right).
\ee

Equation \eqref{kp} implies the existence of momentum gaps: some values of $k$ admit no solution for $\Omega$, much like
some energy bands are forbidden in a semiconductor. Here, however, the roles of momentum and energy are
exchanged, since the periodicity is in time
rather than in space.
The solutions are shown in an $\Omega(k)$ plot in Fig.~\ref{fig:gaps} for two values of the ratio $\eta_0/m_a$.
One of the ratios shown is unreasonably large, in order to
show clearly the existence of the gaps.
\begin{figure}[tbh]
\centering
        \begin{subfigure}[b]{0.35\textwidth}
                \includegraphics[width=\linewidth]{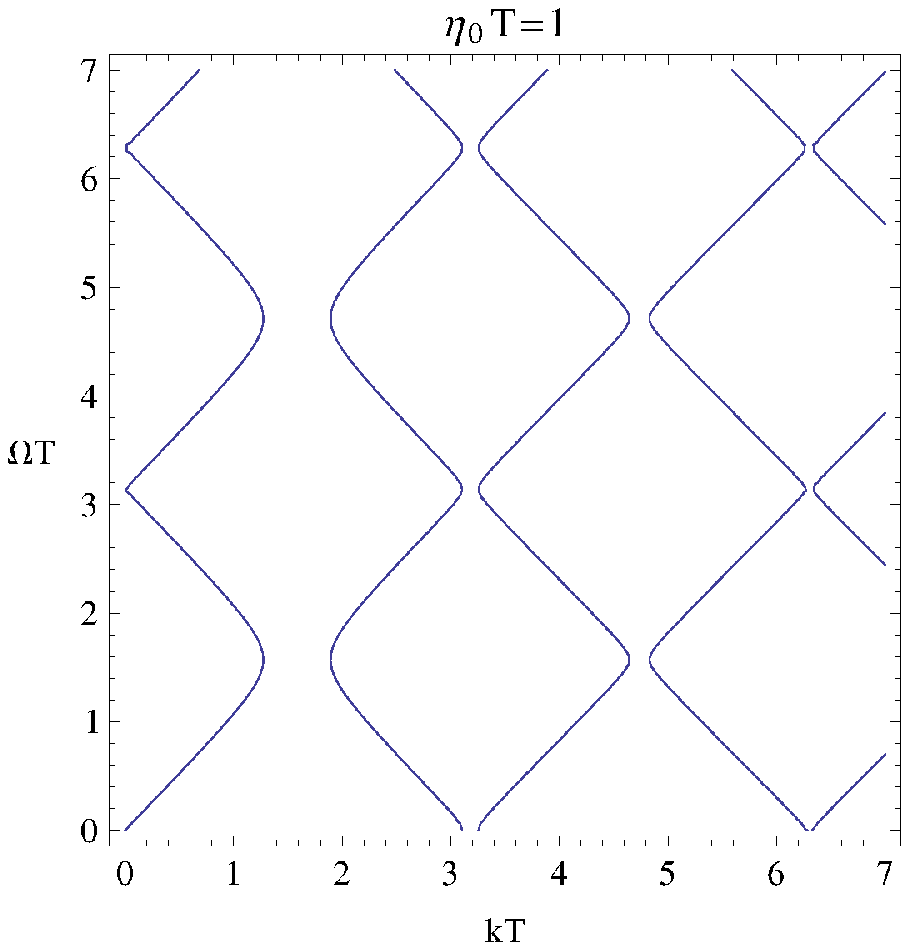}
        \end{subfigure}
        \begin{subfigure}[b]{0.35\textwidth}
                \includegraphics[width=\linewidth]{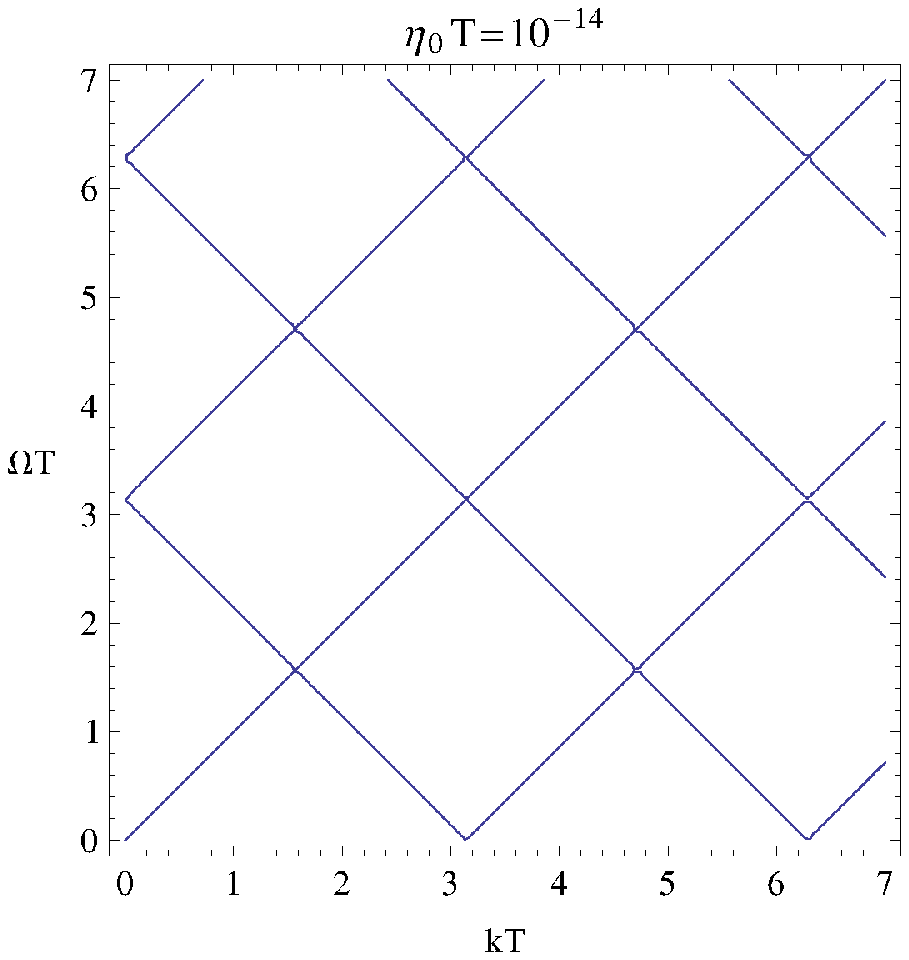}
        \end{subfigure}
        \caption{Plot of the solutions to the gap equation. In the left figure the value for the ratio $\eta_0/m_a$ is unreasonably large
        and it is presented here only to make the gaps in the photon momentum clearly visible.}\label{fig:gaps}
\end{figure}

The first order in $\eta_0$ drops from Eq.~\eqref{kp} but to second order it reads
\begin{equation}
 \cos(2\Omega T)=\cos(2kT)+\frac{\eta_0^2}{4k^2}\left[-1+\cos(2kT)+kT\sin(2kT)\right]
\end{equation}
(recall that $T=\pi/m_a$). There is no solution when the r.h.s. of this expression becomes larger
than one. The gaps are approximately located at
\be
k_n=\frac{nm_a}2,~n\in\mathbb{N}
\ee
and their width is
\begin{equation}
 \Delta k\sim\left\{\begin{array}{cc}
\displaystyle\frac{\eta_0}{n\pi} & \text{for $n$ odd} \\
\\
\displaystyle\frac{\eta_0^2}{2nm_a} & \text{for $n$ even}
                    \end{array}
                    \right..
\end{equation}
These results agree well with the exact results as can be easily seen in the left side of Fig.~\ref{fig:gaps}.
Unfortunately we are not aware of any way of detecting such a tiny forbidden band for the range of values of $\eta_0$ previously
quoted ($10^{-20}$ eV or less) that correspond to the allowed values of $f_a$.

It may be interesting to think what would happen if one attempts to produce a `forbidden' photon, i.e. one
whose momentum falls in one of the forbidden bands. A photon with
such a wave number is `off-shell' and as such it will always decay. For instance, it could decay into three
other photons with appropiately lower energies. However, because the off-shellness is so small (typically
$10^{-20}$ eV or less) it could live for a long time as a metastable state, travelling distances
commensurable with the solar system. For more technical details, see e.g. Ref.~\cite{sasha}.

We realise that the small bandwidth of the forbidden momentum bands make them unobservable
in practice. However
their mere existence is of theoretical interest. Conclusions might be different for other axion-like backgrounds.

\section{Proper modes in a magnetic field and axion background}\label{sec:magnetic}
In the presence of a magnetic field, but no CAB ($b\neq0$, $\eta_0=0$) there is no longer a
time dependence in the coefficients of the equations, so
we can Fourier transform with respect to time as well. We find the following dispersion relations:
\begin{align}\label{dispnoeta}
\omega_a^2&=k^2+\frac{m_a^2+b^2}{2}+\frac12\sqrt{(m_a^2+b^2)^2+4b^2k^2}\approx (k^2 + m_a^2)
\left(1+\frac{b^2}{m_a^2}\right)\cr
\omega_1^2&=k^2+\frac{m_a^2+b^2}{2}-\frac12\sqrt{(m_a^2+b^2)^2+4b^2k^2}\approx k^2 \left(1-\frac{b^2}{m_a^2}\right)\cr
\omega_2^2&=k^2,
\end{align}
where the $\approx$ symbol indicates the limit $\frac{bk}{m_a^2}\ll1$. These results are well known~\cite{mpz}.
We have identified as corresponding to `photons' the two modes that if $b=0$ reduce to the two usual polarisation
modes. The third frequency corresponds predominantly
to the axion (or axion-like particle), but of course it has also a small photon component as
the $\parallel$ polarised photon mixes with the axion.

If laser light of frequency $\omega$ is injected into a cavity, the different components will develop different
wave-numbers resulting in the appeareance of changes in the plane of polarisation (ellipticity
and rotation) unless the photon polarisation is initially exactly parallel or exactly perpendicular to the
magnetic field. We will review these effects later.
From the above expressions it would appear that the relevant figure of merit to observe
distortions with respect the unperturbed photon propagation is the ratio $\frac{b^2}{m_a^2}$ and
this is indeed true at large times or distances (precisely for $x\gg \frac{\omega}{m_a^2}$). This number is of
course very small, typically $10^{-28}$ for the largest conceivable magnetic fields (note that this
ratio is actually independent of $f_a$ and $m_a$ provided that we are considering Peccei-Quinn axions.)

Laser interferometry is extremely precise and Michelson-Morley type experiments
are capable of achieving a relative error as small as $10^{-17}$ using  heterodyne interferometry techniques~\cite{LI1,LI2}
and the PVLAS collaboration  claims that a sensitivity of order $10^{-20}$ in the difference of
refraction indices is ultimately achievable~\cite{pvlas} (see also Ref.~\cite{tamyang}).
In spite of this the above figure seems way too small to be detectable.

Let us now explore the situation where both the CAB and the magnetic field are present. We choose to work with the
linear polarisation basis. Again, in each time interval  we can define $(a,if_\parallel,f_\perp)=e^{i\omega t}(x,iX_\parallel,X_\perp)$.
Then the equations in matrix form are
\begin{equation}
\left(
\begin{array}{ccc}
 -\omega^2+k^2+m_a^2 & \omega b         & 0         \\
 \omega b          & -\omega^2+k^2 & -\eta_0 k                    \\
 0      & -\eta_0 k                & -\omega^2+k^2
\end{array}
\right)
\left(
\begin{array}{c}
x\\ iX_\parallel \\ X_\perp
\end{array}\right)
=
\left(
\begin{array}{c}
 0 \\ 0 \\ 0
\end{array}
\right),
\end{equation}
and involve a full three-way mixing as previously mentioned.
The proper frequencies of the system turn out to be
\begin{align}
\omega_a^2&= k^2+\frac{m_a^2+b^2}3+2\sqrt{Q}\cos\phi,\cr
\omega_1^2&= k^2+\frac{m_a^2+b^2}3-\sqrt{Q}\left(\cos\phi+\sqrt3\sin\phi\right),\cr
\omega_2^2&= k^2+\frac{m_a^2+b^2}3-\sqrt{Q}\left(\cos\phi-\sqrt3\sin\phi\right),
\end{align}
where
\begin{align}
 \phi&=\frac13\arctan\frac{\sqrt{Q^3-R^2}}R,\cr
 Q&=\left(\frac{m_a^2+b^2}3\right)^2+\frac13k^2(b^2+\eta_0^2),\cr
 R&=\frac{m_a^2+b^2}{54}\left[2m^4+b^2(9k^2+4m_a^2+2b^2)\right]-\frac{\eta_0^2k^2}{6}(2m_a^2-b^2).
\end{align}
It can be observed that they depend only on even powers of $\eta_0$, so they are not altered when $\eta(t)$ changes sign.
According to the discussion at the end of Sec.~\ref{sec:eom} the limit $\eta_0\ll b\ll\{m_a,k\}$ is quite reasonable.
The approximate expressions for the proper frequencies in this limit are\footnote{Extreme care has
to be exercised when using approximate formulae based on series expansions in $b$ or $\eta_0$ because there
is a competition among dimensionful quantities, several of which take rather small values.}

\begin{align}
\omega_a^2&\approx (k^2+m_a^2)\left(1+\frac{b^2}{m_a^2}\right),\cr
\omega_1^2&\approx k^2-k\sqrt{\eta_0^2+\left(\frac{b^2k}{2m_a^2}\right)^2
}-\frac{b^2k^2}{2m_a^2},\cr
\omega_2^2&\approx k^2+k\sqrt{\eta_0^2+\left(\frac{b^2k}{2m_a^2}\right)^2
}-\frac{b^2k^2}{2m_a^2}.
\end{align}

Corresponding to each frequency, the eigenvectors that solve the system are
\begin{equation}\label{autovec}
\omega_a:~
\left(
\begin{array}{c}
1\\
\displaystyle \frac{b\sqrt{k^2+m_a^2}}{m_a^2}\\
\displaystyle -\frac{\eta_0bk\sqrt{k^2+m_a^2}}{m_a^4}
\end{array}\right)
,\quad
\omega_1:~
\left(
\begin{array}{c}
\displaystyle-\frac{bk}{m_a^2}\\
1\\
\varepsilon
\end{array}\right),\quad
\omega_2:~
\left(
\begin{array}{c}
\displaystyle\frac{bk}{m_a^2}\varepsilon\\
-\varepsilon\\
1
\end{array}\right),
\end{equation}
where
\be\label{ellipt}
\varepsilon=\frac{\eta_0}{\frac{b^2k}{2m_a^2}+\sqrt{\eta_0^2+\left(\frac{b^2k}{2m_a^2}\right)^2}}.
\ee

Note that the above eigenvectors are written in the basis described in Eq.~\eqref{imaginary} that includes an imaginary unit for the
parallel component. Therefore the eigenvectors for $\omega_{1,2}$ correspond to photon states elliptically polarised with
ellipticity
\footnote{Ellipticity is the ratio of the minor to major axes of an ellipse.} $|\varepsilon|$. In addition, unless exactly aligned
to the magnetic field there will be a change in the angle of polarisation. We will return to this in Sec.~\ref{sec:rotation}.

We also note that the above value for $\varepsilon$ corresponds to the ellipticity of the eigenmodes. In Sec.~\ref{sec:rotation}
we will discuss the
evolution of the ellipticity of photon state that is initially linearly polarised.

Let us now try to get some intuition on the relevance of the different magnitudes entering in the expressions.
There are two different limits we can study, depending on which term in the square root in Eq.~\eqref{ellipt} dominates.
If $\displaystyle\frac{|\eta_0|}{k}\ll \frac{b^2}{2m_a^2}$ we have $\displaystyle\varepsilon\approx\frac{\eta_0m_a^2}{b^2k}$.
The ellipticity of the eigenmodes is small, so the proper modes are almost linearly polarised photons.
In the case $\displaystyle\frac{|\eta_0|}{k}\gg \frac{b^2}{2m_a^2}$ we have
$\displaystyle\varepsilon\approx\text{sign}(\eta_0)\left(1-\frac{b^2k}{2|\eta_0|m_a^2}\right)$.
Now the ellipticity of the eigenmodes is close to $1$ so the proper modes are almost circularly polarised. We see that while the
proper frequencies depend only on the square of $\eta_0$ (and therefore do not change as we go from one
time interval to the next) the eigenvectors do change with its sign.

The discussion on the size of the different parameters done in Sec.~\ref{sec:eom} and also in this section indicates that
the effect from the cold axion background is actually the dominant one for Peccei-Quinn axions, well above the
effects due to the presence of the magnetic field. Unfortunately both are minute.
In the limit where the magnetic field can be neglected, the photon proper frequencies are
\be\label{dispersion}
\omega_\pm^2 = k^2 \pm \eta_0k .
\ee

Axion-like particles are not constrained by the relation $f_a m_a \simeq {\rm constant}$
required of Peccei-Quinn axions and using (somewhat arbitrarily) the
largest value of $b$ discussed and the smallest mass for $m_a$ we get
a value for $b^2/m_a^2$ in the region $\sim10^{-18}$, to be compared with the largest
acceptable value for $\eta_0$ that gives  $\eta_0/k \sim 10^{-20}$ if $k\sim 1$ eV.
Sensitivity to the magnetic field could be enhanced by being able to
reproduce the experiment with even larger magnetic fields.\footnote{Non-destructive magnetic fields close to
$100\text{ T}$ have been achieved. This would enhance the sensitivity by a factor 100.}

\section{Propagator in a magnetic field and an axion background}\label{sec:propagator}
We will now compute the propagator of a photon field with two backgrounds: a cold axion background and
a constant magnetic field. From the interaction term in Eq.~\eqref{prima} we get two relevant terms:
\be\label{vertexs}
\mathcal L_{a\gamma\gamma}\ni\frac12\eta\epsilon^{ijk} A_i\partial_j A_k
+ga\partial_\mu A_\nu \tilde F^{\mu\nu},
\ee
where $\tilde F^{\mu\nu}$ stands for the dual tensor of the external magnetic field: $\tilde F^{0i}=B^i$, $\tilde F^{ij}=0$.

Here we will take $\eta(t)$ to be constant; therefore the results that follow
are valid only if the distance travelled by the photon, $l$, verifies $l< 2\pi/m_a$.

The vertices and Feynman rules corresponding to these terms are shown in Fig.~\ref{vertex}.
\begin{figure}[ht]
\center
\includegraphics[width=0.9\textwidth]{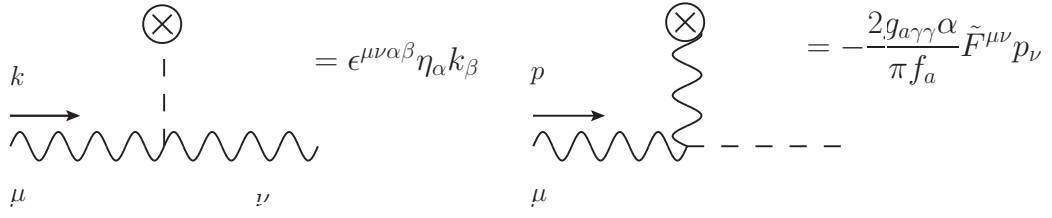}
\caption{The two relevant vertices with their corresponding Feynman rules.}
\label{vertex}
\end{figure}
To compute the propagator, we first sum all the contributions coming from the axion background, (double wavy line, see Fig.~\ref{prop1}). 
Unless otherwise noted, for the discussion of the propagator, the 4-vector notation will be used. $k=(\omega,\vec k)$ and $k^2\equiv k^\mu k_\mu$.
\be
D^{\mu\nu}=-i\left(\frac{X^{\mu\nu}}{k^2}+\frac{P_+^{\mu\nu}}{k^2-\hh |\vec k|}
+\frac{P_-^{\mu\nu}}{k^2+\hh |\vec k|}\right).\label{photonprop}
\ee
Here $X^{\mu\nu}$ is defined as
\be
X^{\mu\nu}=g^{\mu\nu}-\frac{S^{\mu\nu}}{\eta_0^2\vec k^2},
\ee
where $S^{\mu\nu}$ is defined in Eq.~\eqref{S}. Its components are
\be
X^{00}=1,\quad X^{i0}=X^{0i}=0,\quad X^{ij}=-\frac{k^ik^j}{\vec k^2},
\ee 
so this piece will vanish when contracted with a polarisation vector.

The physical polarisations are projected out by $P_\pm^{\mu\nu}$ and exhibit poles at
$\omega^2=\vec k^2 \pm \eta_0 |\vec k|$, as expected. The projectors are defined in Eq.~\eqref{p+-}.

\begin{figure}[ht]
\center
\includegraphics[width=0.9\textwidth]{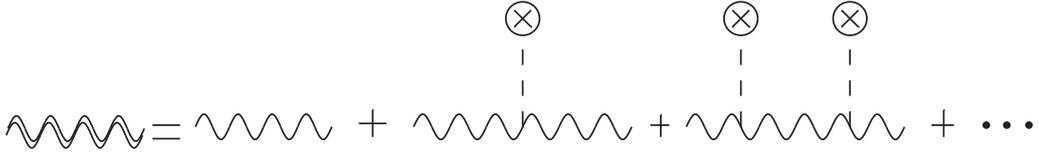}
\caption{Propagator in the axion background.}
\label{prop1}
\end{figure}

We now compute the propagator in the presence of a magnetic field, using the second term in Eq.~\eqref{vertexs}.
In order to do that we use the propagator just found and include the
interactions with the external magnetic field (triple wavy line, see Fig.~\ref{prop2}. The dashed line corresponds to the axion propagator.).
\begin{figure}[ht]
\center
\includegraphics[width=0.9\textwidth]{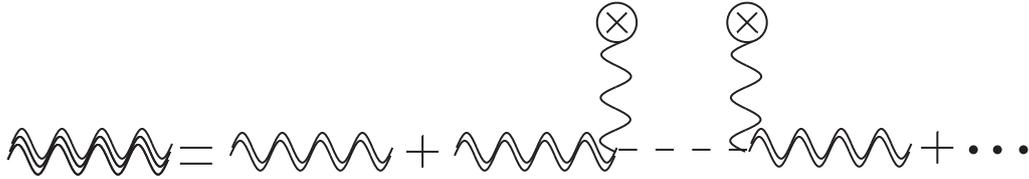}
\caption{Full propagator after resummation of the interactions with the external magnetic field.}
\label{prop2}
\end{figure}
\be
\mathcal D_{\mu\nu}=D_{\mu\nu}+f_\mu h_\nu\frac{-ig^2}{k^2-m_a^2+ig^2K
},
\ee
where
\be
f_\mu=D_{\mu\alpha}\tilde F^{\alpha\lambda}k_\lambda,\quad h_\nu=\tilde F^{\sigma\phi}k_\phi D_{\sigma\nu},\quad
\displaystyle  K=\tilde F^{\beta\rho}k_\rho D_{\beta\gamma}\tilde F^{\gamma\xi}k_\xi.
\ee
In order to simplify the result we shall assume that $\vec k \cdot \vec B=0$, which may correspond
to an experimentally relevant situation. Then we get
\be
f_\mu=i\omega g^i_\mu\frac{k^2B_i-i\hh(\vec B\times\vec k)_i}{(k^2-\hh |\vec k|)(k^2+\hh |\vec k|)},
\ee
\be
h_\nu=i\omega g^j_\nu\frac{k^2B_j+i\hh(\vec B\times\vec k)_j}{(k^2-\hh |\vec k|)(k^2+\hh |\vec k|)},
\ee
\be
K=i\omega^2 B^2\frac{k^2}{(k^2-\hh |\vec k|)(k^2+\hh |\vec k|)},
\ee
and finally, the full propagator is
\ba\label{propagator}
\mathcal D_{\mu\nu}&=&D_{\mu\nu}+\frac{i\omega^2g_\mu^j g_\nu^l}{(k^2-\hh |\vec k|)(k^2+\hh |\vec k|)(k^2-m_a^2)-b^2\omega^2 k^2}\cr
&&\left\{b_jb_l+\frac{i\hh k^2\left(b_jq_l-b_lq_j\right)-
\hh^2 b^2 \vec k^2 X_{jl}}{(k^2-\hh |\vec k|)(k^2+\hh |\vec k|)}\right\},
\ea
where $\vec q=\vec b\times\vec k$.
\section{Polarisation change}\label{sec:rotation}

In this section we will compute the evolution of a photon state in this background. For our purposes it will be useful
to consider the electric field correlator, easily obtained from the resummed  photon propagator
derived the previous section. Since we plan to contract this correlator with polarisation vectors, we will restrict it to the spatial indices and drop any terms
proportional to $k_i k_j$.\\
First, we will work in the case where there is no axion background. Using that $\vec k\cdot \vec B=0$, we get in momentum space
\be
D^E_{ij}(\omega,k)=-\frac{ig_{ij}\omega^2}{\omega^2-k^2}-\frac{i\omega^4 b_i b_j}
{(\omega^2-k^2)[(\omega^2-k^2)(\omega^2-k^2-m_a^2)-\omega^2 b^2]}.
\ee
Notice the rather involved structure of the dispersion relation implied in the second term, which
is only present when $b\neq 0$, while the first piece corresponds to the unperturbed propagator.
For a given value of the wave-number $|\vec k|$ the zeros
of the denominator correspond to the proper frequencies $\omega_a$, $\omega_1$ and $\omega_2$ of Eq.~\eqref{dispnoeta}.
We consider the propagation of plane waves moving in the $\hat x$ direction.
The inverse Fourier transform with respect to the spatial component will describe the space evolution of
the electric field. In order to find it, we decompose
\be
\frac1{(\omega^2-k^2)[(\omega^2-k^2)(\omega^2-k^2-m_a^2)-\omega^2b^2]}=
\frac{A}{k^2-\omega^2}+\frac{B}{k^2-F^2}+\frac{C}{k^2-G^2},
\ee
where $\omega$, $F$ and $G$ are the roots of the denominator
\begin{align}
F^2&=\omega^2-\frac{m_a^2}2+\frac12\sqrt{m_a^4+4\omega^2b^2}\approx\left(1+\frac{b^2}{m_a^2}\right)\omega^2,\cr
G^2&=\omega^2-\frac{m_a^2}2-\frac12\sqrt{m_a^4+4\omega^2b^2}\approx\left(1-\frac{b^2}{m_a^2}\right)\omega^2-m_a^2,
\end{align}
and
\begin{align}
A&=-\frac1{\omega^2-F^2}\frac1{\omega^2-G^2}=\frac1{\omega^2 b^2},\cr
B&=-\frac1{F^2-\omega^2}\frac1{F^2-G^2}\approx-\frac1{\omega^2 b^2}\left(1-\frac{\omega^2b^2}{m_a^4}\right),\cr
C&=-\frac1{G^2-\omega^2}\frac1{G^2-F^2}\approx-\frac1{m_a^4}.
\end{align}
The space Fourier transform of the electric field propagator is then
\begin{align}
D_{ij}^E(\omega,x)&=-g_{ij}\frac{\omega}{2}e^{i\omega x}+
\frac{\omega^4}{2}b_ib_j\left(\frac A\omega e^{i\omega x}+ \frac{B}{F}e^{iFx}+\frac{C}{G}e^{iGx}\right)\cr
&=\frac\omega2e^{i\omega x}\left[-g_{ij}+\omega^3b_ib_j\left(\frac A\omega + \frac{B}{F}e^{i(F-\omega)x}+\frac{C}{G}e^{i(G-\omega)x}\right)\right],
\end{align}
where $x$ is the travelled distance. After factoring out the exponential $e^{i\omega x}$ we consider the
relative magnitude of the differential
frequencies  $F-\omega$ and $G-\omega$. The latter
is much larger and for $m_a^2 x/2\omega \gg 1$ the corresponding exponential can be dropped. This approximation is valid for the range of axion masses 
envisaged here and $\omega\sim 1$ eV when considering all astrophysical
and most terrestrial experiments. As for the exponential containing $F-\omega$, we can safely expand it for table-top experiments
and retain only the first non-trivial term.
In this case, the leading terms in the propagator are
\be
D_{ij}^E(\omega,x)\approx\frac\omega2e^{i\omega x}\left[-g_{ij}+\hat b_i\hat b_j\left(\frac{\omega^2b^2}{m_a^4}-
i\frac{\omega b^2x}{2m_a^2}\right)\right],
\ee
where $\hat b$ is a unitary vector in the direction of the magnetic field.
For very light axion masses, neglecting the $e^{i(G-\omega)x}$ exponential can not be justified for table-top experiments.
In this case, one should use
a slightly more complicated propagator, namely
\be
D_{ij}^E(\omega,x)\approx\frac\omega2e^{i\omega x}\left\{-g_{ij}+\hat b_i\hat b_j
\frac{\omega^2b^2}{m^4_a}\left[ 1-\cos\frac{m_a^2x}{2\omega} +i\left(\sin\frac{m_a^2x}{2\omega}- \frac{m_a^2x}{2\omega}\right)\right]\right\}.
\ee
These expressions agree in the appropriate limits with the ones in Ref.~\cite{mpz}.

When a CAB is considered, the electric field propagator changes to
\ba
D^E_{ij}(\omega,k)&=&-i\omega^2\left(\frac{P_{+ij}}{\omega^2-k^2-\eta_0k}+\frac{P_{-ij}}{\omega^2-k^2+\eta_0k}\right)\cr
&&-i\omega^4\frac{ b_i b_j}
{(\omega^2-k^2)[(\omega^2-k^2)(\omega^2-k^2-m_a^2)-\omega^2 b^2]},
\ea
see App.~\ref{sec:app_prop} for a complete discussion.
The external magnetic field can be set to zero in the previous expressions, if desired.

By projecting on suitable directions and taking the modulus square of the resulting quantity,
the following expression for the angle of maximal likelihood (namely, the one where it is more
probable to find the direction of the rotated electric field) as a function of the distance $x$ can be found
\be\label{likelyrotat}
\alpha(x)=\beta-\frac{\eta_0x}{2}-\frac\epsilon2\sin2\beta,
\ee
where $\beta$ is the initial angle that the oscillation plane of the electric field
forms with the background magnetic field and
\be
\epsilon\approx-\frac{\omega^2b^2}{m_a^4}\left(1-\cos\frac{m_a^2x}{2\omega}\right).
\ee
From the results in App.~\ref{app:photon}, the ellipticity turns out to be
\be
e=\frac12\left|\varphi\sin2\beta\right|,\quad \varphi\approx\frac{\omega^2b^2}{m_a^4}\left(\frac{m_a^2x}{2\omega}-\sin\frac{m_a^2x}{2\omega}\right).
\ee
For small distances, $\frac{m_a^2x}{2\omega}\ll1$, we can expand the trigonometric functions to get
\be
\epsilon\approx-\frac{b^2x^2}{8},\quad\varphi\approx\frac{m^2b^2x^3}{48\omega}.
\ee
If this limit is not valid, we have instead
\be
\epsilon\approx-\frac{\omega^2b^2}{m^4},\quad\varphi\approx\frac{\omega b^2x}{2m^2}.
\ee
It can be noted that the effect of the magnetic field always comes with the factor $\sin2\beta$, which means that
it disappears if the electric field is initially parallel
($\beta=0$) or perpendicular ($\beta=\pi/2$) to the external magnetic field.

The results of Ref.~\cite{mpz}, which we reproduce in the case $\eta_0=0$,  are known to be in agreement
with later studies such as Ref.~\cite{raff&sto}, which has somehow become a standard reference in the
field. However, their
approach is not adequate to deal with time dependent backgrounds and therefore it is
not easy to reinterpret the results derived in the present work when a non-vanishing CAB is present in the
language of Ref.~\cite{raff&sto}.

\section{Measuring the influence of the CAB in polarimetric experiments}\label{sec:polarisation}

If $\eta_0\neq 0$ a rotation is present even in the absence of a magnetic field. This is a characteristic
footprint of the CAB. This `anomalous' rotation attempts to bring the initial polarisation plane to agree
with  one of the two elliptic eigenmodes. In the case where the effect of $\eta_0$ dominates, the eigenmodes are
almost circularly, rather than linearly,
polarised so the changes in the plane of polarisation could be eventually of order one. The effect is independent of the frequency.
Equation~\eqref{likelyrotat} shows however that the process of rotation due to the CAB is very slow, with a
characteristic time $\eta_0^{-1}$.

Typically, in interferometric-type experiments the laser light is made to bounce and is folded many times.
Equation~\eqref{likelyrotat} can be used each time that the light travels back and forth. When this happens, $\beta$ changes
sign and so does $\sin2\beta$. Since $\epsilon$
is always negative, the effect of the magnetic field is always to increase $\beta$ in absolute value
(i.e. moving the polarisation plane away from the magnetic field). So
in this sense, the rotation accumulates.
The situation is different for the CAB term. It does not change sign when $\beta$ does, so its effect
compensates each time the light bounces. However, recall that $\eta_0$
changes sign with a half-period $\pi m_a^{-1}$ so the effect could be accumulated by tuning the length
between each bounce. The range of values of $\pi m_a^{-1}$ makes
this perhaps a realistic possibility for table-top experiments (we are talking here about separations between
the mirrors ranging from millimeters to meters for most accepted values of $m_a$).

It turns out that for Peccei-Quinn axions the effect is actually independent both of the actual values for
$f_a$ and $m_a$, as it depends only on the combination $f_a m_a \simeq 6 \times 10^{15}$ eV$^2$ and the local
axion density. Assuming that the laser beam travels a distance $L=\pi m_a^{-1}$ before bouncing, the total
maximum rotation athat can be observed will be given by $|\eta_0| x$. The total travelled distance will be
$x= {\cal N} L$, where ${\cal N}$ is the total number of turns that depends on the finesse
of the resonant cavity. Replacing the expression for $|\eta_0|$ in the previous expression in terms of the local
dark matter density $\rho$ (that we assume to be 100\% due to axions) we get
$|\eta_0|= g_{a\gamma\gamma} \frac{4\alpha}{\pi^2}\frac{\sqrt{2\rho}}{f_a}$. Then
\be
|\eta_0| x =  g_{a\gamma\gamma} \frac{4\alpha}{\pi}\frac{\sqrt{2\rho}}{6 \times 10^{15}\ {\rm eV}^2} {\cal N}
\simeq 2 \times 10^{-18} \ {\rm eV}^{-2} \times \sqrt\rho \times {\cal N}.
\ee
Plugging in the expected value for the local axion density one gets for every bounce an increment
in the angle of rotation of $2 \times 10^{-20}$. This is of course a very small number and we realise that the chances
of being able to measure this anytime soon are slim. At present there are cavities whose 
reflection losses are below 1 ppm~\cite{supermirrors} but these numbers still fall short.
However this result may be interesting for several reasons.
First of all, it is actually independent of the axion parameters, as long as they are Peccei-Quinn axions, except
for the dependence on $g_{a\gamma\gamma}$, which is certainly model dependent but always close to 1. Second,
in this case it depends directly on the local halo density and nothing else. Third, a positive result obtained by
adjusting the length of the optical path would give an immediate direct measurent of $m_a$ and an indirect
one of $f_a$. There are no hidden or model dependent assumptions, the only ingredient that is needed is QED.

Observing a net rotation of the initial plane of polarisation when the magnetic field is absent
(or very small) would be a clear signal of the collective effect of a CAB. On the contrary, a non-zero value
for $\eta_0$ does not contribute at leading order to a change in the ellipticity (and subleading corrections
are very small). In Ref.~\cite{ahlers} the authors discuss in some detail the different backgrounds, all of wich are very small
with the exception of the dichroism originating from the experimental apparatus itself~\cite{zavattini}. Ways of partially
coping with these experimental limitations are discussed in the previous reference.

Notice that the effect is directly proportional to the distance travelled and
therefore any improvement in the finesse of the cavity directly translates into a longer distance and a better
bound. Recall that in order to measure the rotated angle it is actually much better not to consider an external
magnetic field, making the experimental setup much easier. Incidentally this also liberates us from
the non-linear QED effects discussed in Sec.~\ref{sec:eom}.

Axion-like particles not constrained by the Peccei-Quinn relation $f_am_a \simeq$ constant could be easier to rule out
if they happen to be substantially lighter than their PQ counterparts as cavities in this case can be longer and
one could have longer accumulation times.

\section{Summary}
In this chapter we have analyzed the axion-photon mixing in the presence of an external magnetic and a cold axion background (CAB). 
The mixing is substantially involved and the two photon
polarisations mix even without a magnetic field. In particular in our results we can take the limit where the magnetic field
vanishes, a situation that would make experiments easier even if it would be really challenging to measure the predicted
effects.

We have made one approximation that we believe is not essential, namely we have approximated the assumed sinusoidal
variation in time of the CAB by a piece-wise linear function; resulting in a fully analytically solvable problem. We believe
that this captures the basic physics of the problem and we expect only corrections of ${\cal O}(1)$ in
some numerical coefficients
but no dramatic changes in the order-of-magnitude estimates.

The existence of some momentum gaps due to the periodic time dependence of the CAB and its implications has been reviewed too.
It seems challenging to design experiments to verify or falsify their existence, but in any case they are unavoidable if
dark matter is explained in terms of an axion background; in fact it would possibly be a direct evidence
of the existence of a CAB.

We have obtained the proper modes and their ellipticities and we have analyzed in detail the evolution
of the system. It should be said that CAB-related effects dominate in some regions of the allowed parameter space.
We have also studied the possible presence of accumulative effects that might enhance the rotation
of the instantaneous plane of polarisation. This would also be a genuine CAB effect.

In order to analyze the evolution of the system we have made use of the two point function for the
electric field, that correlates the value at $x=0$ with the one at a given value for $x$. We find this a convenient
and compact way of treating this problem. It is valuable to have this tool at hand as the propagator encompasses
all the information of the travelling photons.

Of course the most relevant question is whether laser experiments
may one day shed light on the existence and properties of the CAB. In both cases the required precision is several
orders of magnitude beyond present accuracy, but progress in this field is very fast.

Apart from the precision issue, there are several caveats to take into account when attempting to
experimentally test the predictions of the present work. For instance, a scan on
$m_a$ (i.e. the mirror separation) has to be performed until a cumulative effect is found, which obviously takes time
(this is somewhat equivalent to the scan on the resonant frequency of the cavity in ADMX).
The total number of reflections
is limited by mirror quality (finesse) and it typically induces a spurious rotation that needs to be disentangled from
the true effect. We do not think that any of the approximations made in this work (basically the piecewise linear
 approximation
for the CAB profile) is experimentally significant provided that the coherence length of the CAB is larger than
the spatial region experimentally probed.

Checking the coherence of a putative cold axion background is not easy because
the physical effects associated to it are subtle and small in magnitude. The present proposal analyzes the consequences of
the existence of a CAB on photon propagation and as we have seen its effects can be of a size comparable
to other phenomena that are being actively investigated in optical experiments. For these reasons we believe it is important
to bring the present analysis to the
atention of the relevant experimental community.
\graphicspath{{4-cosmic/figures/}{figures/}}

\chapter{High-energy cosmic ray propagation in a cold axion background}\label{ch-cosmic}

In the previous chapter, we investigated the influence of a cold axion background on the propagation of photons and derived
some consequences for optical experiments. 
In the present chapter, we continue exploring the effects of the CAB that could lead to its (indirect) detection, turning our attention to cosmic ray propagation.

Cosmic rays are very energetic charged particles reaching Earth from outside. They consist of electrons, protons, and other heavy nuclei. Primary cosmic rays 
are produced at astrophysical sources (e.g. stars), while secondary cosmic rays are particles produced by the interaction of primaries with interstellar gas.
The flux of cosmic rays is defined as the number of cosmic rays with a given energy per unit time, incident on a surface element coming from a certain solid angle
\be
J(E)=\frac{d^4N}{dEdSdtd\Omega}.
\ee
Experimentally, it is seen that this flux depends on the energy of the cosmic ray according to a power law
\be\label{crflux}
J(E)=N_iE^{-\gamma_i},
\ee
where $\gamma_i$ is known as the spectral index and takes different values in several energy intervals.\\

The energy density associated to the cold axion background coming from vacuum misalignment, described in Sec.~\ref{sec:cab}, is very small, see Eq.~\eqref{halodensity}.
One could think that this very diffuse concentration of a pseudoscalar condensate is irrelevant, except for its gravitational effects. 
However, the CMB photon density is also very small and yet it imposes the Greisen-Zatsepin-Kuzmin (GZK) cutoff~\cite{GZK1,GZK2}.
It states that the number of cosmic rays above a certain energy threshold should be very small. 
Above that threshold, cosmic rays particles interact with photons from the Cosmic Microwave Background (CMB) to produce pions:
\be
\gamma_{\rm CMB}+p\longrightarrow p+\pi^0\quad \text{or}\quad \gamma_{\rm CMB}+p\longrightarrow n+\pi^+.
\ee
The energy threshold is about $10^{20}$~eV.
Because of the mean free path associated with these reactions, cosmic rays with energies above the threshold and traveling over distances larger than 50 Mpc 
should not be observed on Earth.

Since a low density background such as the CMB can have sizeable effects, the study of the effect of the axion background 
on highly energetic charged particles also deserves consideration.\\

In Chapter~\ref{ch-photon} we saw that in the presence of a CAB and a magnetic field a three-way mixing between the axion and the two photon polarisation appears. When no
magnetic field is considered, axions no longer mix and photons with definite helicity have a modified dispersion relation given by Eq.~\eqref{dispersion}.
This new dispersion relation allows processes normally forbidden to take place. In this chapter we study one of these processes: the emission of photons by a
charged particle, which is a mechanism by which cosmic rays can lose energy as they travel.\\

This chapter is organized as follows. In Sec.~\ref{sec:qedcab} we study a modification of QED provided by a constant vector. When a cold axion background
is considered, this vector only has a temporal component and its effect is to modify the photon dispersion relation, which makes possible some processes
that are forbidden in regular QED due to momentum conservation.

In Secs.~\ref{sec:kin} and~\ref{sec:ampdecay} we study one of such processes: 
the emission of a photon by a charged particle. We derive its kinematical constraints and find out the energy threshold
above which the process can take place and the range of allowed momenta for the emitted photons. We use standard QED rules to derive the probability amplitude 
related to the process, since the modification provided by the axion background only affects the kinematics. We also compute the differential decay width.

Up to this point, calculations are performed neglecting the time-variation of the cold axion background. In Sec.~\ref{sec:timevar} we prove that the obtained results
are applicable even when the time-variation is considered.

Finally, in Sec.~\ref{sec:enloss} we compute the total energy loss that a cosmic ray experiences due to photon emission and find it to be negligible in the relevant 
energy ranges. However, in Sec.~\ref{sec:radyield}, we derive the energy flux of emitted photons, which may still be measurable.

\section{QED in a cold axion background}\label{sec:qedcab}

A particle travelling at almost the speed of light through a cold axion background will see coherent regions with 
quasi-constant values of the CAB of a size inversely proportional to the axion mass. Although this size may be small, it is very many orders
of magnitude bigger than the wave length of a particle travelling with momentum $\vec p$, characteristic of a very highly energetic cosmic ray. For $|\vec p|\gg m_a$
we can treat this slowly varying term as a constant and use the Lagrangian
\be
\mathcal{L}=-\frac14F_{\mu\nu}F^{\mu\nu}+\frac12m_\gamma^2A_\mu A^\mu+\frac12\eta_\mu A_\nu\tilde F^{\mu\nu}.
\ee
An effective photon mass (equivalent to a refractive index, see Ref.~\cite{raff}) has also been included. It is of order
\be
m_\gamma^2\simeq4\pi\alpha\frac{n_e}{m_e}.
\ee
The electron density in the Universe is expected to be at most $n_e\simeq10^{-7}\rm\,cm^{-3}\simeq10^{-21}\,\rm eV^3$.
This density corresponds to $m_\gamma\simeq10^{-15}\rm\,eV$, but the more conservative limit (compatible with Ref.~\cite{pdg})
$m_\gamma=10^{-18}\,\rm eV$ will be used here.

The constant vector corresponding to a cold axion background is $\eta^\mu=(\eta_0,0,0,0)$, where 
\be
\eta_0=\frac2\pi ga_0m_a=10^{-20} - 10^{-24} \text{ eV},
\ee
as discussed in Sec.~\ref{sec:eom}.

Photons of positive and negative chirality are solutions of the vector field equations if and only if
\be\label{disprel}
k^{\mu}_{\pm}=(\omega_{{\vec k},\pm} , {\vec k})\qquad
\omega_{{\vec k},\pm}=\displaystyle
 \sqrt{\vec k^2+m_{\gamma}^{2}\pm\eta_0|\vec k|},
\ee
see App.~\ref{sec:app:proj} for a complete derivation.

In order to avoid problems with causality we want $k_{\pm}^2 > 0$. For
photons of a given chirality (negative if $\eta_0>0$, positive if $\eta_0<0$) 
this can be if and only if
\be
|\vec k|<\frac{m^{2}_{\gamma}}{|\eta_0|}.
\ee
In fact, for $m_\gamma=0$ these photons cannot exist as physical asymptotic states.
If they are produced, they will eventually decay (to three photons of like
chirality) in a cascade process that leads to a red-shift.

As is known to everyone the processes $e^- \to e^- \gamma$ or 
$\gamma \to e^+ e^-$ cannot occur in vacuum. However, in the present situation, due to the modified dispersion relation in Eq.~\eqref{disprel}, they are allowed. 
For the latter process, energy conservation leads to 
\be
\omega_{\vec k,\pm}=\displaystyle
\sqrt{\vec k^2+m_{\gamma}^2\pm\eta_0|\vec k|}
=\sqrt{\vec p^2+m_e^2} + \sqrt{(\vec p-\vec k)^2+m_e^2}.
\ee
As discussed in \cite{AGS,AEGS}, the process is possible for photons of positive (negative) chirality if $\eta_0>0$ ($\eta_0 <0$) if
\be
|\vec k|\ge\frac{4m_{e}^{2}}{|\eta_0|}\equiv k_{\text{th}}.
\ee
In the subsequent sections we will study in detail the related process $p \to p\gamma$ (although we write $p$ for definiteness, the results apply to electrons as well).

\section{Kinematic constraints}\label{sec:kin}

Having found out the different polarisations and dispersion relations in the  axion background,
let us now turn to kinematical considerations. Let us consider the process
$p({\vec p}) \to p({\vec p}-{\vec k})\,\gamma({\vec k})$ 
with $p^\mu=(E,{\vec p})$, $p=\vert {\vec p} \vert$, 
$k^\mu=(\omega_{\vec k}, {\vec k})$, $k=\vert {\vec k} \vert$ and $\vec p\cdot \vec k=pk\cos\theta$.
Using Eq.~\eqref{disprel}, four-momentum conservation for this process leads to 
\be
\sqrt{E^2 + k^2 - 2 pk \cos\theta} + \sqrt{k^2 - \eta_0 k + m_\gamma^2} - E = 0.
\label{encor}
\ee
For simplicity we have taken the negative sign for the polarisation, understanding that changing the sign
of $\eta_0$ amounts to exchanging positive and negative chiral polarisations for the photon.

Let us first consider the case $m_\gamma=0$. The above energy conservation equation reduces to
\be
k^2 (E^2 - p^2\cos^2\theta + p\eta_0\cos\theta - \eta_0^2/4 ) -kE^2\eta_0=0.
\ee
This equation has the trivial solution $k=0$, where no photon is emitted, and 
\be
k= \frac{E^2\eta_0}{m_p^2+p^2\sin^2\theta + p\eta_0\cos\theta - \eta_0^2/4},
\ee
where we have used $E^2=m_p^2+p^2$.
Since $k$ has to be positive, the process is possible only for a photon of negative chirality if $\eta_0>0$ (positive chirality if $\eta_0<0$). 
Of course, if $\eta_0=0$, the process is impossible (it is the usual QED case).
To find out the kinematical restrictions on $k$ we search for the extrema of the denominator, which are
\be
\cos\theta=\pm1,\quad \cos\theta=\frac{\eta_0}{2p}.
\ee
The last value is providing the minimum value for $k$, 
$k_{\text{min}}=\eta_0$. 
The maximum is found for $\cos\theta=-1$ and corresponds to
\be
k_{\text{max}}= \frac{\eta_0E^2}{m_p^2},
\ee
where we have assumed $p\ll m^2_p/\eta_0$, since $m^2_p/\eta_0$ is well above the GZK cutoff.

Now we turn to $m_\gamma>0$. Retaining only the leading terms in the small quantities $m_\gamma$ and $\eta_0$, this equation can be put in the form
\be
\left(E^2-p^2\cos^2\theta+p\eta_0\cos\theta\right)k^2-\left(E^2\eta_0+m^2_\gamma p\cos\theta\right)k+E^2m_\gamma^2=0,
\ee
which has two roots
\be\label{sqrt}
k_\pm=\frac{E^2\eta_0+pm^2_\gamma\cos\theta\pm 
E\sqrt{E^2\eta_0^2-2m_\gamma^2\left(2E^2-2p^2\cos^2\theta+p\eta_0\cos\theta\right)}}{2\left(m_p^2+p^2\sin^2\theta+p\eta_0\cos\theta\right)}
\ee
The discriminant has to be positive, which imposes the condition
\be\label{disc}
\sin^2\theta\le\frac{\left[\eta_0^2p^2-2m_\gamma^2\eta_0p+m_p^2(\eta_0^2-4m_\gamma^2)\right]}{4p^2m_\gamma^2\left(1-\frac{\eta_0}{4p}\right)}.
\ee
For $\eta_0>2m_\gamma$ (which includes the case $m_\gamma=0$) the numerator of Eq.~\eqref{disc} is always positive and no further restrictions are placed on the cosmic ray momentum $p$.
However, according to the possible values of $m_\gamma$ and $\eta_0$ mentioned in Sec.~\ref{sec:qedcab}, we are in the opposite case.
For $\eta_0<2m_\gamma$, momentum has to be larger than a certain threshold in order to make the r.h.s. of Eq.~\eqref{disc} positive
\be
p>p_{\text{th}}=\frac{m_\gamma^2}{\eta_0}+\frac{2m_\gamma m_p}{\eta_0}\sqrt{1-\frac{\eta_0^2}{4m_\gamma^2}}.
\ee
In terms of the cosmic ray energy, the threshold is
\be
E_\text{th}=\frac{2m_\gamma m_p}{\eta_0}.
\ee
This energy threshold goes to infinity as $\eta_0\to0$ as is expected.
For $p\gg p_{th}$, the maximum value of the angle is given by
\be
\sin^2\theta_{\text{max}} \approx \frac{\eta_0^2}{4m_\gamma^2}.
\ee
We see that photons are emitted in a rather narrow cone $\theta_\text{max}\simeq\frac{\eta_0}{2m_\gamma}$. This justifies {\em a posteriori}
the approximation $\cos\theta \simeq 1- \frac12 \sin^2\theta$ that has been used.

At $\theta_{\text{max}}$ the square root in Eq.~\eqref{sqrt} vanishes and $k_+ = k_-=k(\theta_\text{max})$. Keeping only the leading terms,
\be
k(\theta_{\text{max}})\simeq \frac{2m_\gamma^2}{\eta_0}\left(1 - 3 \frac{p
m_\gamma^2}{E^2\eta_0}\right)\stackrel{p\gg p_\text{th}}{\longrightarrow} \frac{2m_\gamma^2}{\eta_0}.
\ee
From Eq.~(\ref{sqrt}) we work out the value for $\theta=0$, which is the minimum value of $\theta$ from the 
bound in Eq.~(\ref{disc}).  
\be
k_{+}(0)\simeq \frac{E^2\eta_0+pm_\gamma^2 + E\sqrt{E^2\eta_0^2-4m_p^2m_\gamma^2-2p\eta_0 m_\gamma^2}}{2p\eta_0 + 2 m_p^2}
\stackrel{p\gg p_\text{th}}{\longrightarrow} \frac{\eta_0E^2}{m_p^2},
\ee
which is the same result obtained before, and
\be
k_{-}(0)\simeq \frac{E^2\eta_0 +pm_\gamma^2 - E\sqrt{E^2\eta_0^2-4m_p^2m_\gamma^2-2p\eta_0 m_\gamma^2}}{2p\eta_0 + 2 m_p^2}
\stackrel{p\gg p_\text{th}}{\longrightarrow}\frac{m_\gamma^2}{\eta_0}.
\label{tresh}
\ee
Now we notice that $k_-(0)<k(\theta_\text{max})<k_+(0)$. To show that 
$k_\text{min}=k_-(0)$ and $k_\text{max}=k_+(0)$ we have to study the derivative 
of $\theta$ 
versus $k$, namely we should have  
$d\cos\theta/dk<0$ for $k<k(\theta_{max})$ and 
$d\cos\theta/dk>0$ for $k>k(\theta_{max})$.

We isolate $\cos\theta$ from
the energy conservation relation Eq.~\eqref{encor}
\be
\cos\theta = \frac{\eta_0 k-m_\gamma^2 +2 E\sqrt{m_\gamma^2+k^2-\eta_0 k}}{2pk}\label{coss}
\ee
and compute the derivative
\be
\frac{d\cos\theta}{dk}= \frac{m_{\gamma}^2}{2 k^2 p} - 
\frac{E}{2 k^2 p} \frac{2 m_{\gamma}^2-\eta_0 k}{\sqrt{m_\gamma^2+k^2-\eta_0 k}}
\stackrel{p\gg p_\text{th}}{\longrightarrow} 
-\frac{2 m_{\gamma}^2-\eta_0 k}{2 k^2 \sqrt{m_\gamma^2+k^2-\eta_0 k}}.
\ee
For $k\to k(\theta_\text{max})$ $d\cos\theta/dk\to 0$ and it is the only zero which of course
means that this value of $k$ corresponds to a minimum of $\cos \theta$ (i.e. to
a maximum of $\sin\theta$). On the other hand,  
for $k<k(\theta_\text{max})$, $d\cos\theta/dk<0$ and for
$k>k(\theta_\text{max})$, $d\cos\theta/dk>0$.
Then $k_\text{min}=k_-(0)$ and $k_\text{max}=k_+(0)$.\\

To summarize, photons are emitted in a narrow cone of angle
\be
\theta_\text{max}\simeq\frac{\eta_0}{2m_\gamma}
\ee
and their momenta lie between
\be\label{kminmax}
k_\text{min}=\frac{m_\gamma^2}{\eta_0} \quad\text{and}\quad k_\text{max}=\frac{\eta_0E^2}{m_p^2}.
\ee

\section{Amplitude and differential decay width}\label{sec:ampdecay}

The next step in studying the process of photon emission is to compute the relevant matrix element. 
Since the process takes place through the usual QED interaction, the calculation is rather straightforward. Using the standard Feynman, rules we get 
\be
i\mathcal{M}=\bar u(q)ie\gamma^\mu u(p)\varepsilon^{-}_\mu(k)^*,
\ee
with the polarisation vector defined in Eq.~\eqref{polvec}.
Now we take the square of $i\mathcal{M}$ and sum and average over the final and initial proton helicities, respectively. Note that we do not average
over photon polarisations, since the process is possible only for one of them. Using also 4-momentum
conservation and Eq.~\eqref{orto} we get
\be
\overline{|\mathcal{M}|^2}=2e^2\left\{-p^\mu
k_\mu+\left[\varepsilon^-_\mu(k)^*\varepsilon^-_\nu(k)+\varepsilon^-_\mu(k)\varepsilon^-_\nu(k)^*\right]p^\mu
p^\nu\right\}.
\ee
Last we use the closure relation of Eq.~\eqref{closure}, Eq.~\eqref{S} and the value of the angle found in Eq.~\eqref{coss} to write
\be
\overline{|\mathcal{M}|^2}=2e^2\left(\frac{\eta_0k-m_\gamma^2}{2}+p^2\sin^2\theta\right).
\ee
Recalling the minimum value for the photon momentum in Eq.~\eqref{kminmax}, the first term is
\be
\eta_0k-m_\gamma^2=\eta_0\left(k-k_\text{min}\right),
\ee
so $\overline{|\mathcal{M}|^2}$ is clearly positive.\\
The differential decay width of the proton is given by
\be
d\Gamma=(2\pi)^4\delta^{(4)}(q+k-p)\frac1{2E}\overline{|\mathcal{M}|^2}dQ,
\ee
where $dQ$ refers to the final state phase space
\be
dQ=\frac{d^3q}{(2\pi)^32E(q)}\frac{d^3k}{(2\pi)^32\omega(k)}.
\ee
We use three of the momentum conservation Dirac deltas to fix $\vec q=\vec p-\vec k$. Then, we write $d^3k=k^2dkd(\cos\theta)d\varphi$, use the remaining delta to
fix $\cos\theta$ to the value given in Eq.~\eqref{coss} and integrate over $\varphi$, which gives a factor of $2\pi$, since nothing depends on it. Finally,
\be
\frac{d\Gamma}{dk}=\frac{\alpha}{2}\frac{k}{Ep\omega}\left(\frac{\eta_0k-m_\gamma^2}{2}+p^2\sin^2\theta\right),
\quad \alpha=\frac{e^2}{4\pi}.
\ee
An alternative form for the differential decay rate, useful for future calculations, is
\be\label{dgamma}
\frac{d\Gamma}{dk}=\frac{\alpha}{8\omega k}\left[A(k)+B(k)E^{-1}+C(k)E^{-2}\right],
\ee
with
\begin{align}
A(k)&=4(\eta_0k-m_\gamma^2),\quad B(k)=-4\omega(\eta_0k-m_\gamma^2),\cr
C(k)&=2k^2(\eta_0k-m_\gamma^2-\eta_0^2-4m_p^2)+m_\gamma^2(2\eta_0k-m_\gamma^2).
\end{align}

\section{Incorporating the time variation of the background}\label{sec:timevar}

We have argued that taking $\eta(t)$ to be constant is a good approximation for high values of the momentum. In this section we will compute the process $p\to p\gamma$
with a square-wave function for $\eta$, see Eq.~\eqref{square}. To do this we write the photon field as
\be
A_\mu(t,\vec x)=\int \frac{d^3k}{(2\pi)^3}
\sum_\lambda\left[a(\vec k,\lambda)g(t,\vec k,\lambda)\varepsilon_\mu(\vec k,\lambda)e^{-ikx}
+a^\dag(\vec k,\lambda)g^*(t,\vec k,\lambda)\varepsilon^*_\mu(\vec k,\lambda)e^{ikx}\right],
\ee
where we have included both polarisations, denoted by $\lambda$. $a^{(\dag)}$ is an annihilation (creation) operator and $kx\equiv\Omega t-\vec k\cdot\vec x$. Recall that 
$g(t)=e^{i\Omega t}f(t)$, and $f(t)$ is piecewise defined, according to the discussion on Sec.~\ref{sec:gaps}.

Now we want to compute $\langle f|S|i\rangle$ for an
initial state $|i\rangle$ of one proton of momentum $p$
and a final state $|f\rangle$ of a proton of momentum $q$ and a photon of momentum $k=p-q$
\be
\langle f|S|i\rangle= ie\varepsilon^*_\mu(\vec k,\lambda)\bar u_q\gamma^\mu u_p(2\pi)^3\delta^{(3)}
(\vec k+\vec q-\vec p)\int dt g^*(t,\vec k,\lambda)e^{i(\Omega+E_q-E_p)t}.
\ee
If we take $\eta(t)$ constant, $g(t,\vec k,\lambda)=1$ and we have the usual result
\be
\langle f|S|i\rangle=ie\varepsilon^*_\mu(\vec k,\lambda)\bar u_q\gamma^\mu u_p(2\pi)^4\delta^{(4)}(k+q-p).
\ee
In the square wave approximation, the time integration yields
\begin{align}
\langle f|S|i\rangle
={}&ie\bar u_q\gamma^\mu u_p\varepsilon^*_\mu(\vec k,\lambda)(2\pi)^3\delta(\vec k+\vec q-\vec p)\pi\cr
&\left[A\delta(\omega_++E_q-E_p)
+B \delta(\omega_-+E_q-E_p)\right.\cr
&\left.+ A'\delta(-\omega_++E_q-E_p)-B'\delta(-\omega_-+E_q-E_p)\right]\label{AB}\\
\approx{}&ie\bar u_q\gamma^\mu u_p\varepsilon^*_\mu(\vec k,\lambda)\left
(1+\frac{\eta_0}{4k}\right)(2\pi)^3\delta(\vec k+\vec q-\vec p)\pi\cr
&\left\{\left(1-\frac{\eta_0}{2k}\right)\delta(\omega_++E_q-E_p)\right.
+ \delta(\omega_-+E_q-E_p)\cr&\left.+ \frac{\eta_0}{4k}
\left[\delta(-\omega_++E_q-E_p)+\delta(-\omega_-+E_q-E_p)\right]\right\},
\end{align}
where the coefficients $A$, $A'$, $B$ and $B'$ are defined in Eqs.~\eqref{coefficients} and~\eqref{BB}.
Equation~\eqref{AB} holds for any value of $k$. The $\approx$ symbol indicates the use of the approximate values for the coefficients in Eq.~\eqref{limit}.
It turns out that at the leading order in the $\eta_0$ expansion this expression agrees exactly
with the one obtained assuming that $\eta(t)$ was constant except for the fact
that for each value of the polarisation only one of the two delta functions
that are not suppressed by terms of the form $\eta_0/k$ can be simultaneously satisfied; namely
the one that implies that $\omega_+$ or $\omega_-$ equals $\sqrt{k^2-|\eta_0|k}$, contributing with a factor $1/2$ with respect
to what is found for constant $\eta$ to the amplitude. Thus in the transition
reduced matrix element $i\mathcal M$ one gets for each polarisation exactly one half of what
is obtained for $\eta(t)$ constant. Since in the present case both polarisations contribute, finally we get
$(1/2)^2+(1/2)^2=1/2$ of the result obtained with constant $\eta(t)$.

\section{Energy loss}\label{sec:enloss}

Having established that the result obtained with $\eta(t)$ constant is indeed a good approximation, in this section 
we want to compute the energy lost by a cosmic ray due to the photon emission process $p\to p\gamma$. Since each photon takes
away an energy $\omega$, the energy lost per unit length is
\be
\frac{dE}{dx}=\frac{dE}{dt}\frac{dt}{dx}=\frac1v\left(-\int\omega d\Gamma\right)
\ee
Using the expression for $d\Gamma$ from Eq.~\eqref{dgamma} and $v=p/E\approx1$ and with the integration limits in Eq.~\eqref{kminmax}, the leading term is
\be\label{Eloss}
\frac{dE}{dx}\approx-\frac{\alpha\eta_0^2 E^2}{4m^2_p}.
\ee

At this point it becomes obvious why we have bothered to keep the proton mass $m_p$ all along the calculation.
There are two key scales in this problem. The first one is the energy threshold 
where the process $p \to p \gamma$ becomes kinematically possible in the presence of a pseudoscalar 
background represented by $\eta_0\neq 0$, namely $E_\text{th}= 2m_\gamma m_p/\eta_0$.
The other relevant scale is $m_p^2/\eta_0$. The energy loss
per unit length below this energy is effectively proportional to $\eta_0^2$, as seen in Eq.~\eqref{Eloss}.
Above this second scale, energy loss is proportional to $\eta_0$. Since this second scale is far above the GZK cutoff, we have not discussed
the results here, but they can be found in Ref.~\cite{shield}.
Therefore, even if we are talking about very energetic particles, the mass is a relevant parameter
when Lorentz violating interactions are present.

In the relevant $E\ll m_p^2/\eta_0$ regime, the expression for $E(x)$ is
\be
E(x)= \frac{E(0)}{1+\frac{\alpha\eta_0^2}{4m_p^2} E(0) x}.
\ee
It depends on the very small combination $\frac{\alpha\eta_0^2}{4m_p^2}$.
Thus, the slowing of cosmic rays due to a cold axion background is completely negligible. However, the emitted photons could still be detectable. 
We study them in the following section.

\section{Radiation yield}\label{sec:radyield}

Let us turn to the radioemission
due to the axion-induced Bremsstrahlung. We will study photon emission by protons and electrons due to the process described in the previous sections.
For primary protons, using $\eta_0=10^{-20}$~eV and $m_\gamma=10^{-18}$~eV as indicative values and keeping in mind the GZK cutoff
there would be electromagnetic activity in the region of the spectrum
\be
10^{-16}\text{ eV}
< k < 100\text{ eV}
.
\ee
As we have seen, the maximum photon momentum is supressed by the charged particle mass, recall Eq.~\eqref{kminmax}. Therefore, 
electrons or positrons radiate more than protons. For primary electrons/positrons, we would expect activity in the range
\be
10^{-16}\text{ eV}< k  < 400 \text{ eV},
\ee
where we have assumed a cutoff similar to that of protons. This is very questionable (see Refs.\cite{electrons1,electrons2,electrons3}). It is however unimportant, 
as the intensities of electrons at such energies is very small.\\

The number of cosmic rays with a given energy crossing a surface element per unit time in a certain direction is
\be
d^4N=J(E)dEdSdt_0d\Omega,
\ee
where $J(E)$ is the cosmic ray flux. These cosmic rays will radiate at a different time $t$. The number of
photons is given by
\be
d^6 N_\gamma = d^3 N \frac{d\Gamma}{dk} dk dtd\Omega= J(E) \frac{d\Gamma}{dk} dE dk dt_0 dS dtd\Omega.
\ee
Assuming that the cosmic ray flux does not depend on time, we integrate over $t_0$ obtaining a
factor $t(E)$: the age of the average cosmic ray with energy $E$. Since we do not care
about the energy of the primary cosmic ray (only that of the photon matters),
we integrate also over $E$, starting from $E_\text{th}$, the energy threshold. Therefore,
the flux of photons is
\be\label{yield}
\frac{d^4 N_\gamma}{dk dS dt d\Omega}=\int_{E_\text{th}}^\infty dE t(E) J(E)\frac{d\Gamma}{dk}\theta\left(\frac{\eta_0E^2}{m_{p,e}^2}-k\right). 
\ee
Notice the step function $\theta$, which appears because a photon of momentum $k$ can only be radiated from a cosmic ray of sufficiently high energy, see Eq.~\eqref{kminmax}.
This means that the lower limit of the integral is effectively 
\be\label{Emink}
E_\text{min}(k)=m_{p,e}\sqrt{\frac{k}{\eta_0}},
\ee
if $E_\text{th}<E_\text{min}(k)$.
Finally, the photon energy flux is obtained by multiplying by the energy carried by each photon
\be
I_\gamma(k)=\omega(k)\int_{E_\text{min}(k)}^\infty dE t(E) J(E)\frac{d\Gamma}{dk}.
\ee
Since both the average lifetime $t(E)$ and the cosmic ray flux $J(E)$ depend on whether we are working with protons or electrons, we will compute 
both contributions separately.

\subsection{Proton primaries}

As mentioned in the beginning of the chapter, the flux of cosmic rays follows a power law, Eq.~\eqref{crflux}. 
The flux of protons is known with good precision (see Refs.~\cite{rays1,rays2}).  
The spectral indices are as indicated in the following table.
\begin{center}
\begin{tabular}{|c|c|}
  \hline
  Energy (eV) & Spectral index $\gamma$ \\
  \hline
  $10^9\le E<4\cdot10^{15}$ & $2.68$ \\
  $4\cdot10^{15}\le E<4\cdot10^{18}$ & $3.26$ \\
  $4\cdot10^{18}\le E<2.9\cdot10^{19}$ & $2.59$ \\
  $E\ge2.9\cdot10^{19}$ & $4.3$ \\
  \hline
\end{tabular}
\end{center}

The important suppression at higher energies ($\gamma=4.3$ for $E>2.9\cdot10^{19}$~eV) can be attributed to the GZK cutoff.
We will use the following parametrisation for the proton flux
\be\label{flux2}
J_p(E)=
\left\{\begin{array}{ll}
5.87 \cdot 10^{19} E^{-2.68} &  10^9\le E\le4\cdot10^{15} \\
6.57 \cdot 10^{28}E^{-3.26} & 4\cdot10^{15}\le E\le4\cdot10^{18} \\
2.23  \cdot 10^{16}E^{-2.59} & 4\cdot10^{18}\le E\le2.9\cdot10^{19} \\
4.22  \cdot 10^{49}E^{-4.3} &  E\ge2.9\cdot10^{19}
\end{array}\right.,
\ee
with $E$ in eV and $J_p(E)$ in eV$^{-1}$m$^{-2}$s$^{-1}$sr$^{-1}$.\\
~\\
\indent Because cosmic rays are deflected by magnetic fields, they follow a nearly random trajectory within
the Galaxy. Collisions of cosmic rays of large atomic number with the interstellar medium
sometimes produce lighter unstable radioactive isotopes. By measuring their abundance we know that on average a hadronic
cosmic ray spends about 10 million years in the Galaxy before escaping into intergalactic space. We will therefore 
use the following approximation for protons
\be
t_p(E)=T=10^7\text{ yr}.
\ee
Recalling the expression for the differential decay width in Eq.~\eqref{dgamma}, the photon energy flux due to proton primaries is
\begin{align}
I_\gamma^p(k)&=\omega\int_{E_\text{min}(k)}^\infty  T N_iE^{-\gamma_i}\frac{\alpha}{8\omega k}\left[A(k)+B(k)E^{-1}+C(k)E^{-2}\right]dE\cr
&=\frac{\alpha T}{8k}\sum_iN_i\left[A(k)\frac{E^{1-\gamma_i}}{1-\gamma_i}+B(k)\frac{E^{-\gamma_i}}{-\gamma_i}+C(k)\frac{E^{-(1+\gamma_i)}}{-(1+\gamma_i)}
\right]_{E_i^{initial}}^{E_i^{final}}.
\end{align}
In this expression the labels ``initial'' and ``final'' refer to the
successive energy ranges where the different parameters $N_i$ and $\gamma_i$ change values, as per Eq.~\eqref{flux2}. 
For each $k$, only energy ranges above $E_\text{min}(k)$ contribute, and the smaller of the initial values is replaced by
$E_\text{min}(k)$, as this is the starting point of the integral. Numerically, it is straightforward to see that the leading 
contribution is given by $A(k)=4(\eta_0k-m_\gamma^2)\approx4\eta_0k$ and is dominated by the initial point $E_\text{min}(k)$.
We then get the rather simple expression
\be\label{approx2}
I_\gamma^p(k)=\frac{\alpha\eta_0T}{2}\frac{E_\text{min}(k)J_p[E_\text{min}(k)]}{\gamma_\text{min}-1},
\ee
where the value $\gamma_\text{min}$ is to be read from Eq.~\eqref{flux2}, depending
on the range where $E_\text{min}(k)$ falls.

\subsection{Electron primaries}

The electron flux is typically around 1\% of the proton flux and is much less well measured. Following Refs.~\cite{electrons1,electrons2,electrons3},
we use the parametrisation
\be
J_e(E)=\left\{\begin{array}{cc}
                     5.87 \cdot 10^{17} E^{-2.68} & E\le5\cdot10^{10} \\
                     4.16 \cdot 10^{21} E^{-3.04} & E\ge5\cdot10^{10}
                   \end{array}
\right.
\label{cosmic2}
\ee
Units are eV$^{-1}$ m$^{-2}$ s$^{-1}$ sr$^{-1}$ for $J_e(E)$ and eV for $E$, as before. The electron flux is poorly known. This is quite regrettable,
as electrons radiate more than protons, and our ignorance about the flux impacts our estimation for the radiation yield.\\

Electron cosmic rays travel for approximately 1 kpc on average before being slowed down and trapped, which
corresponds to a typical age of an electron cosmic ray of about $10^5$ yr~\cite{kin}, a lot less than protons.
In addition, the lifetime of an electron cosmic ray depends on its energy
\be\label{ageee}
t_e(E)\simeq 5 \cdot 10^5 \left(\frac{1\text{ TeV}}{E}\right)\text{ yr}=\frac{\chi}E,\quad \chi=2.4\cdot10^{40}.
\ee
To complicate matters further, it has been argued that the local interstellar flux of electrons is not
even representative of the flux in the Galaxy and may reflect the electron debris from a nearby supernova, about $10^4$ years ago~\cite{age}. 

The calculation of the photon energy flux from electrons is completely analogous to that of photons except that, due to the $1/E$ dependence of the lifetime,
there is one less power of $E$
\be
I_\gamma^e(k)\simeq \frac{\alpha \eta_0 \chi}{2} \frac{J_e[E_\text{min}(k)]}{\gamma_\text{min}}\label{approx1},
\ee
where, again, $\gamma_\text{min}$ is to be read from Eq.~\eqref{cosmic2}.

\subsection{Combined yield and discussion}

Although our expressions of the photon energy flux in Eqs.~\eqref{approx2} and~\eqref{approx1} are well defined through the use of the cosmic ray
fluxes in Eqs.~\eqref{flux2} and~\eqref{cosmic2} and the minimum energy from Eq.~\eqref{Emink}, a closed analytical form in terms of $k$ and $\eta_0$ would be very
cumbersome, especially because $E_\text{min}(k)$ depends on $\eta_0$. 
We can however give approximate expressions valid in a certain range. For photon momenta close to $k=10^{-7}$~eV and $\eta_0$ of about $10^{-20}$~eV, we see that the
dominant contribution comes from electrons
\be
I_\gamma^e(k)\simeq
3\times 10^2 \times \left(\frac{\eta_0}{10^{-20}~{\rm eV}}\right)^{2.52}
  \left(\frac{k}{10^{-7}\ {\rm eV}}\right)^{-1.52}
{\rm ~m}^{-2}\, {\rm s}^{-1}\, {\rm sr}^{-1},\label{approxyield}
\ee
while for protons we get
\be
I_\gamma^p(k)\simeq
6  \times \left( \frac{T}{10^7 \ {\rm yr}}\right) \left(\frac{\eta_0}{10^{-20}~{\rm eV}}\right)^{1.84}
 \left(\frac{k}{10^{-7}\ {\rm eV}}\right)^{-0.84}
{\rm ~m}^{-2}\, {\rm s}^{-1}\, {\rm sr}^{-1}.\label{approxyield1}
\ee
We take $k=10^{-7}$~eV as the reference scale because this is approximately
the minimum wave vector at which the atmosphere is transparent to electromagnetic
radiation, even though the signal is higher for lower frequency photons. This corresponds to
30 MHz, a band in which an extensive antenna array (LWA) is already being commisioned~\cite{lwa}.
In the same range of extremely low frequencies the Square Kilometer Array (SKA) project 
could cover a the range from 70 to 10,000 MHz with enormous sensitivity (see below)~\cite{SKA}.

In a way it is unfortunate that the dominant contribution comes from electron cosmic rays
because they are still poorly understood.
Note that $I(k)$ has the dimensions of energy per unit wave vector per unit surface
per unit time. In radioastronomy the intensity, or energy flux density, is commonly
measured in Jansky
(1 Jy = 10$^{-26}$ W Hz$^{-1}$ m$^{-2}$ sr$^{-1}$ $\simeq 1.5\times 10^7$ eV eV$^{-1}$  m$^{-2}$ s$^{-1}$ 
sr$^{-1}$).

The numerical treatment of the expressions, however, poses no problems and the combined 
photon energy flux can be found in Fig.~\ref{photonflux} for a very wide range of wave vectors 
(many of them undetectable) and for the reference value $\eta_0=10^{-20}$~eV.\\
\begin{figure}[ht]
\center
\includegraphics[width=0.9\textwidth]{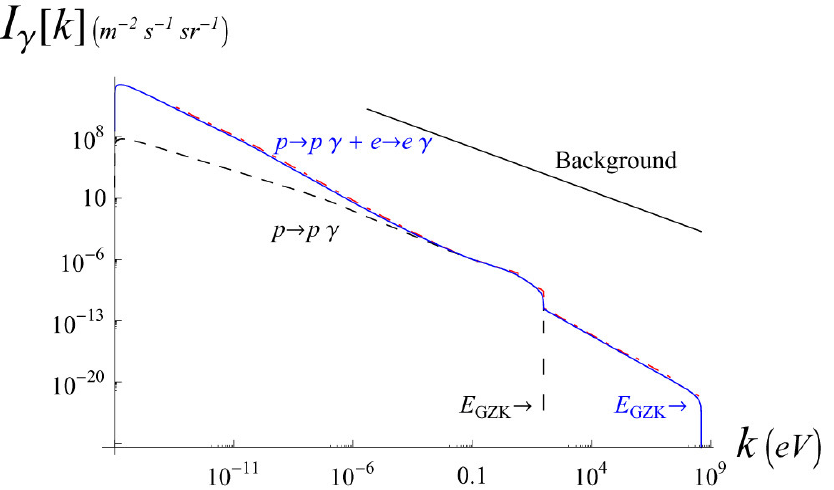}
\caption{Total intensity $I_\gamma(k)= I_\gamma^p(k) + I_\gamma^e(k)$ expected to be measured as a consequence of the axion Bremsstrahlung
effect discussed here. The total yield is the
external envolvent and it is dominated by electrons for a wide range of frequencies. The figure is plotted adding Eqs.~\eqref{flux2} and~\eqref{cosmic2} (solid line).
The proton contribution is shown separately (dashed line). For comparison, the approximate galactic radio background
(basically from electron synchrotron radiation, see Ref.~\cite{kraus}) is shown. Note that the radio background is 
not well measured at present below
10 MHz but there are indications suggesting a marked decrease below 3 MHz. In the 100 MHz region the axion induced signal
is about nine orders of magnitude smaller than the background.}\label{photonflux}
\end{figure}

If the power spectrum of the cosmic rays is characterized by an exponent $\gamma$ then the produced radiation has a spectrum
$k^{-\frac{\gamma-1}{2}}$ for proton primaries, which becomes $k^{-\frac{\gamma}{2}}$ for electron primaries.
The dependence on the key parameter $\eta_0\propto \sqrt{\rho_\text{DM}}/f_a$
comes with the exponent $\eta_0^{\frac{1+\gamma}{2}}$ and $\eta_0^{\frac{2+\gamma}{2}}$ for protons and electrons, respectively.
However, for the regions where the radiation yield is largest, electrons amply dominate. 

It should be noted that the cosmic ray fluxes used in Eqs.~\eqref{flux2} and~\eqref{cosmic2} are values measured locally in the inner Solar system. It is known that
the intensity of cosmic rays increases with distance from the Sun because the modulation due to the Solar wind makes more difficult for them to reach us,
particularly so for electrons. Therefore the above values have to be considered as lower bounds for the cosmic ray flux,
which may be up to 10 times larger in the nearby interstellar medium. This would of course increase $I(k)$ above our estimations.\\

In order to see whether this flux is measurable from Earth or not one has to determine the diffuse noise perceived by the receiver in the appropriate wavelength, 
know identified background sources, and of course take into account the atmosphere transparency at that radiation wavelength. As it is well
known (see Refs.~\cite{opacity,kraus}) the atmosphere is transparent to radiation in the terrestrial microwave window ranging from approximately 6 mm (50 GHz,  $2
\times10^{-4}$ eV) to 20 m (15 MHz, $6 \times10^{-8}$ eV), becoming opaque at some water vapor and oxygen bands and less transparent as
frequency increases up to 1 THz. The current technology allows for radio detection from space up to 2 THz (e.g. with the Herschel Space
Observatory~\cite{Graauw}) but the low receiver sensitivity at frequencies in the submillimeter band ($>300$ GHz) could be an issue.
There are further considerably narrower windows in the near infrared region from 1 $\mu$m (300 THz, 1.2 eV) to around 10 $\mu$m (30 THz, 0.12 eV).
This region can be explored by space missions. The atmosphere blocks out completely the emission in the UV and X-Ray region
corresponding to $\lambda < 600$ nm ($k > 80$ eV), a region that is actively being explored by spaceborne missions.

If  $\lambda > 2.5$ m (0.8 GHz, $3 \times 10^{-6}$eV), the galactic synchroton radiation noise increases rapidly difficulting
the detection of any possible signal. Note however that while the power spectrum of the axion-related radiation from proton primaries is the same as
the one from the synchrotron radiation they produce~\cite{synchro1,synchro2}, the bulk of the Galaxy synchroton radiation is due to electrons whose
spectral power law describing the axion-induced Bremsstrahlung is different. In addition there would be a difference between the galactic and the axion based
synchrotron emission anyway. In fact\footnote{We thank P. Planesas for pointing out this
possibility to us.}, in areas of high galactic latitude, where no local features superpose the broad galactic emission,
the measured spectral index is  $\sim - 0.5$~\cite{rr}.
Instead, the axion induced effect has a power $\sim - 1.5$ if we assume $\gamma\sim 3$.

The maximum observed values~\cite{db1,db2,db3,db4} for the intensities are: 10$^4$ m$^{-2}$s$^{-1}$sr$^{-1}$ in
the X-Ray region and up to 10$^{10} - 10^{14}$ m$^{-2}$s$^{-1}$sr$^{-1}$
in the radio, IR and UV regions but the sensitivity of antenna arrays at very low frequencies such as the LWA~\cite{lwa} can be as low
as 0.1 mJy$\simeq10^3$m$^{-2}$s$^{-1}$sr$^{-1}$ or even less.
The sensitivy that can be reached in the SKA antenna is of particular interest for our purposes. It
can be estimated~\cite{condon} assuming an integration time of 50 hours at the lowest frequency
to be 650 nJy. This is clearly several orders below the expected size of the effect, even
assuming the worst possible case for the electron flux. Therefore, while the effect is below the
sensitivity of existing antennas it will be within the reach of several projects in construction
or under consideration\footnote{It may be worth noticing that the long standing 
project of setting up an antenna on the far side 
of the Moon~\cite{moon} could reach sensitivities of $10^{-5}$ Jy or 
less, also providing enough sensitivity even for
pessimistic values of the electron flux. Such an antenna would of 
course not be limited by atmosphere opacity,
being sensitive -in principle- to even longer wavelengths.}.

Once it is clear that antennas can measure fluxes twelve orders below the dominant Galaxy synchrotron
radiation in the galactic plane, it is obvious that 
sensitivity to the axion signal (`only' nine orders below the average galactic noise) is not an issue, 
the real difficulty is to disentangle the effect from the background or foreground. For this
purpose the rather different power dependence should prove essential. The difference in power spectrum 
between the expected signal and the background is
even more marked for regions of high galactic latitude as already mentioned. Good angular
resolution will be essential too as observers looking for this signal will probably be interested
in focusing their instruments in regions with low magnetic fields\footnote{Note that the
Galaxy magnetic field varies by about three orders of magnitude, from $\mu$G to mG.}, where 
synchrotron radiation
will be at a minimum  gaining several orders of magnitude in the signal-to-noise
ratio\footnote{The synchrotron radiation depends quadratically on the
magnetic field, hence a change of two orders of magnitude in the magnetic 
field represents a variation of four orders
in the amount of the synchrotron radiation background.}.

While it is obviously beyond the scope of this work (and the expertise of the authors)
to present a definite proposal to measure the tiny axion-induced Bremsstrahlung predicted, we do conclude that it 
is conceivably within the reach of a new generation of instruments specifically designed for exploration
of the long wavelength region. We do not exclude that it can be found in the exploration
of close extragalactic sources either. In both cases the main unknown is a detailed understanding
of the nature and spectrum of electron cosmic rays, an issue worth investigating by itself for a variety of
reasons.

\section{Summary}

In this chapter we have considered the effect of a mildly time dependent pseudoscalar background on charged particles. A physical
situation worth exploring is the influence of a diffuse relic cold axion background on cosmic ray propagation. The effect does not depend on the particular
axion model. It is universal and can be computed unambiguously.

The effect is completely calculable in great detail because particle propagation is governed by a modification of QED that is
exactly solvable. We have determined the kinematical constraints, the characteristics of the emitted radiation and the rate of energy loss of charged particles 
moving in such a medium. Some rather non-intuitive features appear and the results, we believe, are interesting per se.

The effects depend, as in Ch.~\ref{ch-photon}, on the key quantity $\eta_0\propto\sqrt{\rho_a}/f_a$. 
However, the rate of energy loss, for primaries with energies that survive the GZK cut-off, depends on $\eta_0^2$ and the effect of the ``axion shield''
is completely negligible. Nevertheless, the ``Bremmstrahlung'' photons emitted may be measurable. We have computed the photon energy flux, which depends on that
of cosmic rays. Our estimate is not very precise, as the main contribution comes from electrons, and their spectrum is not very well known. 

Other comments pertinent here are the following. First, one should note that the effect discussed here is a collective one. 
This is at variance with the GZK effect alluded in the introduction - the CMB radiation is not coherent over large scales. 
For instance, no similar effect exists due to hot axions produced thermally. A second observation is that some of the scales that play a role in the present discussion
are somewhat non-intuitive (for instance the ``cross-over'' scale $m_p^2/\eta_0$ or the threshold scale $m_\gamma m_p/\eta_0$). This is due to
the non Lorentz-invariant nature of this effect. Also, it may look surprising at first that an effect that has such a low probability may give a small but
not ridiculously small contribution. The reason why this happens is that the number of cosmic rays is huge. It is known that they contribute to the energy density
of the Galaxy by an amount similar to the Galaxy's magnetic field~\cite{adr}.

\chapter{Conclusions}\label{ch-concl}

Axions solve two of the puzzles that the Standard Model of particle physics still faces. On the one hand, they emerge in a solution to the strong CP problem:
the fact that QCD appears to respect the CP symmetry despite there being no apparent reason as to why that should happen. 
The Peccei-Quinn solution adds a new approximate chiral symmetry, involving some new fields not present in the current Standard Model. 
When this symmetry is spontenously broken, it necessarily gives rise to a pseudo-Goldstone boson: a very light pseudoscalar particle known as the axion.

On the other hand, axions provide a still viable candidate to explain the dark matter in the Universe, through the process of vacuum misalignment. 
Thus, the interest in studying axions is twofold.\\

In the first part of this thesis we have studied a specific model in which an axion appears, the Dine-Fischler-Srednicki-Zhitnitsky (DFSZ) model. 
It is a rather simple extension of the popular two-Higgs-doublet model (2HDM).
An extra scalar field, singlet under the SM gauge group, is added. This field, however, does carry a PQ charge, and so do
the usual SM quarks. Unlike in the other popular model (the Kim-Shifman-Vainshtein-Zakharov, KSVZ, model), axions have a tree-level coupling to leptons in the DFSZ model.

In the second part of the thesis, we have explored ways to discern whether a cold axion background (CAB), produced via misalignment, 
is responsible for the dark matter content of the Universe. Rather than attempt to directly detect the axion background, we have followed a more indirect path: 
the study of photon propagation in its presence. We propose two different ways to approach the subject: 
the study of the evolution of a photon wave that mixes with axions in the presence of a magnetic field and a CAB and the emission of photons by a fast-moving 
charged particle, a process not possible in usual QED but that can take place when a CAB is considered.\\

In Chapter~\ref{ch-2hdm} we study in detail the DFSZ axion model, an extension of the usual 2HDM that adds an extra scalar singlet, which
is coupled only to the Higgs fields through the scalar potential. The chiral PQ symmetry is 
realised with this singlet, the Higgs fields and the quarks. The Higgs fields get the usual vevs at the electroweak scale $v$, 
while the extra scalar gets a vev that must be at a much higher scale, since it is proportional to the axion decay constant $f_a$. Between the Higgs doublets and the 
singlet there are ten degrees of freedom, of which three are ``eaten'' by the gauge bosons, as usual. Of the remaining seven particles, 
one is necessarily massless (since the PQ
symmetry is spontaneously broken) and is identified as the axion. Because the axion is a combination of the phases of the singlet and the Higgs fields, it has couplings to
leptons at tree level in this model. Another state has a mass at the electroweak scale and can be readily identified as the $126$~GeV particle found at the LHC, while
a third one always has a very high mass, proportional to $f_a$. The remaining four particles are a neutral scalar, a neutral pseudo-scalar and two charged particles (the
usual ``extra Higgses'' of the 2HDM). They all have similar masses, proportional to the axion decay constant, 
but also to the coupling between the singlet and the Higgs doublets. We study different scenarios, where their masses range from values slightly above the 
Higgs discovered at the LHC to the order of $f_a$, far beyond the reach of current experiments. We also examine the mass spectrum in
the case where the potential has an additional symmetry known as custodial symmetry.

We derive the effective Lagrangian at low energies, which contains only the three Goldstone bosons (which give mass to the gauge bosons), the light Higgs and the axion.
Higgs and axion couplings are also derived, the former being different in the DFSZ model than those of regular 2HDM models.

Finally, we compute the electroweak parameter $\Delta\rho$, which is measured experimentally with great accuracy and derive the constraints it imposes on the parameters
of the model, in the different proposed scenarios.\\

The following two chapters are devoted to the study of a cold axion background, produced via vacuum misalignment, as the sole contributor to the dark matter
content of the Universe. All the results depend on the key quantity $\eta_0$ that is related to both the dark matter density and the strength of axion interacions,
$\eta_0\propto\sqrt{\rho_{DM}}/f_a$. Since this quantity is very small (at most, we expect $\eta_0\sim10^{-20}$~eV), all the effects are tiny.\\

In Chapter~\ref{ch-photon} we study the axion-photon system in the presence of two backgrounds: an external magnetic field $\vec B$ and a cold axion background (CAB), 
which we assume to be entirely responsible for dark matter. 
First we consider the case where only the CAB is present and find out that some frequencies are forbidden due to the oscillatory nature of the background, an effect
similar to the appearance of energy gaps in a semiconductor. The CAB has the effect of mixing the two photon polarisations, 
while a magnetic field mixes the axion with just one of them (the one parallel to $\vec B$). When both backgrounds are considered, the three states are mixed and
we compute the eigenvectors and the proper frequencies of the system.

Afterwards, we compute the propagator of the photon field in this setting and use it to derive the evolution of a photon wave that is initially linearly polarised
at an angle with respect to the external magnetic field. Such a wave will change its polarisation over time, due to the CAB and its mixing with axions. 
We find out how polarisation evolves and separate the effects due to the magnetic field and the CAB, which are fundamentally different. Experiments related to photon
polarisation usually consist of two nearly parallel mirrors. Light is made to bounce back and forth and its polarisation is studied. The magnetic field tends to increase
the angle of polarisation and its effects accumulate each time light bounces. The cold axion background just induces a net rotation of the polarisation. However,
its effects tend to cancel unless the mirror separation is tuned to the inverse of the axion mass. When this happens, the effect is accumulated and could maybe
be measured, which would demonstrate the presence of the CAB. In order to perform this kind of search, a scan over axion masses would have to be implemented, much like 
in other experiments such as ADMX. The measurement of this polarisation change is of course very challenging, as axion interactions are very weak.\\

In Chapter~\ref{ch-cosmic} we use the results obtained in Ch.~\ref{ch-photon} for vanishing $\vec B$ to conclude that only circularly polarised photons are
energy eigenstates in the presence of a CAB. The dispersion relation is modified to include a term proportional to $\eta_0$. One polarisation has a energy slightly above 
a usual QED photon, while the other is a bit less energetic. This means that a charged particle, such as a cosmic ray, can emit one of these photons 
(the one with slightly lower energy), a process forbidden in regular QED on momentum conservation grounds. 
We study in detail this process, establishing kinematical constraints on the cosmic ray energy and the range of emitted photon momenta. The possibility of this process
enables a mechanism for cosmic ray energy loss. However, this energy loss depends not on $\eta_0$, but on its square, making its detection hopeless.

Nevertheless, the amount of cosmic rays is huge, so the number of emitted photons must be consequently large. We compute this energy flux, using a cosmic ray spectrum
of the form $J\propto E^{-\gamma}$. Even though the flux of electrons is smaller than that of protons, they contribute more to the photon energy flux,
due to their smaller mass. We find out that the electron contribution is proportional to $\eta_0^{(2+\gamma)/2}k^{-\gamma/2}$, while the proton contribution goes as
$\eta_0^{(1+\gamma)/2}k^{(1-\gamma)/2}$.
Although the major contribution comes from electrons and their flux is much less well understood, we provide an estimate of the photon energy flux, which may be
within reach of upcoming experiments.\\

\chapter{List of publications}\label{ch-pub}

\section{Publications in refereed journals}

\begin{itemize}
 \item {D. Espriu and A. Renau, \emph{Photons in a cold axion background and strong magnetic fields: polarimetric consequences}, Int.J.Mod.Phys. A30 (2015) 1550099,
 arXiv:1401.0663 [hep-ph].}
 \item {D. Espriu and A. Renau, \emph{Photon propagation in a cold axion background with and without magnetic field}, Phys.Rev. D85 (2012) 025010,
 arXiv:1106.1662 [hep-ph].}
 \item {D. Espriu, F. Mescia and A. Renau, \emph{Axions and high-energy cosmic rays: Can the relic axion density be measured?}, JCAP 1108 (2011) 002,
 arXiv:1010.2589 [hep-ph].}
 \item {A. A. Andrianov, D. Espriu, F. Mescia and A. Renau, \emph{The Axion shield}, Phys.Lett. B684 (2010) 101-105,
 arXiv:0912.3151 [hep-ph].}
\end{itemize}

\section{Other publications}

\begin{itemize}
 \item {D. Espriu and A. Renau, \emph{Photon propagation in a cold axion condensate}, Proceedings of the Patras Workshop 2013, Mainz, June 2013,
 arXiv:1309.6948 [hep-ph].}
 \item {D. Espriu and A. Renau, \emph{How a cold axion background influences photons}, Proceedings of the 7th Patras Workshop on Axions, WIMPs and WISPs (2011),
 arXiv:1111.3141 [hep-ph].}
 \item {D. Espriu and A. Renau, \emph{Axions and Cosmic Rays}, Proceedings of the Quarks 2010 International Seminar, Kolomna, Russia,
 arXiv:1010.3580 [hep-ph].}
\end{itemize}
\section{Publication pending}

\begin{itemize}
 \item {D. Espriu, F. Mescia and A. Renau, \emph{Axion-Higgs interplay in the two Higgs-doublet model}, 2015. Submitted to Phys.Rev. D.
 arXiv:1503.02953 [hep-ph].}
\end{itemize}

\graphicspath{{6-resum/figures/}{figures/}}

\chapter{Resum en català}\label{chap:resum}

\section{Introducció}

El Model Estàndard de la física de partícules és una teoria quàntica de camps que descriu el món de les partícules subatòmiques i les seves interaccions:
les forces nuclears forta i feble i la interacció electromagnètica.

En la seva forma actual, el Model Estàndard conté 17 partícules fonamentals amb les seves antipartícules corresponents. Hi ha dotze fermions, amb spin 1/2:
els quarks i els leptons. Els quarks tenen càrrega de color i, per tant, senten la for\c{c}a nuclear forta,
mentre que els leptons només interactuen mitjan\c{c}ant la interacció feble
i l'electromagnètica. També hi ha els transmissors de les interaccions: els bosons gauge, de spin 1. Per últim, el bosó de Higgs, amb spin 0, és el responsable
de donar massa a les partícules.\\

Malgrat l'enorme èxit teòric i experimental del Model Estàndard, encara hi ha preguntes que no respon, com per exemple:

\begin{itemize}
 \item{No incorpora la interacció gravitatòria.}
 \item{No contempla la massa dels neutrins (se sap que són massius, ja que hi ha diferència de massa entre les diferents espècies de neutrí).}
 \item{La matèria és molt més abundant que l'antimatèria, fet que el Model Estàndard no és capa\c{c} d'explicar.}
 \item{El problema de CP en la interacció forta (\emph{strong CP problem}): per què la cromodinàmica quàntica (QCD) no trenca la simetria CP?}
 \item{El model estàndard cosmològic requereix l'existència de matèria fosca i energia fosca, 
però el Model Estàndard de física de partícules no conté cap candidat per explicar aquests fenòmens.}
\end{itemize}

En aquesta tesi es tracten els dos últims punts. Proposem que els axions, una solució de l'\emph{strong CP problem}, són responsables de la matèria fosca.\\

\subsection{Matèria fosca}

Mesures recents indiquen que la matèria ordinària (o bariònica) constitueix només el 5\% del total de l'energia de l'Univers. 
El 27\% es troba en forma de matèria fosca, que no interactua de manera significativa amb la matèria ordinària, excepte gravitatòriament.
El 68\% restant és energia fosca. 

La naturalesa de la matèria fosca és encara desconeguda. La seva existència s'infereix a partir dels seus efectes gravitatoris: 
alguns objectes astronòmics es comporten com si tinguessin més massa que la que es pot calcular a partir dels seus components lluminosos, 
com les estrelles o el gas interestel$\cdot$lar~\cite{halo}.

\subsubsection{Evidències}

L'existència de la matèria fosca està ben fonamentada amb nombroses observacions. Aquests en són alguns exemples:
\begin{itemize}
\item{\textbf{Corbes de rotació de les galàxies}. Un objecte en una òrbita circular té una velocitat radial $v(r)=GM(r)/r$, 
on $M(r)$ és la massa continguda a l'interior de l'òrbita de radi $r$. Per tant, 
la velocitat de les estrelles a la regió exterior d'una galàxia espiral hauria de decréixer segons $v(r)\propto1/\sqrt r$, 
ja que la massa interior és aproximadament constant. En canvi, s'observa que el perfil de velocitats esdevé constant, 
la qual cosa permet concloure que hi ha un ``halo'' de matèria fosca.}
\item{\textbf{Lents gravitatòries}. Malgrat que els fotons no tenen massa, d'acord amb la teoria de la Relativitat General 
la gravetat afecta la llum. Quan un raig de llum passa a prop d'un objecte massiu, 
la seva trajectòria queda alterada, i la posició aparent de la font que el va emetre és modificada. 
Això dóna lloc al fenomen de les lents gravitatòries: un cos massiu situat entre un emissor i un receptor corba els raigs de llum, 
de la mateixa manera que ho fa una lent òptica. L'estudi de les lents gravitatòries permet deduir la massa de l'objecte que actua com a lent. 
En molts casos, s'observa que aquesta massa és superior a la que s'infereix degut a la lluminositat de l'objecte. La massa que falta es troba en forma de matèria fosca.}
\item{\textbf{Radiació còsmica de fons}. Es tracta de radiació provinent de l'Univers primitiu. 
Segueix un espectre de cos negre, amb una temperatura mitjana $\langle T\rangle=2.7$~K. Tot i això, 
la temperatura presenta petites variacions que permeten extreure informació cosmològica. En particular, 
es poden fer servir per determinar la quantitat de matèria bariònica i la quantitat total de matèria de l'Univers. Aquestes dues quantitats no coincideixen.
Per tant, hi ha matèria fosca.}
\end{itemize}

\subsubsection{Candidats a matèria fosca}

Un bon candidat a matèria fosca ha de ser una partícula neutra (si fos carregada, emetria llum i no seria ``fosca'') i no-relativista 
(la seva energia cinètica ha de ser molt menor a la seva energia en repòs), d'acord amb la formació d'estructura observada a l'Univers.

D'entre les partícules del Model Estàndard, l'única possibilitat són els neutrins, però no són prou abundants per explicar el total de la matèria fosca.
Per tant, els candidats a matèria fosca han de provenir de teories més enllà del Model Estàndard.

En aquesta tesi ens centrem en l'estudi d'un d'aquests possibles candidats, l'axió, que apareix en una de les solucions de l'\emph{strong CP problem}. 
No és, però, l'única opció. Entre els altres candidats, trobem els neutrins estèrils i les partícules massives feblements interactuants, 
que provenen de teories supersimètriques o teories amb dimensions extra.

\subsection{L'\emph{strong CP problem}}

L'existència d'unes certes configuracions de buit fa que aparegui un terme extra al lagrangià de QCD, anomenat terme $\theta$:
\be
\mathcal{L}_\theta=\theta\frac{g_s^2}{32\pi^2}G_a^{\mu\nu}\tilde G_{a\mu\nu}.
\ee
Aquest terme viola la simetria CP i indueix un moment dipolar elèctric per al neutró d'ordre~\cite{nedm}
\be
d_n\simeq\frac{e\theta m_q}{m_N^2}\simeq10^{-16}\theta \,e\,\text{cm}.
\ee
El límit experimental per a aquesta quantitat~\cite{baker} implica un límit sobre el paràmetre $\theta$:
\be\label{catnedm}
|d_n|<2.9\cdot10^{-26} e\,\text{cm}\quad\longrightarrow\quad\theta<10^{-9}.
\ee
En això consisteix l'\emph{strong CP problem}: per què és tan petit aquest paràmetre, en principi, arbitrari?

\subsubsection{Una solució al problema}

Una solució natural al problema consisteix a introduir una nova simetria quiral global, coneguda com a $U(1)_{\text {PQ}}$, degut a R. Peccei i H. Quinn.
Quan aquesta simetria es trenca espontàniament, apareix un bosó de Goldstone anomenat axió. 
L'axió té associat un potencial que, en ser minimitzat, substitueix el paràmetre $\theta$ pel camp dinàmic de l'axió, de manera
que l'\emph{strong CP problem} desapareix. Els ingredients mínims per a la seva resolució són un camp escalar que adquireix un valor esperat en el buit i un quark, 
que pot ser un dels quarks del Model Estàndard o un de diferent.

\subsection{Axions}

L'axió és, doncs, un producte de la solució de Peccei i Quinn a l'\emph{strong CP problem}. Es mescla amb el pió neutre i el mesó $\eta$ i adquireix una massa
\be\label{cataxionmass1}
m_a=\frac{z^{1/2}}{1+z}\frac{f_\pi}{f_a}m_\pi,\quad z=\frac{m_u}{m_d},
\ee
inversament proporcional a la constant de desintegració $f_a$.

Independentment del model, els axions sempre tenen un acoblament amb dos fotons
\be\label{catprima}
\mathcal L_{a\gamma\gamma}=g_{a\gamma\gamma}\frac\alpha{2\pi}\frac a{f_a}F^{\mu\nu}\tilde F_{\mu\nu}=\frac g4aF^{\mu\nu}\tilde F_{\mu\nu},
\ee
on $F_{\mu\nu}$ és el tensor electromagnètic. La constant d'acoblament, $g_{a\gamma\gamma}$, depèn del model, però és sempre d'ordre $\mathcal{O}(1)$.

\subsubsection{Models}

La solució de Peccei i Quinn involucra necessàriament física més enllà del Model Estàndard, 
ja que aquest conté un sol doblet de Higgs que s'acobla tant als quarks tipus
up com als quarks tipus down.\\

El model original, conegut com a model de Peccei, Quinn, Weinberg i Wilczek (PQWW~\cite{PQ1,PQ2,PQ3}), conté dos doblets de Higgs i els quarks del Model Estàndard. 
Ara bé, com
que l'axió està totalment contingut en els camps de Higgs, el valor de la constant de desintegració és proporcional a l'escala feble  
\be
f_a\propto v=246\text{ GeV}.
\ee
Un axió amb aquestes característiques interactua massa fortament, i està descartat experimentalment. 
La constant de desintegració ha de prendre un valor molt superior.

En canvi, altres models amb $f_a\gg v$ es coneixen com a models amb axions ``invisibles'' i encara són experimentalment viables.\\

Un d'aquests és el model de Kim-Shifman-Vainshtein-Zakharov (KVSZ~\cite{models3,models4}). Aquest model incorpora un nou singlet escalar i un nou quark pesant.
Com que cap camp del Model Estàndard duu càrrega de PQ en el model KSVZ, els axions no tenen cap acoblament amb els quarks i els leptons a nivell arbre.

Un altre model popular és el model de Dine-Fischler-Srednicki-Zhitnisky (DFSZ~\cite{dfs,models2}). \'Es un model amb dos doblets de Higgs, com el model de PQWW,
però afegeix un singlet escalar extra. El nou singlet i els dos doblets de Higgs es transformen sota la simetria PQ, així com els quarks i els leptons. 
Com que l'axió es troba
tant als doblets com al singlet, té acoblaments amb quarks i leptons a nivell arbre.

\subsubsection{El fons fred d'axions}

A més de ser part d'una solució de l'\emph{strong CP problem}, els axions també resolen el problema de la matèria fosca.

Després del trencament de simetria PQ, però a temperatures per sobre de l'escala de QCD, no hi ha potencial per al camp de l'axió i, 
per tant, pot prendre qualsevol valor. Quan la temperatura cau per sota de l'escala de QCD, els instantons indueixen un potencial sobre l'axió. 
Quan això passa, l'axió no es troba necessàriament al mínim del potencial i, per tant, oscil$\cdot$la. 
Aquestes oscil$\cdot$lacions es coneixen amb el nom de realineació del buit i emmagatzemen una densitat d'energia que pot
ser responsable de la matèria fosca de l'Univers.\\

Tot i que l'axió ha de ser necessàriament molt lleuger (per satisfer els límits experimentals), 
aquest fons fred d'axions té una dispersió de velocitats molt petita,
així que les partícules són no-relativistes.

\subsubsection{Límits astrofísics i cosmològics}

Tant la intensitat de les interaccions de l'axió com la seva massa són inversament proporcionals a la constant de desintegració $f_a$. 
Es poden fer servir consideracions astrofísiques i cosmològiques per posar límits sobre aquest paràmetre (i, per tant, sobre la massa).

Per exemple, els axions poden produir-se a l'interior de les estrelles i ser radiats cap a l'exterior. Això constitueix un mecanisme de pèrdua d'energia en 
les estrelles. Conseq\"uentment, els axions han d'interactuar prou feblement per tal que aquest mecanisme no afecti l'evolució estel$\cdot$lar observada.
D'aquesta manera s'obté un límit inferior a la constant de desintegració~\cite{stars}
\be
f_a>10^7\text{ GeV}, 
\ee
que es tradueix en un límit superior per a la massa
\be
m_a<0.1\text{ eV}.
\ee
Imposant que la densitat d'axions no excedeixi la densitat observada de matèria fosca, s'obté un límit superior per a la constant de desintegració
\be
f_a<10^{11}\text{ GeV}
\ee
o, en termes de la massa,
\be
m_a>10^{-5}\text{ eV}.
\ee

Combinant els dos límits obtenim la finestra de valors dels paràmetres per als quals els axions poden explicar la matèria fosca
\be\label{catbounds}
10^7\text{ GeV}<f_a<10^{11}\text{ GeV},\quad 10^{-5}\text{ eV}<m_a<0.1\text{ eV}.
\ee

\subsubsection{Cerques experimentals}

L'acoblament d'un axió amb dos fotons és el que proporciona les millors opcions de detectar experimentalment els axions.
Un fotó polaritzat paral$\cdot$lelament a un camp magnètic extern pot convertir-se en un axió i viceversa, mentre que els fotons polaritzats perpendicularment no es
veuen afectats pel camp magnètic. Aquest fet és la base de la majoria d'experiments que intenten detectar axions.

Curiosament, la majoria d'experiments d'axions tenen una característica en comú. El possible senyal d'aquests experiments augmenta considerablement quan una combinació
de paràmetres coincideix amb la massa de l'axió. Per tant, l'estratègia consisteix a fer un escombrat, passant per diferents valors de $m_a$.\\

Els experiments d'axions solars, com el CERN Axion Solar Telescope~\cite{cast} i l'International Axion Observatory~\cite {iaxo}, 
pretenen detectar axions provinents directament del nucli solar. Mitjan\c{c}ant un camp magnètic, 
converteixen els axions en fotons, que són recollits per un detector de raigs X.

Els haloscopis, com l'Axion Dark Matter Experiment~\cite{admx}, 
intenten detectar la presència d'axions a l'halo de matèria fosca de la Galàxia mitjan\c{c}ant una cavitat
ressonant, que pot convertir axions en fotons quan la seva freq\"uència ressonant coincideix amb la massa de l'axió.

Per últim, els experiments de regeneració de fotons, com l'Optical Search for QED Vacuum Birefringence, Axions and Photon Regeneration~\cite{osqar} i
l'Any Light Particle Search~\cite{alps}, envien un feix de fotons a una paret absorbent, que n'impedeix el pas. Col$\cdot$locant un camp magnètic davant i darrera de la
paret, pretenen convertir fotons en axions, que travessen la paret lliurement i, en ser reconvertits en fotons, poden ser detectats.

\section{Axió i Higgs en el model de dos doblets}

Al Capítol~\ref{ch-2hdm} estudiem en detall el model DFSZ, una extensió del model de dos doblets de Higgs que afegeix un nou singlet escalar. 
Els doblets de Higgs adquireixen els valors esperats en el buit usuals, a l'escala electrofeble, 
mentre que el nou singlet té un valor esperat en el buit molt més gran, de
l'ordre de la constant de desintegració de l'axió, $f_a$.

Presentem el model i n'estudiem les simetries, en particular la simetria PQ, 
necessària per resoldre l'\emph{strong CP problem}. L'espectre d'aquesta teoria conté una partícula sense massa, 
l'axió, i una partícula molt pesant, associada al mòdul
del singlet. Un dels escalars neutres pot identificar-se amb el Higgs de 126 GeV recentment descobert a l'LHC. Les quatre partícules restants
(un escalar neutre, dos
escalars carregats i una partícula pseudoescalar) són els ``Higgs extra'' usuals dels models de dos doblets. 
Tots ells tenen masses similars, que poden ser lleugerament 
superiors a la del Higgs més lleuger o bé molt més pesants, segons el valor dels acoblaments entre el singlet i els dos doblets. 
Considerem quatre possibles escenaris,
segons els diferents valors que poden prendre aquests paràmetres.
També examinem la simetria custòdia, que restringeix el valors d'alguns dels paràmetres del model. \\

Estudiem el lagrangià efectiu a baixes energies, que conté els bosons de Goldstone (que donen massa als bosons gauge),
l'axió i el Higgs més lleuger, integrant la resta de camps pesants. També trobem els acoblaments efectius de l'axió i el Higgs en aquest model.

Per últim, imposem les restriccions provinents dels paràmetres electrofebles sobre les correccions obliq\"ues, 
en particular sobre $\Delta\rho$, continuant el treball de la
Referència~\cite{ce}. 

\section{Propagació de fotons en un fons fred d'axions i un camp magnètic}

Al Capítol~\ref{ch-photon} considerem els efectes d'un fons fred d'axions i un camp magnètic sobre la propagació de fotons.
Trobem les equacions de moviment per al sistema i veiem que els tres estats (axió i les dues polaritzacions del fotó) queden mesclats. 
El camp magnètic mescla l'axió amb una de les components del fotó, mentre que el fons d'axions mescla les dues polaritzacions del fotó.\\

En el cas que no hi ha camp magnètic, observem que algunes freq\"uències del fotó estan prohibides. 
L'efecte és similar al que succeeix en un semiconductor, on la periodicitat
espacial del potencial fa que alguns valors de l'energia dels electrons no siguin possibles.
En el nostre cas, el fons d'axions és periòdic en el temps, així que apareixen
bandes prohibides en el moment. Calculem la posició i l'amplada de les bandes, que són, probablement, massa estretes com per ser detectades.\\

Més endavant, trobem els estats propis i les freq\"uències pròpies del sistema, així com el propagador del camp del fotó. 
Fent ús d'aquest propagador, calculem l'evolució d'una ona inicialment polaritzada de manera lineal. 
Estudiem el canvi en la seva polarització i l'aparició d'una certa el$\cdot$lipticitat.
Separem els efectes del camp magnètic i el fons d'axions i proposem una manera de detectar el fons fred d'axions en experiments polarimètrics.
Concretament, estudiem el cas en què una ona de llum es fa rebotar entre dos miralls paral$\cdot$lels. Mentre que el camp magnètic té un efecte que s'acumula cada
vegada que la llum rebota, el fons d'axions és més complicat de detectar, ja que és oscil$\cdot$lant. Ara bé, si s'ajusta la separació entre miralls de manera que 
coincideixi amb l'invers de la massa de l'axió (que determina el període d'oscil$\cdot$lació), se'n poden acumular els efectes i, finalment, detectar-lo.

\section{Propagació de raigs còsmics altament energètics en un fons fred d'axions}

Al Capítol~\ref{ch-cosmic} fem servir els resultats anteriors per estudiar la propagació de raigs còsmics. 
Els raigs còsmics són partícules carregades (electrons, protons i altres nuclis) altament energètiques que arriben a la Terra des de l'exterior. 

El fons fred d'axions modifica la relació de dispersió dels fotons i fa que només les polaritzacions circulars siguin estats propis de l'energia.
Una de les dues polaritzacions té una energia lleugerament superior a la d'un fotó lliure, mentre que l'altra la té lleugerament inferior.
Això permet l'emissió de fotons per part d'una partícula carregada, un procés que està
prohibit en l'electrodinàmica quàntica degut a la conservació de l'energia i el moment. Degut a l'emissió de fotons, els raigs còsmics perden energia. 
Calculem el ritme
de pèrdua d'energia, que resulta ser completament indetectable. Malgrat això, el nombre de raigs còsmics és molt gran, 
així que els fotons emesos sí que podrien ser detectats.

Calculem també l'espectre d'emissió dels fotons. Depèn del flux de raigs còsmics, que segueix una llei de potències.
L'espectre de fotons també segueix una llei de potències, però amb un exponent diferent. La major contribució prové dels electrons,
ja que la seva massa és menor. Ara bé, el flux d'electrons no està gaire ben entès a hores d'ara,
així que el càlcul és poc precís. Malgrat això, donem una estimació de l'espectre de fotons, que podria estar a l'abast d'experiments futurs.

\section{Conclusions}

Els axions representen una bona solució al problema de violació de CP en la interacció forta, un dels enigmes encara no resolts del Model Estàndard.
La solució de Peccei-Quinn afegeix una nova simetria al Model Estàndard que, en ser trencada espontàniament, produeix un bosó de Goldstone, l'axió.
En una finestra de l'espai de paràmetres que encara no està exclosa, els axions són un candidat viable a ser la matèria fosca de l'Univers. Per tant, l'interès
a estudiar-los és doble.\\

Actualment hi ha dos models de referència, anomenats KSVZ i DFSZ. En la primera part de la tesi recuperem el model DFSZ i l'estudiem a fons, tenint en compte el 
descobriment recent del bosó de Higgs a l'LHC.\\

La majoria d'experiments d'axions només intenten esbrinar si l'axió existeix com a partícula, però no consideren la seva rellevància com a matèria fosca. 
Per tant, és important dissenyar nous mètodes per explorar aquesta possibilitat. Aquest és el segon objectiu de la tesi.
Proposem maneres indirectes de detectar la presència d'un fons d'axions mitjan\c{c}ant els seus efectes en la propagació de fotons i raigs còsmics. 
En particular, ens centrem en el canvi de polarització que experimenta una ona electromagnètica que travessa un fons d'axions i 
en l'emissió de fotons per part dels raigs còsmics, també en presència del fons d'axions.

\section{Publicacions}

\subsection{Publicacions en revistes}

\begin{itemize}
 \item {D. Espriu and A. Renau, \emph{Photons in a cold axion background and strong magnetic fields: polarimetric consequences}, Int.J.Mod.Phys. A30 (2015) 1550099,
 arXiv:1401.0663 [hep-ph].}
 \item {D. Espriu and A. Renau, \emph{Photon propagation in a cold axion background with and without magnetic field}, Phys.Rev. D85 (2012) 025010,
 arXiv:1106.1662 [hep-ph].}
 \item {D. Espriu, F. Mescia and A. Renau, \emph{Axions and high-energy cosmic rays: Can the relic axion density be measured?}, JCAP 1108 (2011) 002,
 arXiv:1010.2589 [hep-ph].}
 \item {A. A. Andrianov, D. Espriu, F. Mescia and A. Renau, \emph{The Axion shield}, Phys.Lett. B684 (2010) 101-105,
 arXiv:0912.3151 [hep-ph].}
\end{itemize}

\subsection{Altres publicacions}

\begin{itemize}
 \item {D. Espriu and A. Renau, \emph{Photon propagation in a cold axion condensate}, Proceedings of the Patras Workshop 2013, Mainz, June 2013,
 arXiv:1309.6948 [hep-ph].}
 \item {D. Espriu and A. Renau, \emph{How a cold axion background influences photons}, Proceedings of the 7th Patras Workshop on Axions, WIMPs and WISPs (2011),
 arXiv:1111.3141 [hep-ph].}
 \item {D. Espriu and A. Renau, \emph{Axions and Cosmic Rays}, Proceedings of the Quarks 2010 International Seminar, Kolomna, Russia,
 arXiv:1010.3580 [hep-ph].}
\end{itemize}
\subsection{Treballs pendents de publicació}

\begin{itemize}
 \item {D. Espriu, F. Mescia and A. Renau, \emph{Axion-Higgs interplay in the two Higgs-doublet model}, 2015. Enviat a Phys.Rev. D.
 arXiv:1503.02953 [hep-ph].}
\end{itemize}


\begin{appendices}

\graphicspath{{appendices/figures/}{figures/}}

\chapter{Technical aspects of Chapter~\ref{ch-2hdm}}\label{app:2hdm}
\section{Minimisation conditions of the potential}
\label{sec:min}
The minimisation conditions for the potential \eqref{potential} are
\be
\lambda_1\left(2v^2c_\beta^2-V_1^2\right)+\lambda_3\left(2v^2-V_1^2-V_2^2\right)+\frac{v_\phi^2}2\left(a+c\tan\beta\right)=0,
\ee
\be
\lambda_2\left(2v^2s_\beta^2-V_2^2\right)+\lambda_3\left(2v^2-V_1^2-V_2^2\right)+\frac{v_\phi^2}2\left(b+\frac{c}{\tan\beta}\right)=0,
\ee
\be
\lambda_\phi\left(v_\phi^2-V_\phi^2\right)+2v^2\left(ac_\beta^2+bs_\beta^2-cs_{2\beta}\right)=0.
\ee
These allow us to eliminate the dimensionful parameters
$V_\phi$, $V_1$ and $V_2$ in favor of the different couplings, $v$ and $v_\phi$. In the case
where $\lambda_\phi=0$ it is also possible to eliminate $c$. 

\section{\texorpdfstring{$0^+$}{\space} neutral sector mass matrix}
\label{app:B}
The $3\times 3$ mass matrix\footnote{Being a mass matrix, $M_{HS\rho}$ is symmetrical. Terms above the diagonal have been omitted.} is
\small\be
M_{HS\rho}=4
\left(
\begin{array}{ccc}
8v^2\left(\lambda_1c_\beta^4+\lambda_2s_\beta^4+\lambda_3\right) &  
&  \\
 4v^2\left(-\lambda_1c_\beta^2+\lambda_2s_\beta^2\right)s_{2\beta}  & \frac{2cv_\phi^2}{s_{2\beta}}+2v^2\left(\lambda_1+\lambda_2\right)s_{2\beta}^2 
 &  \\
2vv_\phi\left(ac_\beta^2+bs_\beta^2-cs_{2\beta}\right) & -vv_\phi\left[\left(a-b\right)s_{2\beta}+2cc_{2\beta}\right] & \lambda_\phi v_\phi^2
\end{array}
\right)
\ee\normalsize
This is diagonalised with a rotation
\be
\left(\begin{array}{c}
H\\
S\\
\rho
\end{array}\right)=R
\left(\begin{array}{c}
h_1\\h_2\\h_3
\end{array}\right).
\ee
We write the rotation matrix as
\be
R=\exp\left(\frac v{v_\phi}A+\frac{v^2}{v_\phi^2}B\right),\quad A^T=-A,\quad B^T=-B
\ee
and work up to second order in ${v}/{v_\phi}$. We find
\be
A_{12}=B_{13}=B_{23}=0,
\ee
so the matrix is
\be
R=\left(
\begin{array}{ccc}
1-\frac{v^2}{v_\phi^2}\frac{A_{13}^2}{2} & -\frac{v^2}{v_\phi^2}\frac{A_{13}A_{23}-2B_{12}}{2} & \frac{v}{v_\phi}A_{13}\\
-\frac{v^2}{v_\phi^2}\frac{A_{13}A_{23}+2B_{12}}{2} & 1-\frac{v^2}{v_\phi^2}\frac{A_{23}^2}{2} & \frac{v}{v_\phi}A_{23}\\
-\frac{v}{v_\phi}A_{13} & -\frac{v}{v_\phi}A_{23}  &  1-\frac{v^2}{v_\phi^2}\frac{A_{13}^2+A_{23}^2}{2}
\end{array}
\right),
\ee
with
\be
A_{13}=\frac2{\lambda_\phi}\left(ac_\beta^2+bs_\beta^2-cs_{2\beta}\right),\:
A_{23}=\frac{(a-b)s_{2\beta}+2cc_{2\beta}}{\frac{2c}{s_{2\beta}}-\lambda_\phi},
\ee
\be
B_{12}=
-\frac2cs^2_{2\beta}\left(\lambda_1c_\beta^2-\lambda_2s_\beta^2\right)
+\frac{s_{2\beta}}{\lambda_\phi c}\frac{c-\lambda_\phi s_{2\beta}}{2c-\lambda_\phi s_{2\beta}}
\left(ac_\beta^2+bs_\beta^2-cs_{2\beta}\right)\left[(a-b)s_{2\beta}+2cc_{2\beta}\right]
\ee
In the case of section \ref{sec:mam} when the breaking of 
custodial symmetry is $SU(2)\times SU(2)\to U(1)$ the mass matrix is
\be
M_{HS\rho}=4\left(\begin{array}{ccc}
     8v^2\left[\lambda_3+\lambda(s^4_\beta+c^4_\beta)\right] & -2\lambda v^2 s_{4\beta} & 2vv_\phi\left(a+2\lambda\frac{v^2}{v_\phi^2} s^2_{2\beta}\right)  \\
     -2\lambda v^2 s_{4\beta} & -4\lambda v^2 c^2_{2\beta} & 2\lambda\frac{v^3}{v_\phi} s_{4\beta} \\
     2vv_\phi\left(a+2\lambda\frac{v^2}{v_\phi^2} s^2_{2\beta}\right) & 2\lambda\frac{v^3}{v_\phi} s_{4\beta} & \lambda v_\phi^2
   \end{array}\right).
\ee
For case 4 of section \ref{sec:mam} the rotation matrix is
\be
R=\left(
\begin{array}{ccc}
\cos\theta & -\sin\theta & 0\\
\sin\theta & \cos\theta & 0\\
0 & 0 & 1
\end{array}
\right),\quad
\tan2\theta=
-\frac{\left(\lambda_1c_\beta^2-\lambda_2s_\beta^2\right)s_{2\beta}}{\left(\lambda_1c_\beta^2-\lambda_2s_\beta^2\right)c_{2\beta}+\lambda_3-
{cv_\phi^2}/(4v^2s_{2\beta})}.
\ee

\section{The limit \texorpdfstring{$\lambda_\phi=0$}{\space}}
\label{app:C}
The eigenvalues of  the $3\times 3$ mass matrix in the $0^+$ sector are
\begin{align}
m_{h_1}^2&=32v^2\left(\lambda_1c^4_\beta+\lambda_2s^4_\beta+\lambda_3\right)\cr
m_{h_2}^2&=\frac{v_\phi^2}2s^2_{2\beta}(ac^2_\beta+bs^2_\beta)+\mathcal{O}(v^2)\cr
m_{h_3}^2&=-8v^2\frac{(ac_\beta-bs_\beta)^2}{ac^2_\beta+bs^2_\beta}
\end{align}
Either $m_2^2$ or $m_3^2$ is negative. Note that the limit of $a,b$ small can not be taken directly in this case.

\section{Vacuum stability conditions and mass relations}
\label{sec:stab}
Vacuum stability implies the following conditions  on the
parameters of the potential~\cite{2HDM2}:
\be\nonumber
\lambda_1+\lambda_3>0,\quad \lambda_2+\lambda_3>0,\quad
2\lambda_3+\lambda_4+2\sqrt{(\lambda_1+\lambda_3)(\lambda_2+\lambda_3)}>0,
\ee
\be 
\lambda_3+\sqrt{(\lambda_1+\lambda_3)(\lambda_2+\lambda_3)}>0.
\ee
In the case of custodial symmetry except for $\lambda_{4B}\ne 0$, 
these conditions reduce to
\be
\lambda+\lambda_3>0,\quad \lambda+2\lambda_3>0,\quad 
4\lambda+4\lambda_3+\lambda_{4B}>0
\ee
and assuming $a,\,b,\,c$ very small (e.g. case 4) they impose two conditions on the masses:
\be
m_{A_0}^2+m_{h_1}^2-m_{h_2}^2>0,\quad m_{H_\pm}^2+m_{h_1}^2-m_{A_0}^2>0.
\ee

\section{Vertices and Feynman rules}
In the limit of $g'=0$, all the diagrams involved in the calculation of $\varepsilon_1$ are of the type of Fig.~\ref{vvxy}. 
All the relevant vertices are of the type seen in Fig.~\ref{vxy}, with all momenta assumed to be incoming. 
The relevant Feynman rules are as follows:
\begin{figure}[ht]
\center
\includegraphics[scale=0.6]{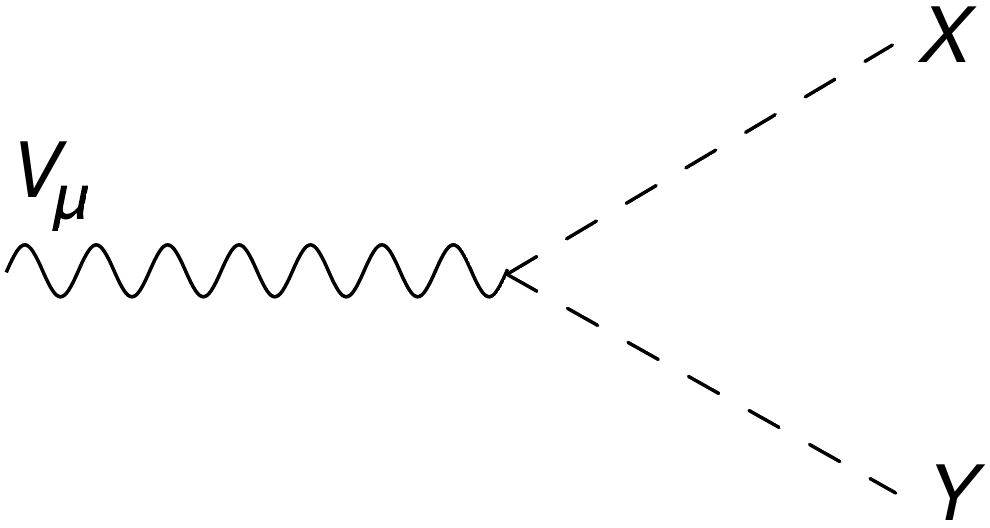}
\caption{Two scalars and a gauge boson.}\label{vxy}
\end{figure}
\begin{center}
\begin{tabular}{|c|c|}
\hline
\textbf{Interaction term} & \textbf{Feynman Rule for the vertex}\\
\hline
$\lambda V^\alpha X\partial_\alpha Y$ & $\lambda p_Y^\mu$\\
$\lambda V^\alpha X\der_\alpha Y$ & $\lambda(p_Y-p_X)^\mu$\\
\hline
\end{tabular}
\end{center}
To compute $\Pi_{ZZ}$ entering Eq.~(\ref{eq:deltarho}), we need diagrams like Fig.~\ref{vvxy} with $V=Z$. The $X,Y$ pairs are 
\begin{center}
\begin{tabular}{|c|c|c|c|}
\hline
\textbf{$X$} & \textbf{$Y$}&\textbf{Interaction term}&\textbf{Feynman Rule for the vertices}\\
\hline
$H_+$ & $H_-$        & $-\frac i2gW_3^\alpha H_+\der_\alpha H_-$ & $\frac{g^2}4(2p+q)_\mu(2p+q)_\nu$ \\
$S$   & $A_0$ & $\frac g2\frac{v_\phi}\vf W_3^\alpha S\der_\alpha A_0$ & $-\frac{g^2}4\frac{v_\phi^2}\vvf(2p+q)_\mu(2p+q)_\nu$ \\
$S$   & $a_\phi$        & $g\frac{v\sin2\beta}\vf W_3^\alpha S\partial_\alpha a_\phi$ & $-g^2\frac{v^2\sin^22\beta}{\vvf}(p+q)_\mu(p+q)_\nu$ \\
$H$   & $G_0$        & $-gW^\alpha_3H\partial_\alpha G_0$ & $-g^2(p+q)_\mu(p+q)_\nu$ \\
$G_+$ & $G_-$        & $\frac i2gW_3^\alpha G_+\der_\alpha G_- $& $\frac{g^2}4(2p+q)_\mu(2p+q)_\nu$ \\
\hline
\end{tabular}
\end{center}
To compute $\Pi_{WW}$ entering Eq.~(\ref{eq:deltarho}), we need diagrams like Fig.~\ref{vvxy} with $V=W_+$. The $X,Y$ pairs are 
\begin{center}
\begin{tabular}{|c|c|c|c|}
\hline
\textbf{$X$} & \textbf{$Y$}&\textbf{Interaction term}&\textbf{Feynman Rule for the vertices}\\
\hline
$H_+$ & $S$          & $\frac i2gW_+^\alpha H_-\der_\alpha S$ & $\frac{g^2}4(2p+q)_\mu(2p+q)_\nu$ \\
$H_+$ & $A_0$ & $\frac g2\frac{v_\phi}\vf W_+^\alpha A_0\der_\alpha H_--$ & $-\frac{g^2}4\frac{v_\phi^2}\vvf(2p+q)_\mu(2p+q)_\nu$ \\
$H_+$ & $a_\phi$        & $-g\frac{v\sin2\beta}\vf W_+^\alpha H_-\partial_\alpha a_\phi $ & $-g^2\frac{v^2\sin^22\beta}{\vvf}(p+q)_\mu(p+q)_\nu$ \\
$H$   & $G_+$        & $-gW^\alpha_+H\partial_\alpha G_-$ & $-g^2(p+q)_\mu(p+q)_\nu$ \\
$G_+$ & $G_0$        & $\frac i2gW_+^\alpha G_0\der_\alpha G_- $& $\frac{g^2}4(2p+q)_\mu(2p+q)_\nu$ \\
\hline
\end{tabular}
\end{center}
\graphicspath{{appendices/figures/}{figures/}}

\chapter{Technical aspects of Chapters~\ref{ch-photon} and~\ref{ch-cosmic}}\label{app:photon}

\section{Projectors and polarisation vectors}\label{sec:app:proj}

Consider QED in the presence of a constant vector $\eta^\mu$, which can bue due to a cold axion background, as argued in Chs.~\ref{ch-photon} and~\ref{ch-cosmic},
and with an effective photon mass
\be
\mathcal{L}=-\frac14F_{\mu\nu}F^{\mu\nu}+\frac12m_\gamma^2A_\mu A^\mu+\frac12\eta_\mu A_\nu\tilde F^{\mu\nu}.
\ee
The equations of motion are, in momentum space,
\be\label{eqA}
\left[g^{\mu\nu}(k^2-m_\gamma^2)+i\epsilon^{\mu\nu\alpha\beta}\eta_\alpha k_\beta\right]\equiv K^{\mu\nu}\tilde A_\nu(k)=0.
\ee
We now define
\be\label{S}
S^{\mu}_{~\nu}\equiv\epsilon^{\lambda\mu\alpha\beta}\eta_{\alpha}k_\beta\epsilon_{\lambda\nu\rho\sigma}\eta^\rho k^\sigma.
\ee
Using the contraction property $\epsilon_{\mu\lambda\rho\sigma}\epsilon^{\mu\nu\alpha\beta}=-3!g_{[\lambda}^\nu g_\rho^\alpha g_{\sigma]}^\beta$ it can be written
more conveniently as
\be
S^{\mu\nu}=\left[\left(\eta\cdot k\right)^2-\eta^2 k^2\right]g^{\mu\nu}
- \left(\eta\cdot k\right)\left(\eta^\mu k^\nu + k^\mu\eta^\nu \right)
+ k^2\eta^\mu\eta^\nu + \eta^2 k^\mu k^\nu.
\ee
It has the following properties
\be
S^\mu_{~\nu}\eta^\nu=S^\mu_{~\nu}k^\nu=0,\quad S=S^\mu_{~\mu}=2\left[\left(\eta\cdot k\right)^2-\eta^2 k^2\right],\quad
S^{\mu\nu}S_{\nu\lambda}=\frac S2S^\mu_{~\lambda}.
\ee
For the case in which the constant vector only has a time component, $\eta_\mu=(\eta,0,0,0)$, we have $S=2\eta^2\vec k^2$. 
This is the interesting case for a cold axion background.

With the help of $S^{\mu}_{~\nu}$ we can define two projectors
\be\label{p+-}
P^{\mu\nu}_\pm=\frac{S^{\mu\nu}}S\mp\frac i{\sqrt{2S}}\epsilon^{\mu\nu\alpha\beta}\eta_\alpha k_\beta,
\ee
which have the following properties:
\begin{align}\label{properties}
P^{\mu\nu}_{\pm}\eta_{\nu}=P^{\mu\nu}_{\pm}k_{\nu}=0,\quad
g_{\mu\nu}P^{\mu\nu}_{\pm}=1,\quad  (P_\pm^{\mu\nu})^*=P_\mp^{\mu\nu}=P_\pm^{\nu\mu} ,\cr
P^{\mu\lambda}_{\pm}P_{\pm\lambda\nu}=P^{\mu}_{\pm\nu},\quad P^{\mu\lambda}_{\pm}\,P_{\mp\lambda\nu} = 0,\quad P^{\mu\nu}_{+} + P^{\mu\nu}_{-}= \frac{2}{S}S^{\mu\nu}.
\end{align}
With these projectors, we can build  a pair of polarisation vectors to solve \eqref{eqA}. We start from a space-like unit vector, for example $\epsilon=(0,1,1,1)/\sqrt3$. Then, we project it:
\be
\tilde\varepsilon^\mu=P_\pm^{\mu\nu}\epsilon_\nu.
\ee
In order to get a normalized vector, we need
\begin{align}
(\tilde\varepsilon^\mu_\pm)^*\tilde\varepsilon_{\pm\mu}&=P_\pm^{\nu\mu}\epsilon_\nu P_{\pm\mu\lambda}\epsilon^\lambda=P^\nu_{\pm\lambda}\epsilon_\nu\epsilon^\lambda=\frac{S^{\nu\lambda}\epsilon_\nu\epsilon_\lambda}S\cr
&=\frac{S/2\epsilon^\mu\epsilon_\mu+\eta^2(\epsilon\cdot k)^2}S
=-\frac12+\frac{(\epsilon\cdot k)^2}{2\vec k^2}
\end{align}
(this is of course negative because $\epsilon$ is space-like).
Then, the polarisation vectors are
\be\label{polvec}
\varepsilon^\mu_\pm=\frac{\tilde\varepsilon^\mu_\pm}{\sqrt{-\tilde\varepsilon^\nu_\pm\tilde\varepsilon^*_{\pm\nu}}}=
\left[\frac{\vec k^2-(\epsilon\cdot k)^2}{2\vec k^2}\right]^{-1/2}P_\pm^{\mu\nu}\epsilon_\nu.
\ee
These polarisation vectors satisfy
\be\label{orto}
g_{\mu\nu}\varepsilon_\pm^{\mu*}\varepsilon_\pm^\nu=-1,\quad g_{\mu\nu}\varepsilon_\pm^{\mu*}\varepsilon_\mp^\nu=0
\ee
and
\be\label{closure}
\varepsilon_\pm^{\mu*}\varepsilon_\pm^\nu+\varepsilon_\pm^\mu\varepsilon_\pm^{\nu*}=-\frac2SS^{\mu\nu}=-\frac{S^{\mu\nu}}{\eta^2\vec k^2}
\ee
With the aid of the projectors, we can write the tensor in \eqref{eqA} as
\be
K^{\mu\nu}=g^{\mu\nu}(k^2-m_\gamma^2)+\sqrt{\frac S2}\left(P_-^{\mu\nu}-P_+^{\mu\nu}\right).
\ee
Then we have for $k=(\omega_\pm,\vec k)$
\begin{align}
K^\mu_\nu\varepsilon^\nu_\pm&=\left[(k^2-m_\gamma^2)\mp\sqrt{\frac S2}\right]\varepsilon^\nu_\pm=\left(k^2-m_\gamma^2\mp\eta|\vec k|\right)\varepsilon^\mu_\pm\cr
&=\left(\omega_\pm^2-\vec k^2-m_\gamma^2\mp\eta|\vec k|\right)\varepsilon^\mu_\pm.
\end{align}
Therefore, $\tilde A^\mu=\varepsilon_\pm^\mu$ is a solution of \eqref{eqA} if and only if
\be
\omega_\pm(\vec k)=\sqrt{m_\gamma^2\pm\eta|\vec k|+\vec k^2}.
\ee

\section{Propagator}\label{sec:app_prop}
Considering only the spatial components, Eq. \eqref{propagator} becomes:
  \begin{align}
  \mathcal{D}^{ij}(\omega,k)={}&D^{ij}+i\omega^2\Bigg\{\frac{b^ib^j}{(k^4-\eta_0^2\vec k^2)(k^2-m_a^2)-\omega^2k^2b^2}\cr
  &+\frac{i\eta_0k^2(b^iq^j-q^ib^j)}{(k^4-\eta_0^2\vec k^2)[(k^4-\eta_0^2\vec k^2)(k^2-m_a^2)-\omega^2k^2b^2]}\Bigg\},
  \end{align}
  where
  \begin{equation}
  D^{ij}=-i\left(\frac{P_+^{ij}}{k^2-\eta_0|\vec k|}+\frac{P_-^{ij}}{k^2+\eta_0|\vec k|}\right),\quad \vec q=(\vec b\times\vec k)
  \end{equation}
and the projectors $P_\pm$ have been defined in Eq.~\eqref{p+-}.
Terms proportional to $k^ik^j$ have been dropped, since we are interested in contracting the propagator with a photon polarisation vector.
The roots of the denominators are $|\vec k|=F_j$, with
\begin{align}
F^2_{1,2}&=\omega^2+\frac{\eta_0^2}2\mp\frac{\eta_0}2\sqrt{4\omega^2+\eta_0^2}\approx\omega^2\mp\omega\eta_0,\cr
F^2_{3,4}&=\omega^2-\frac{m_a^2-\eta_0^2}{3}+\sqrt{W}(\cos\chi\mp\sqrt3\sin\chi),\cr
F^2_5&=\omega^2-\frac{m_a^2-\eta_0^2}{3}-2\sqrt{W}\cos\chi.
\end{align}
\begin{align}
W&\approx\left(\frac{m_a^2}3\right)^2\left(1+\frac{3\omega^2b^2}{m_a^4}\right),\cr
\chi&\approx\frac{1}{m_a^2}\sqrt3\omega\xi,\cr
\xi&\approx\left(1+\frac{9\omega^2b^2}{2m_a^4}\right)^{-1}\sqrt{\eta_0^2+
\left(\frac{\omega b^2}{2m_a^2}\right)^2+\left(\frac{\omega^2 b^3}{m_a^4}\right)^2}.
\end{align}
$F_1$ and $F_2$ correspond to the pieces with $P_+$ and $P_-$, respectively.
The piece proportional to $b^ib^j$ has poles at $F^2_{3,4,5}$ and the last piece contains all five poles.
We decompose the denominators in simple fractions:
\begin{equation}
\frac1{(k^4-\eta_0^2\vec k^2)(k^2-m_a^2)-\omega^2k^2b^2}=\sum_{l=3}^5\frac{A_l}{\vec k^2-F_l^2},
\end{equation}
with
\begin{equation}
A_l=\frac{-1}{\prod_{m\neq l,1,2}(F_l^2-F^2_m)},\quad l=3,4,5
\end{equation}
and
\begin{equation}
\frac{k^2}{(k^4-\eta_0^2\vec k^2)[(k^4-\eta_0^2\vec k^2)(k^2-m_a^2)-\omega^2k^2b^2]}=\sum_{l=1}^5\frac{\tilde A_l}{\vec k^2-F_l^2},
\end{equation}
with
\begin{equation}
\tilde A_l=\frac{-(\omega^2-F_l^2)}{\prod_{m\neq l}(F_l^2-F^2_m)},\quad l=1,...,5.
\end{equation}
Then,
\begin{align}
\mathcal{D}^{ij}(\omega,\vec k)={}&i\left(\frac{P_+^{ij}}{\vec k^2-F^2_1}+\frac{P_-^{ij}}{\vec k^2-F^2_2}\right)\cr
&+i\omega^2b^2\left[\hat b^i\hat b^j\sum_{l=3}^5\frac{A_l}{\vec k^2-F^2_l}+
i\eta_0(\hat b^i\hat q^j-\hat q^i\hat b^j)\sum_{l=1}^5\frac{|\vec k|\tilde A_l}{\vec k^2-F^2_l}\right]
\end{align}
We choose the axes so that
\begin{equation}
\hat k=(1,0,0),~\hat b=(0,1,0),~\hat q=(0,0,-1).
\end{equation}
The propagator in position space is, after dropping an overall factor,
\begin{align}
d^{ij}(\omega,x)\approx{}&(P_+^{ij}+P_-^{ij})\cos\left(\frac{\eta_0x}2\right)+i(P_+^{ij}-P_-^{ij})\sin\left(\frac{\eta_0x}2\right)\cr
&+\hat b^i\hat b^j\sum_{l=3}^5a_le^{i\alpha_lx}-i(\hat b^i\hat q^j-\hat q^i\hat b^j)\sum_{l=1}^5\tilde a_le^{i\alpha_lx},
\end{align}
where
\begin{equation}
a_l=\frac{\omega^3b^2A_l}{F_l},\quad\tilde a_l=\omega^3b^2\eta_0\tilde A_l,\quad\alpha_l=F_l-\omega.
\end{equation}
All the $\alpha_l$ are proportional to $\eta_0$ or $b^2$, except for $\alpha_5\approx-\frac{m_a^2}{2\omega}$.
Restricting ourselves only to $y-z$ components, we can write $d(\omega,x)$ in matrix form.
\begin{equation}
P^i_{+j}+P^i_{-j}=
\left(
  \begin{array}{cc}
    1 & 0 \\
    0 & 1 \\
  \end{array}
\right),
\end{equation}
\begin{equation}
i(P^i_{+j}-P^i_{-j})=
\left(
  \begin{array}{cc}
    0 & 1 \\
    -1 & 0 \\
  \end{array}
\right),
\end{equation}
\begin{equation}
\hat b^i\hat b_j=
\left(
  \begin{array}{cc}
    -1 & 0 \\
    0 & 0 \\
  \end{array}
\right),
\end{equation}
\begin{equation}
-i(\hat b^i\hat q_j-\hat q^i\hat b_j)=
\left(
  \begin{array}{cc}
    0 & -i \\
    i & 0 \\
  \end{array}
\right).
\end{equation}
If we write
\begin{equation}
\sum_l a_le^{i\alpha_lx}=-(\epsilon+i\varphi),\quad i\sum_l \tilde a_le^{i\alpha_lx}=-(\tilde\epsilon+i\tilde\varphi),
\end{equation}
we have
\begin{equation}\label{propmatrix}
d^i_j(\omega,x)=
\left(
  \begin{array}{cc}
    \cos\frac{\eta_0x}{2}+\epsilon+i\varphi & \sin\frac{\eta_0x}{2}+\tilde\epsilon+i\tilde\varphi \\
    -\left(\sin\frac{\eta_0x}{2}+\tilde\epsilon+i\tilde\varphi\right) & \cos\frac{\eta_0x}{2}\\
  \end{array}
\right)
\end{equation}

\section{Ellipticity and rotation}
The quantities appearing in Eq.~\eqref{propmatrix} are
\be
\epsilon\approx-\frac{\omega^2b^2}{m_a^4}\left(1-\cos\frac{m_a^2x}{2\omega}\right),\quad
\varphi\approx\frac{\omega^2b^2}{m_a^4}\left(\frac{m_a^2x}{2\omega}-\sin\frac{m_a^2x}{2\omega}\right),
\ee
while $\tilde\epsilon$ and $\tilde\varphi$ are both proportional to $\eta_0b^2$, so they are negligible.\\
In the limit $\frac{m_a^2x}{2\omega}\ll1$ we have
\begin{equation}\label{lim1}
\epsilon\approx -\frac{b^2x^2}8,\quad\varphi\approx \frac{m_a^2b^2x^3}{48\omega}
\end{equation}
whereas if $\frac{m_a^2x}{2\omega}\gg1$ the trigonometric functions oscillate rapidly and can be dropped:
\begin{equation}\label{lim2}
\epsilon\approx-\frac{\omega^2b^2}{m_a^4},\quad\varphi\approx\frac{\omega b^2x}{2m_a^2}.
\end{equation}
Equation~(\ref{lim1}) agrees with Eq.~16 of Ref.~\cite{mpz}  (although their $k^2$ in the denominator should be only $k$,
the dimensions do not fit otherwise).
Equation~(\ref{lim2}) agrees with their Eq.~(20,21), at least to second order in $b$.

If we start with a polarisation $\vec n_0=(\cos\beta,\sin\beta)$, after a distance $x$ we have
\begin{equation}
n^i_x=d^i_j(x)n^j_0=\left(\begin{array}{c}
\cos(\beta-\frac{\eta_0x}{2})+(\epsilon+i\varphi)\cos\beta\\
\sin(\beta-\frac{\eta_0x}{2})
                                  \end{array}\right)
\end{equation}
Following Section~1.4 of Ref.~\cite{bornwolf}, this vector describes a polarisation at an angle
\begin{equation}
\alpha\approx\beta-\frac{\eta_0x}{2}-\frac\epsilon2\sin2\beta
\end{equation}
and with ellipticity
\begin{equation}
e=\frac12\left|\varphi\sin2\beta\right|.
\end{equation}
This ellipticity differs from the one in Ref.~\cite{mpz} by the factor of $\sin2\beta$.

Quantum mechanically the quantity that is relevant is not the amplitude itself, but the modulus
squared of it. From this, the
probability of finding an angle $\alpha$ given an initial angle $\beta$ will be
\begin{equation}
P(\alpha,\beta)=\left|\epsilon'_id^{ij}\epsilon_j\right|^2\approx\cos^2\left(\alpha-\beta+\frac{\eta_0x}{2}\right)
+2\epsilon\cos\left(\alpha-\beta+\frac{\eta_0x}{2}\right)\cos\alpha\cos\beta.
\end{equation}
The angle of maximum probability, satisfying $\partial_\alpha P(\alpha,\beta)=0$ is also, to first order,
\be
\alpha=\beta-\frac{\eta_0x}{2}-\frac\epsilon2\sin2\beta.
\ee

\end{appendices}

\begin{spacing}{0.9}

\bibliographystyle{newutphys}
   \cleardoublepage
\bibliography{biblio/biblio}

\end{spacing}

\end{document}